\documentclass[preprint,aps,superscriptaddress,nofootinbib,tightenlines,floatfix]{revtex4}

\usepackage{epsfig}
\usepackage{bm}
\def\OMIT#1{{}}

\newcommand{\beq}{\begin{equation}}
\newcommand{\eeq}{\end{equation}}
\newcommand{\bea}{\begin{eqnarray}}
\newcommand{\eea}{\end{eqnarray}}

\newcommand{\benn}{\begin{displaymath}}
\newcommand{\eenn}{\end{displaymath}}
\newcommand{\gsim}{\raisebox{-0.7ex}{$\stackrel{\textstyle >}{\sim}$ }}

\begin{document}

\begin{figure}[!t]
\vskip -1.5cm
\leftline{
{\epsfxsize=1.5in \epsfbox{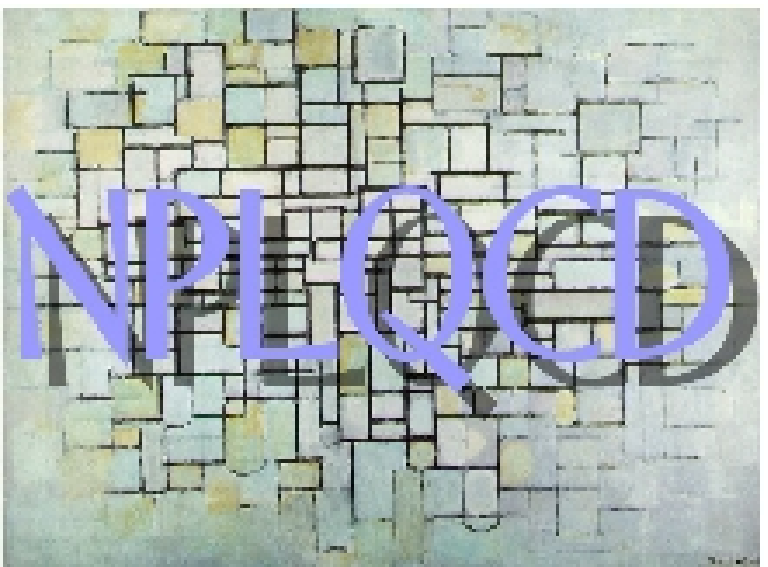}}}
\end{figure}

\preprint{\vbox{
\hbox{NT@UW-07-05}
\hbox{JLAB-THY-07-625}
}}

\vphantom{}
\title{\bf \LARGE 
$BB$ Potentials in Quenched Lattice QCD
}
\author{William Detmold}
\affiliation{Department of Physics, University of Washington, 
Seattle, WA 98195-1560.}
\author{Kostas Orginos}
\affiliation{Jefferson Laboratory, 12000 Jefferson Avenue, 
Newport News, VA 23606.}
\affiliation{Department of Physics, College of William and Mary, Williamsburg,
  VA 23187-8795.}
\author{Martin J.~Savage}
\affiliation{Department of Physics, University of Washington, 
Seattle, WA 98195-1560.}
\collaboration{ NPLQCD Collaboration }
\noaffiliation
\vphantom{}
\vskip 0.8cm
\begin{abstract} 
\vskip 0.5cm
\noindent  
The potentials between two $B$-mesons are computed in the heavy-quark
limit using quenched lattice QCD at $m_\pi\sim 400~{\rm MeV}$.
Non-zero central potentials are clearly evident in all four
spin-isospin channels, $(I,s_l) = (0,0) , (0,1) , (1,0) , (1,1)$,
where $s_l$ is the total spin of the light degrees of freedom.  At
short distance, we find repulsion in the $I\ne s_l$ channels and
attraction in the $I=s_l$ channels.  Linear combinations of these
potentials that have well-defined spin and isospin in the $t$-channel
are found, in three of the four cases, to have substantially smaller
uncertainties than the potentials defined with the $s$-channel
$(I,s_l)$, and allow quenching artifacts from single hairpin exchange
to be isolated.  The $BB^\ast\pi$ coupling extracted from the
long-distance behavior of the finite-volume $t$-channel potential is
found to be consistent with quenched calculations of the matrix
element of the isovector axial-current.  The tensor potentials in both
of the $s_l = 1$ channels are found to be consistent with zero within
calculational uncertainties.

\end{abstract}
\maketitle

%\tableofcontents
\vfill\eject

%%%%%%%%%%%%%%%%%%%%%%%%%%%%%%%%%%%%%%%%%%%%%%%
\section{Introduction}

Nuclei and nuclear processes can be described with remarkable
precision by treating the nucleons as non-relativistic particles
interacting via a local potential.  The wealth of nucleon-nucleon (NN)
scattering data has enabled the construction of precise
phenomenological potentials, defined up to unitary transformations,
which are used to calculate the spectra of the light nuclei.  As
expected from QCD, the two-body potentials alone are insufficient to
reproduce the spectra, but when supplemented with three- and four-
(and higher) body interactions successfully reproduce the structure of
light nuclei \cite{Pieper:2001mp,Navratil:2007we}.  During the last 15
years or so, immense effort has been put into developing an effective
field theory (EFT) to describe the interactions between nucleons
\cite{Beane:2000fx,Bedaque:2002mn}, and to allow for a systematic
improvement of nuclear physics phenomenology using the approximate
chiral symmetry of QCD.  The values of the counterterms that appear at
a given order in the EFT expansion have to be obtained from experiment
or lattice QCD.  In situations where experiments are not possible,
lattice QCD is the only rigorous calculational technique with which to
determine such counterterms.  Recently, NN
scattering~\cite{Beane:2006mx} and hyperon-nucleon
scattering~\cite{Beane:2006gf} have been calculated with
fully-dynamical lattice QCD\footnote{Recently, it has been claimed
  that the nucleon-nucleon potential has been calculated with quenched
  lattice QCD.  We believe that the arguments used to extract the
  potential are flawed, and discuss this further in
  Appendix~\ref{sec:ill-def-pot}.} by measuring the finite-volume
shifts of two particle energies \cite{Luscher:1986pf,Luscher:1990ux}.
Due to limited computational resources, the calculations were
performed at unphysical pion masses, $m_\pi\gsim 350~{\rm MeV}$, at
the upper limits of the range of applicability of the EFTs.  Until the
computational resources for lattice QCD calculations are significantly
greater than presently available, it will not be possible to calculate
NN scattering parameters at a large number of different energies and
then construct NN potentials in the same way that NN cross-section
measurements are processed.  However, in addition to extracting the NN
phase-shifts at various moment, one can hope to learn qualitative
information about the EFT describing NN interactions by performing
lattice QCD calculations of systems that are similar.

In this work we study the potential between two $B^{(\ast)}$ mesons in
the heavy-quark 
limit~\cite{Isgur:1989vq,Isgur:1989ed,Eichten:1989zv,Georgi:1990um}~\footnote{The
  $B$ and $B^\ast$ mesons are
  degenerate in the heavy-quark limit and henceforth we will use
  $B$-meson to denote the $B$ and $B^\ast$ super-multiplet.}, a limit
in which the potential is a well-defined object.  $B$-mesons are
isospin-${1\over 2}$ hadrons, and in the heavy-quark limit the spin of
the light degrees of freedom (ldof) becomes a good quantum number,
$s_l={1\over 2}$, as spin-dependent interactions with the heavy-quark
are suppressed by $1/m_b$.  At distances which are large compared with
the chiral symmetry breaking scale, $\Lambda_\chi$, the EFT describing
the interactions between two $B$-mesons is the same as that between
two nucleons as the isospin-spin quantum numbers are the same.  The
differences between the two EFTs are in the values of the
counterterms.  At distances that are of order, or shorter than
$\Lambda_\chi$ the interactions between two $B$-mesons will be
arbitrarily different from that between two-nucleons as the structure
of the hadrons are very different, and in particular that
strong-Coulomb interaction between the heavy-quarks becomes dominant,
behaving as $\sim \alpha(r^{-1})/r$ as the separation, $r\rightarrow
0$.

In addition to providing insight into the NN potential, the potentials
between B-mesons are interesting in their own right.  A precise
determination of the potentials between two B-mesons will allow for
investigations of possible shallow bound states.  These would be
molecular tetra-quark states, similar to the deuteron in the
NN-sector.  The location of such molecular states would be very
sensitive to the potentials (due to the fine-tunings) and as such,
quenched calculations at the unphysical pion mass would in general
provide unreliable results.

Lattice calculations of the potentials between two $B$-mesons in the
heavy-quark limit have been performed
previously~\cite{Richards:1990xf, Mihaly:1996ue, Stewart:1998hk,
  Michael:1999nq, Pennanen:1999xi, Green:1999mf, Fiebig:2001mr,
  Fiebig:2001nn, Cook:2002am, Takahashi:2006er, Doi:2006kx}.  However,
given the large statistical uncertainties in those calculations, the
potentials remain largely unexplored.  We have chosen to work in a
relatively small lattice volume in order to explore the intermediate
and short-distance components of the potential, but this is at the
expense of having contributions from image pairs somewhat mask the
long-distance component.  Further, we have attempted to extract the
tensor potentials in the $s_l=1$ channels, but have not found results
that are statistically different from zero.  Due to limited
computational resources, we have performed quenched calculations.  It
is important to stress that the long-distance component of the
potential computed in quenched QCD is polluted by the presence of
``one-hairpin-exchange'' (OHE), as discussed in
Ref.~\cite{Beane:2002nu}, which becomes dominant at large distances
due to its exponential fall-off, as opposed to the Yukawa-type
behavior of one pion exchange (OPE).  However, the OHE contribution
can be isolated by defining potentials with well-defined $t$-channel
spin-isospin quantum numbers. In three of the $t$-channel potentials,
quenching artifacts are expected to be higher order in the quenched
EFT expansion. Using these finite-volume $t$-channel potentials, we
are able to investigate the long range part of the infinite-volume
potential and extract a $BB^\ast\pi$ coupling consistent with that
measured in quenched lattice calculations of axial matrix elements.

The outline of our paper is as follows. In Section \ref{sec:lattice},
we discuss the numerical implementation of our calculation. Sections
\ref{sec:hadron-spectrum} and \ref{sec:exotics} present our single
particle results for the heavy hadron spectrum, including exotic
states. Section \ref{sec:potentials} presents the main results of our
work, the $BB$ potentials in the various channels. These results are
then discussed in Section~\ref{sec:resdisc}. In Appendix
\ref{sec:ill-def-pot}, we discuss the issue of extraction of nucleon
potentials from lattice QCD wavefunctions, and in Appendix \ref{ap:A}
we provide details of the perturbative lattice calculations needed in
this work.

%%%%%%%%%%%%%%%%%%%%%%%%%%%%%%%%%%%%%%%%%%%%%%%
\section{Details of the Lattice Calculation}
\label{sec:lattice}

Our calculations were performed using 284 quenched configurations of
dimension $16^3\times 32$ generated with the DBW2 action
\cite{Takaishi:1996xj,deForcrand:1999bi} for $\beta=1.04$, giving a
lattice spacing of $b=0.0997\pm 0.0015$~\cite{Aoki:2002vt}. On each of
these configurations, eight Wilson light-quark propagators, equally
spaced in the time-direction and offset in space, were generated from
smeared sources to determine the light hadron spectrum.  The
light-quark mass selected gave rise to a pion mass of $m_\pi =
402.5\pm 6.7~{\rm MeV}$, and the other hadron masses that are shown in
Table.~\ref{tab:single}.  The finite lattice spacing and finite-volume
effects~\cite{Colangelo:2003hf} have not been removed from these
masses and are expected to be a few percent.
\begin{table}[!h]
\begin{ruledtabular}
\begin{tabular}{ccccc}
&Quantity  &  $b \ M $  & $M$ [MeV] \\
\hline 
&$m_\pi$  & $0.2034\pm 0.0015$ & $402.5\pm 6.7$ \\
&$m_\rho$  & $0.3754\pm 0.0080$ & $743\pm 19$ \\
&$m_N$  & $0.5756\pm 0.0080$ & $1139\pm 23$ \\
&$m_\Delta$  & $0.6770\pm 0.0095$ & $1340\pm 28$ \\
&$m_\Delta-M_N$ & $0.102\pm 0.014$ & $201\pm 27$ \\
\end{tabular}
\end{ruledtabular}
 \caption{The masses of the light hadrons at finite lattice spacing and finite volume.}
\label{tab:single}
\end{table}

In order to compute the potential between two $B$-mesons separated by
lattice vectors ${\bf r}=b(n,0,0),\,b(0,0,n)$ for $n=0,\ldots,8$ and
additionally by ${\bf r}=b(1,1,0),\, b(2,1,0)$, Wilson light-quark
propagators were generated from smeared sources on one time-slice
located at each point in the $x$-$z$ plane on two adjacent
spatial-slices in the y-direction on each gauge configuration.
Therefore, a total of $\sim 1.4\times 10^5$ light-quark propagators
were generated.  This choice of lattice separation vectors was
dictated by the available computational time, and not by physics.  The
calculations were performed on a 16-node dual-Xeon cluster and a
number of workstations. The total computational cost of this work was
$\sim$~40 Gflop-yrs.

In the heavy-quark limit, the heavy-quark propagator is the tensor
product of a Wilson line and a positive-energy projector,
\begin{eqnarray}
  \label{eq:heavyprop}
  S_Q({\bf x},t;t_0)=\left(\frac{1+\gamma_4}{2}\right)\prod_{t^\prime=t_0}^{t} 
U_4({\bf x},t^\prime)\,,
\end{eqnarray}
where $U_\mu(x)$ are the gauge link variables and the product is
time-ordered. Our Dirac matrices use the Euclidean Dirac convention.
The light quark propagator, $S({\bf x},t;{\bf y},t_0)$, is generated
with the unimproved Wilson action, thereby introducing ${\cal O}(b)$
discretisation errors.  It is generated from a gauge-invariant
Gaussian smeared-source.

To determine the single particle energies of the heavy hadrons, the
correlators
\begin{eqnarray}
  \label{eq:5}
  C_B(t,t_0) &=& \sum_{{\bf x}} {\rm tr}\left[S_Q({\bf x},t;t_0)
   S^\dagger({\bf x},t;{\bf x},t_0) \right]
\ H({\bf x})\,,
\\
  C_{\Sigma_b/\Lambda_b}(t,t_0) &=& \sum_{{\bf x}} 
    S^{k^\prime k}_{Q;\sigma\rho}({\bf x},t;t_0)\epsilon^{ijk}
    \epsilon^{i^\prime j^\prime k^\prime}
    \left(S^{i i^\prime}({\bf x},t;{\bf x},t_0) \Gamma\right)_{\rho\alpha} 
    \left(\Gamma S^{j j^\prime}({\bf x},t;{\bf x},t_0)\right)_{\sigma\alpha}  
H({\bf x})
\, ,
\nonumber\\
\end{eqnarray}
were computed, where the Dirac matrices are $\Gamma = C \gamma_5,\, C
\gamma_i$ for the $\Lambda_b$ and $\Sigma_b^i$ respectively.  The
trace is over color and spinor indices, and the function $H({\bf x})$
is unity if a heavy-quark source was placed at the point ${\bf x}$,
and vanishes elsewhere.  In the baryon correlators, upper indices
label color and lower Greek indices label spin.

\begin{figure}[!t]
  \centering
  \includegraphics[width=0.35\columnwidth]{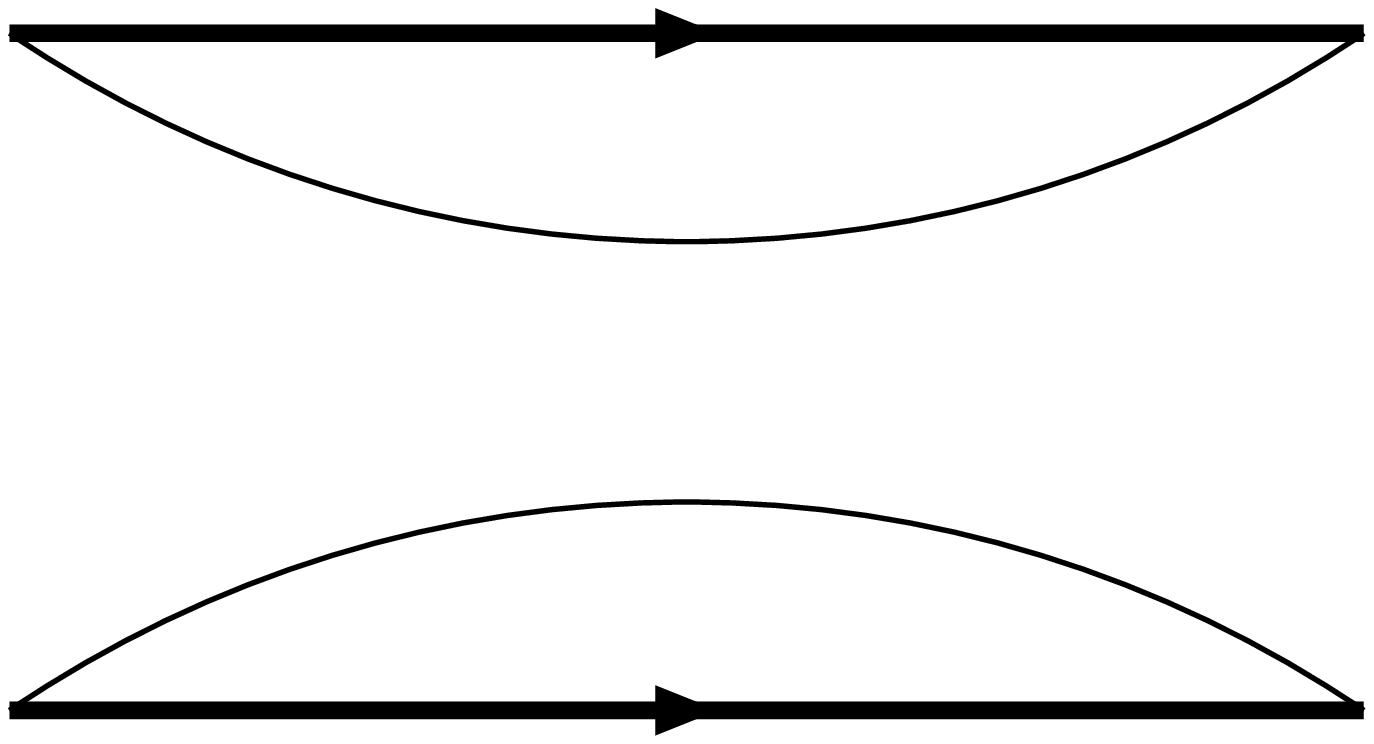}
\hspace{8mm}
  \includegraphics[width=0.35\columnwidth]{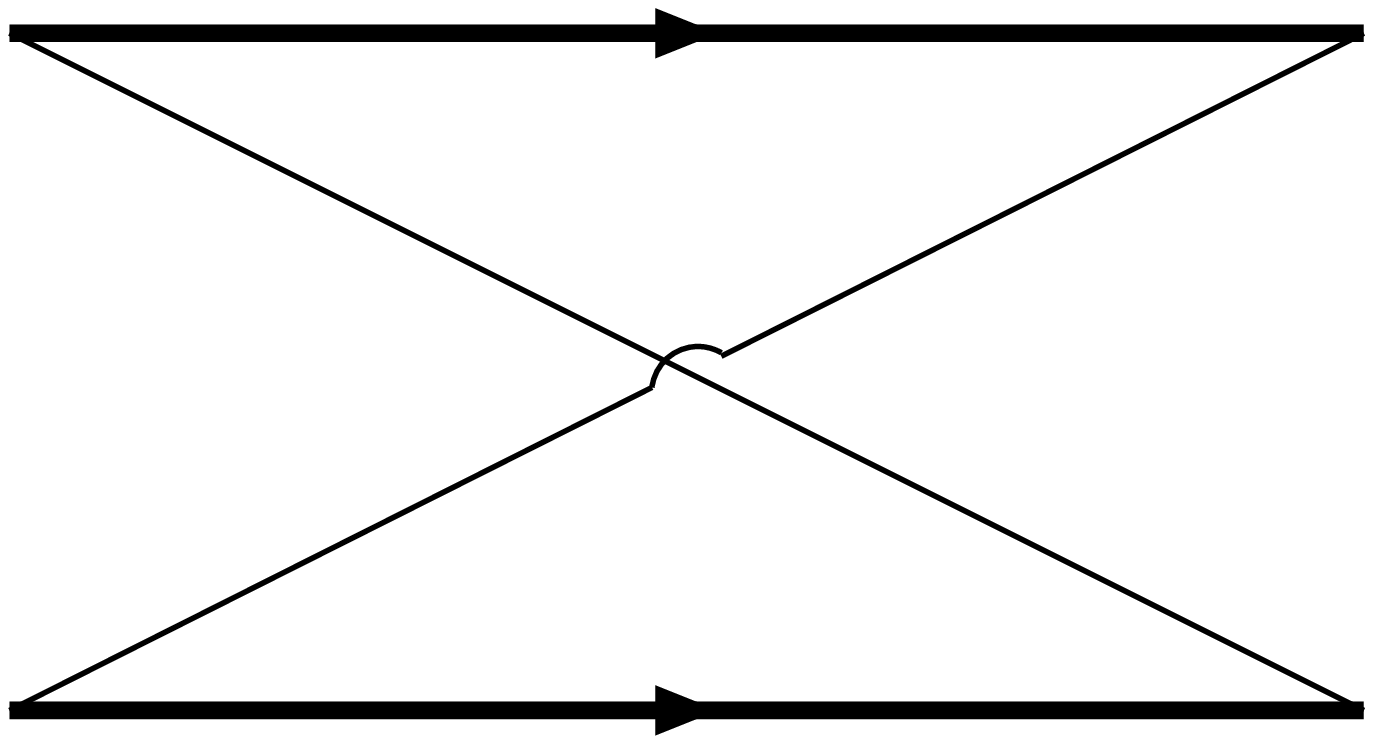}\\
(a)\hspace{6.3cm}(b)
  \caption{Disconnected and connected quark contractions contributing
    to the correlation functions in
    Eqs.~\protect{(\ref{eq:4})}--\protect{(\ref{eq:4444})}. Heavy and
    light lines correspond to the heavy- and light-quark propagators
    respectively.}
  \label{fig:corrfunc}
\end{figure}

In order to measure the potential, we computed the correlators
$C_{I,s_l}(t,t_0,\tilde{\vec{r}})$ given by
\begin{eqnarray}
  \label{eq:4}
  C_{0,0}(t,t_0,\tilde{{\bf r}}) &=& \sum_{\bf x} 
  \left[A_0^{(1)}-A_0^{(2)} \right]\ H({\bf x})\,,
\\
  C_{1,0}(t,t_0,\tilde{{\bf r}}) &=& \sum_{\bf x} 
  \left[A_0^{(1)}+A_0^{(2)} \right]\ H({\bf x})\,,
\\
  C_{0,1}(t,t_0,\tilde{{\bf r}}) &=& \frac{1}{2}\sum_{\bf x} 
  \left[A_+^{(1)} + A_-^{(1)} - A_+^{(2)} - A_-^{(2)} \right]\ H({\bf x})\,,
\\
\label{eq:4444}
  C_{1,1}(t,t_0,\tilde{{\bf r}}) &=& \frac{1}{2}\sum_{\bf x} 
  \left[A_+^{(1)} + A_-^{(1)} + A_+^{(2)} + A_-^{(2)} \right]\ H({\bf x})\,,
\end{eqnarray}
where 
\begin{eqnarray}
  \label{eq:6}
  A_0^{(1)} &=& 
{\rm tr_D}\Big[
{\rm tr_C}\left(S^\dagger({\bf x},t;{\bf x}+{\bf r},t_0) 
S_Q({\bf x},t;t_0)\right) \Gamma_0
\\&&\hspace*{3cm}\times
{\rm tr_C}\left[{\cal S}\left\{S^\dagger({\bf x}+{\bf r},t;{\bf x},t_0) 
S_Q({\bf x}+{\bf r},t;t_0)\right\}\right]^T
\Gamma_0^T
\Big]\,,
\nonumber
\\
  A_0^{(2)} &=& 
-{\rm tr}\Big[
{\cal S}\left\{S^\dagger({\bf x},t;{\bf x}+{\bf r},t_0)
S_Q({\bf x}+{\bf r},t;t_0)\right\}
\\&&\hspace*{3cm}\times
\Gamma_{0} 
\left[S^\dagger({\bf x}+{\bf r},t;{\bf x},t_0) \right]^T
S_Q({\bf x},t;t_0)
\Gamma_0
\Big]\,,\nonumber
\\
  A_\pm^{(1)} &=& {\rm tr}\left[
\Gamma_\pm S_Q({\bf x},t;t_0)
S^\dagger({\bf x},t;{\bf x},t_0)\right]
\\&&\hspace*{3cm}\times
{\rm tr}\left[
\Gamma_\pm 
{\cal S}\left\{S^\dagger({\bf x}+{\bf r},t;{\bf x}+{\bf r},t_0) 
S_Q({\bf x}+{\bf r},t;t_0)\right\}
\right]\,, \nonumber
\\
  A_\pm^{(2)} &=& - {\rm tr}\left[
S_Q({\bf x},t;t_0)
{\cal S}\left\{S^\dagger({\bf x},t;{\bf x}+{\bf r},t_0)
S_Q({\bf x}+{\bf r},t;t_0) \right\}\Gamma_\pm 
S^\dagger({\bf x}+{\bf r},t;{\bf x},t_0) 
\Gamma_\pm
\right]\,,\hspace*{7mm}
\end{eqnarray}
$\Gamma_{0}=\frac{1}{2}(1+\gamma_4)\gamma_5$ and
$\Gamma_\pm=\frac{1}{2}(1\pm i\gamma_1\gamma_2) (1+\gamma_4)$ and
${\rm tr_C}$, ${\rm tr_D}$ and ${\rm tr}$ indicate traces over color,
spin and both.  Here ${\cal S}\{\ldots\}$ indicates translation of the
propagators by the lattice vector $-{\bf r}$ and the transpose $^T$
denotes spin transpose only.  The different contributions to the
various correlators, $A_{0,\pm}^{(1)}$ and $A_{0,\pm}^{(2)}$
correspond to the two contractions shown in Fig.~\ref{fig:corrfunc}.

For each correlator, we determine the ground state energy by seeking
plateaus in the ensemble jackknife average of the effective energy,
\begin{eqnarray}
  \label{eq:effE}
  b \ E_{I,s_l}(t-t_0)=\log\left[\frac{C_{I,s_l}(t-1,t_0)}
{C_{I,s_l}(t,t_0)}\right] 
\  .
\end{eqnarray}

%%%%%%%%%%%
\section{The Heavy Hadron Spectrum}
\label{sec:hadron-spectrum}

In the heavy quark limit, the mass of the $B$ meson is
\begin{eqnarray}
M_B & = & m_b \ +\ \overline{\Lambda}_{{1\over 2},{1\over 2}}\ +\ {\cal O}(1/m_b)
\ \ ,
\label{eq:MBexp}
\end{eqnarray}
where $m_b$ is the heavy quark mass and $\overline{\Lambda}_{I,s_l}$
denotes the energy of the ldof with total isospin $I$ and spin $s_l$.
To determine $\overline{\Lambda}_{{1\over 2},{1\over 2}}$ from lattice
calculations, the energy, ${\cal E}_{{1\over 2},{1\over 2}}$, of a
meson composed of a Wilson-line (static color source) and a light
anti-quark is computed using Eqs.~(\ref{eq:5}) and (\ref{eq:effE}).
This by itself does not isolate $\overline{\Lambda}_{{1\over
    2},{1\over 2}}$, as the interactions of the static source with the
gauge fields generate a residual mass~\cite{Falk:1992fm} for the
heavy-quark, $\delta m$, which while vanishing in dimensional
regularization, is non-zero on the lattice and scales as $1/b$ with
the lattice spacing \cite{Duncan:1994uq}.  Therefore, both ${\cal
  E}_{{1\over 2},{1\over 2}}$ and $\delta m$ diverge as $1/b$, but the
difference between them is finite in the continuum limit, and is
$\overline{\Lambda}_{{1\over 2},{1\over 2}}$,
\begin{eqnarray}
\overline{\Lambda}_{{1\over 2},{1\over 2}} 
& = & {\cal E}_{{1\over 2},{1\over 2}} - \delta m
\ \ \ ,
\label{eq:residual}
\end{eqnarray}
and more generally, $\overline{\Lambda}_{I,s_l}={\cal E}_{I,s_l} -
\delta m$.

The residual mass of the static source has been computed previously
out to the two-loop level in quenched lattice perturbation theory for
the Wilson action~\cite{Martinelli:1998vt}.  At the one loop level,
the residual mass for a gauge action, $f$, is given by
\begin{eqnarray}
\delta m_f^{(\alpha)} & = & 
{\overline{\alpha}(\mu)\over 3\pi^2 b}\ 
\int_{-\pi}^\pi d\tilde q_x \ 
\int_{-\pi}^\pi d\tilde q_y\ 
\int_{-\pi}^\pi d\tilde q_z \ 
G_{00}^{(f)}\left(\hat q_x, \hat q_y,\hat q_z,0\right)
\nonumber\\
& = & {\overline{\alpha}(\mu)\over b}\ {\cal S}^{(f)}
\ \ \ ,
\label{eq:deltamoneloop}
\end{eqnarray}
where $\hat q_\gamma = 2\sin \left(\tilde q_\gamma/2\right)$, and
$\tilde q = q b$ is dimensionless.  $G_{00}^{(f)}(\hat k_x, \hat k_y,
\hat k_z, \hat k_t)$ is the lattice gluon propagator for the
particular gauge action, $f$. This has the form
$G_{00}^{(Wilson)}(\hat k_x, \hat k_y, \hat k_z, \hat k_t) = 1/{\hat
  k}^2$ for the Wilson action, but is considerably more complicated
for improved actions, such as the
L\"uscher-Weisz (LW)~\cite{Luscher:1984xn,Luscher:1985zq}, 
DBW2~\cite{Takaishi:1996xj,deForcrand:1999bi}, and Iwasaki~\cite{Iwasaki}
actions. For these actions, the form of the propagator was presented
in Ref.~\cite{Weisz:1983bn} and involves an improvement coefficient,
$c_1$, with $c^{\rm LW}_1=-1/12$, $c^{\rm DBW2}_1 = -1.40686$,
and $c_1^{\rm Iwasaki}= -0.331$ (for a recent review of lattice
perturbation theory, see Ref.~\cite{Capitani:2002mp}).  The quantity
${\cal S}^{(f)}$ has been computed previously for the Wilson action,
${\cal S}^{(Wilson)}= 2.1173$, and for the improved actions we find
that ${\cal S}^{(DBW2)} = 0.6921$, ${\cal S}^{(LW)} = 1.8335$ and
${\cal S}^{(Iwasaki)}=1.3598$ (these numbers differ from those
presented in Ref.~\cite{Aoki:2002iq} where an incorrect improved gluon
propagator was used).  At the one-loop-level, the choice of scale is
not well-defined, but as the only scale in the lattice calculation is
the lattice spacing, it is convenient to use
$\mu=1/b$.\footnote{Perhaps a better estimate would be $\mu=\pi/b$ as
  that is the maximum momentum in the one-loop diagram.  The typical
  momenta in the one-loop diagram to be somewhat less than this.}
Therefore, at the one-loop level, and using $\overline{\alpha}(2~{\rm
  GeV}) = 0.299\pm 0.015$~\footnote{ The strong coupling on these DBW2
  lattices has been determined to be $\overline{\alpha}(b^{-1}) =
  0.154$ in the $\overline{MS}$-scheme~\cite{Aoki:2002iq},
  significantly smaller than the experimentally constrained value of
  $\overline{\alpha}(2~{\rm GeV}) = 0.299\pm 0.015$.  This suggest
  that the perturbative relation between the value of the plaquette
  and the strong coupling is only slowly convergent.  }, we have
$\delta m^{(\alpha)}_{\rm DBW2} = 410\pm 21~{\rm MeV}$ for the DBW2
action.

One can make an improved estimate of the residual mass term by using
the Brodsky-Lepage-Mackenzie (BLM) scale-setting
procedure~\cite{Brodsky:1982gc}, which includes the part of the
two-loop contribution arising from the running of the strong coupling
over the momenta in the one-loop diagram. This BLM improved residual
mass is given by\footnote{Our definition of the improved residual mass
  is different than that in Ref.~\cite{Duncan:1994uq} but agrees with
  Ref.~\cite{Martinelli:1998vt}.}
\begin{eqnarray}
\label{eq:BLM}
\delta m_f^{(\alpha+\alpha^2\beta)}
& = & 
{\overline{\alpha}(\mu)\over b}\ {\cal S}^{(f)}
\ -\ 
{\overline{\alpha}^2(\mu)\beta\over 4\pi b}\ {\cal T}^{(f)}(\mu)\,,
\end{eqnarray}
where
\begin{eqnarray}
{\cal T}^{(f)}(\mu)
& = & 
{1\over 3\pi^2}
\int_{-\pi}^\pi d\tilde q_x \ 
\int_{-\pi}^\pi d\tilde q_y\ 
\int_{-\pi}^\pi d\tilde q_z \ 
G_{00}^{(f)}\left(\hat q_x, \hat q_y,\hat q_z,0\right)
\log\left[ { \hat{q}_x^2+\hat{q}_y^2+\hat{q}_z^2\over \mu^2 b^2}
\right] 
\ \ \ ,
\label{eq:deltama2b}
\end{eqnarray}
and $\beta=11$ for quenched QCD. The $\overline{\alpha}^2\beta$ term
in Eq.~(\ref{eq:BLM}) can be perturbatively removed by defining the
BLM scale, $q^{\ast(f)}$, such that $\log \left(q^{*(f)} b\right) =
{\cal T}^{(f)}(b^{-1})/( 2 {\cal S}^{(f)})$, and therefore,
\begin{eqnarray}
\delta m_f^{(\alpha+\alpha^2\beta)}
& = & {\overline{\alpha}(q^{*(f)}) \over b}\ {\cal S}^{(f)}
\ \ \ .
\label{eq:deltamBLM}
\end{eqnarray}
For the Wilson action, we find ${\cal T}^{(Wilson)}(b^{-1}) = 1.562$
which produces a BLM scale $q^{*(Wilson)}=1.446/b$ that has previously
been shown to accurately estimate the full two-loop result for the
residual mass using the $\overline{\rm MS}$
coupling~\cite{Martinelli:1998vt}.\footnote{At finite lattice spacing,
  the definition of the BLM scale becomes ambiguous as the two loop
  contributions that the BLM procedure is attempting to resum become
  dependent on the details of the discretisation.  In particular the
  continuum $\log(|\vec{q}|^2)$ becomes a complicated function of the
  lattice momenta and improvement coefficients.  A full two-loop
  calculation will be required to determine the efficacy of our BLM
  estimate using $\overline{\alpha}(q^*)$ for the improved actions and
  {\it a priori} there is no reason to assume the agreement found for
  the unimproved Wilson action persists in these cases. As an
  indication of possible lattice artifacts in the definition of ${\cal
    T}$ we have also computed $\tilde{q}^{*(f)}$ by replacing
  $\log(|\hat{q}|^2) \to \log(|\vec{q}|^2)$ in
  Eq.~(\ref{eq:deltama2b}), finding $\tilde{q}^{*(Wilson)}=1.671/b$,
  $\tilde{q}^{*(DBW2)} = 0.918/b$, $\tilde{q}^{*(LW)}=1.552/b$,
  and $\tilde{q}^{*(Iwasaki)} = 1.318/b$.  Similar perturbative shifts
  in the BLM scale are induced by self-consistently evaluating ${\cal
    T}$ at the scale $q^{*(f)}$.}  For the improved actions, we find
that ${\cal T}^{(DBW2)}(b^{-1}) = -0.239$, which leads to $q^{*(DBW2)} =
0.841/b$, ${\cal T}^{(LW)}(b^{-1}) = 1.121$, which leads to
$q^{*(LW)} = 1.358/b$ and for the Iwasaki action, ${\cal
  T}^{(Iwasaki)}(b^{-1}) = 0.437$, which leads to $q^{*(Iwasaki)} =
1.174/b$.  Therefore, using $\overline{\alpha}(q^{*(DBW2)}) = 0.326\pm
0.018$, the BLM improved estimate of the residual mass for our DBW2
lattices is $\delta m^{(\alpha+\alpha^2\beta)}_{\rm DBW2} = 447\pm
25~{\rm MeV}$.

\begin{figure}[!t]
  \centering
  \includegraphics[width=0.95\columnwidth]{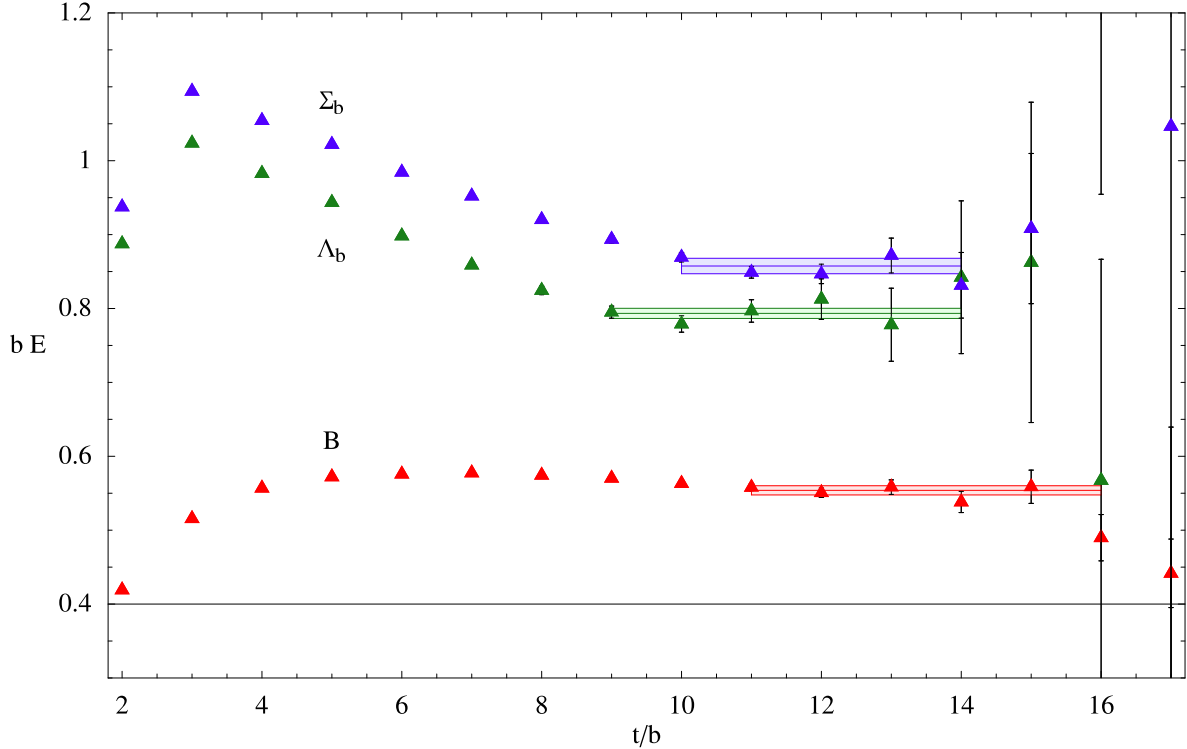}
  \caption{Effective mass ratios of the correlation
    functions for the $B$ (red), $\Lambda_b$ (green) and $\Sigma_b$
    (violet).  The shaded regions correspond to the fit range, central
    value and uncertainty (statistical and systematic errors are added
    in quadrature) for each hadron.}
  \label{fig:effmass}
\end{figure}

In Table~\ref{tab:singleHeavies} we present the extracted lattice
energies and resultant energies of the ldof for the $B$-meson and the
$\Lambda_b$ and $\Sigma_b$ heavy baryons. The effective mass ratios
corresponding to these measurements are shown in
Fig.~\ref{fig:effmass}.  We have performed both correlated and
uncorrelated single and double exponential fits to the correlation
functions in order to extract the ground state energy and to estimate
the uncertainty in the extraction.  The differences between the
extracted ground state energies from these procedures is encapsulated
in the systematic error.  The fitting range quoted in
Table~\ref{tab:singleHeavies} (and all other tables) is that used in
fitting a single exponential to the correlation function.  The
statistical errors are determined by the Jackknife procedure, omitting
a single configuration at each evaluation.  Further, a Bootstrap
analysis was also performed on the data, with both techniques
providing similar central values and uncertainties.  To eliminate
uncertainties common to the energies of two different hadrons, we
formed the correlated differences between the energies of the ldof in
the various systems. The residual masses of the static sources cancel
in these combinations and the results are displayed in Table
\ref{tab:singleHeaviesdiffs}.

\begin{table}[htbp]
\begin{ruledtabular}
\begin{tabular}{ccccc}
Hadron  &  fit range &  $b \ {\cal E}^{(DBW2)}$  & ${\cal E}^{(DBW2)}$ [MeV] & $
\overline{\Lambda}_{DBW2}^{(\alpha+\alpha^2\beta)}$ [MeV] \\
\hline 
$B$ & $11\rightarrow 15$ & $0.5539(37)(50)$ & $1096(18)(10)$ & $ 649(31)(10) $\\
$\Lambda_b$ & $9\rightarrow 14$ & $0.7934(65)(22)$ & $1570(27)(04)$ &  $ 1123(36)(04) $\\
$\Sigma_b $ & $10\rightarrow 14$& $0.8575(74)(74)$ & $1697(29)(15)$ &  $ 1250(38)(15) $\\
\end{tabular}
\end{ruledtabular}
 \caption{
   The spectrum of hadrons comprised of one static color ${\bf 3}$ 
source and light (anti-)quarks, for a light quark mass
   giving the light hadron spectrum shown in Table.~\ref{tab:single}.
   ${\cal E}^{(DBW2)}$ is the mass of the hadron determined in the
   lattice calculation.  
 $\overline{\Lambda}^{(\alpha+\alpha^2\beta)}_{DBW2} = {\cal
     E}^{(DBW2)} - \delta m^{(\alpha+\alpha^2\beta)}_{DBW2}$
    is the BLM-improved determination of the energy of the ldof.
   The first uncertainty is statistical  and the second  uncertainty is
   systematic.
}
\label{tab:singleHeavies}
\end{table}
\begin{table}[htbp]
\begin{ruledtabular}
\begin{tabular}{ccccc}
& Hadrons  &  $b\, {\cal E} = b \, \delta \overline{\Lambda}$ &  $\delta \overline{\Lambda}$ [MeV] &
\\
\hline 
& $\Lambda_b-B$ &  $0.2395(75)(72)$ & $474(15)(14)$ & \\
& $\Sigma_b-B$ & $0.304(08)(12)$ & $601(16)(24)$ & \\
& $\Sigma_b-\Lambda_b$ &  $0.0641(98)(96)$ & $126(19)(19)$ & \\
\end{tabular}
\end{ruledtabular}
 \caption{The mass differences between the hadrons comprised of one static
   color ${\bf 3}$ source and light (anti-)quarks. The residual mass of the static source cancels in
   the differences, leaving the differences between the energy of the light
   degrees of freedom, denoted by $\delta \overline{\Lambda}$.
The first uncertainty is statistical and the second is systematic.
  }
\label{tab:singleHeaviesdiffs}
\end{table}
%

%%%%%%%%%%%%%%%%%%%%%%%%%%%%%%%%%%%%%%%%%%
\section{Exotic Baryons}
\label{sec:exotics}

\begin{figure}[!th]
  \centering
  \includegraphics[width=0.95\columnwidth]{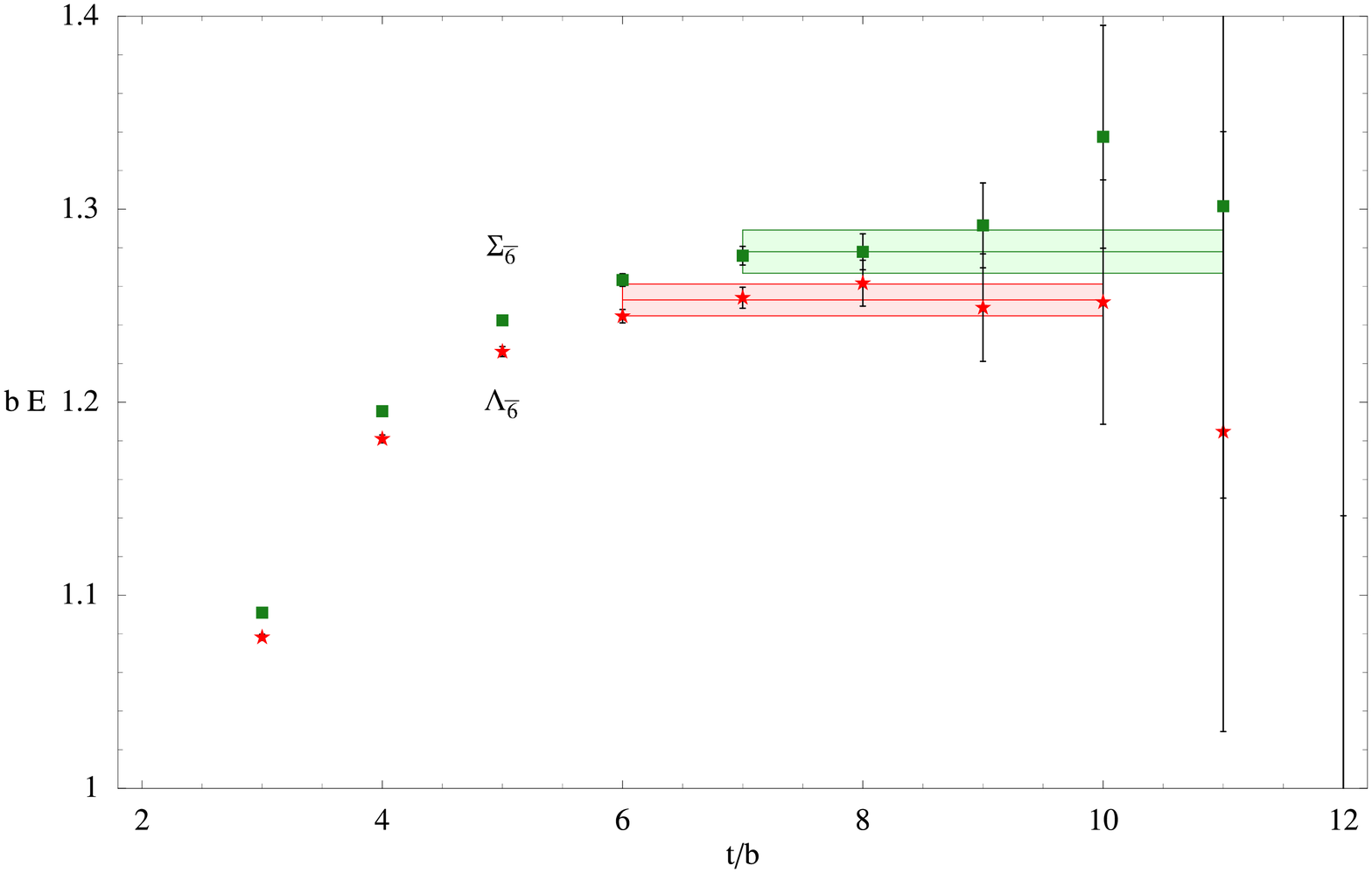}  
  \caption{The effective mass plots for the exotic hadrons,
    $\Lambda_{\overline{\bf 6}}$ and $\Sigma_{\overline{\bf 6}}$.}
  \label{fig:effmassexotic}
\end{figure}
\begin{table}[!tbp]
\begin{ruledtabular}
\begin{tabular}{ccccc}
Hadron  &  fit range &  $b \ {\cal E}^{(DBW2)}$  & ${\cal E}^{(DBW2)}$ [ MeV ] &
$\overline{\Lambda}_{\overline{\bf 6}; DBW2}^{(\alpha+\alpha^2\beta)}$ [ MeV ]\\
\hline 
$\Lambda_{\overline{\bf 6}}$ & $6\rightarrow 10$ & $1.253(08)(02)$ &
$2480(16)(04) $ & $1364(64)(04))$ 
\\
$\Sigma_{\overline{\bf 6}}$ & $7\rightarrow 11$ & $1.278(10)(05)$ &
$2529(20)(10)$ & $1413(65)(10)$\\ 
\hline
& & $b\,\delta {\cal E}^{(DBW2)}$ &\multicolumn{2}{c}{$\delta\overline{\Lambda}$}\\
\hline$\Sigma_{\overline{\bf 6}} - \Lambda_{\overline{\bf 6}}$ & -- & $0.025(13)(05)$
& \multicolumn{2}{c}{$49(25)(11)$}  \\ 
\end{tabular}
\end{ruledtabular}
 \caption{The spectrum of exotic baryons comprised of a static color
   ${\bf 6}$ source and light anti-quarks, for a light quark mass
   giving the light hadron spectrum shown in Table.~\ref{tab:single}.
   ${\cal E}^{(DBW2)}$ is the mass of the hadron determined in the
   lattice calculation.
Using the BLM-improved one-loop lattice perturbation theory
   calculation of the residual mass, as discussed in the text, we
   present results for
   $\overline{\Lambda}^{(\alpha+\alpha^2\beta)}_{\overline{\bf 6};
     DBW2} = {\cal 
     E}^{(DBW2)} - \delta
   m^{(\alpha+\alpha^2\beta)}_{{\bf 6}; DBW2}$, where the
   residual mass term is evaluated at the scale $q^{*(DBW2)}$ given in
   the text. The difference between the masses of the exotic states is
 given in the last line.}
\label{tab:singleExotics}
\end{table}
As a by-product of computing the potential between two $B$-mesons, we
have computed the masses of baryons formed from a static-source
transforming in the ${\bf 6}$ of color and two light anti-quarks with
$(I,s_l)=(0,1)$, which we denote as $\Lambda_{\overline{\bf 6}}$, or
$(I,s_l)=(1,0)$, which we denote as $\Sigma_{\overline{\bf 6}}$.
These are calculated by putting two static sources, each in the ${\bf
  3}$ of color at the same point in space, and requiring the light
anti-quarks to have the appropriate values of $(I,s_l)$.  As the spin of
the ldof decouples from the static source(s) and is a good quantum
number, the ldof form a color $\overline{\bf 6}$.  While the Casimir
of the $\overline{\bf 3}$ representation is $C(\overline{\bf 3})=4/3$,
it is $C({\bf 6})=10/3$ for the ${\bf 6}$ representation, and the
residual mass that must be subtracted from the energy calculated on
the lattice is $\delta m_{{\bf 6};{\rm
    DBW2}}^{(\alpha+\alpha^2\beta)}=1116\pm 62~{\rm MeV}$.  The
effective mass plots for these hadrons are shown in
Fig.~\ref{fig:effmassexotic} and the energies of the ldof are shown in
Table~\ref{tab:singleExotics}. These states are considerably more
massive than the non-exotic hadrons and achieve plateaus at earlier
times. As such exotic baryons have not been observed and heavy quarks
transforming in the ${\bf 6}$ (or other higher representations) of
color are not required in nature (see
Refs.~\cite{Wilczek:1976qi,Karl:1976jk,Marciano:1980zf} for a number
of proposals) and are not expected to be found, we do not dwell on
these results further.

%%%%%%%%%%%%%%%%%%%%%%%%%%%%%%%%%%%%%%%%%%%%%%%%%%%%
\section{B-meson Potentials}
\label{sec:potentials}

In the heavy-quark limit, the separation between the two $B$-mesons is
a good quantum number and the potential, $V({\bf r})$, is simply
defined by the difference between the energy of the two $B$-mesons
separated by a displacement vector ${\bf r}$ (defined by the
separation of the static ${\bf 3}$ color sources) and the energy of
two infinitely separated $B$-mesons,
\begin{eqnarray}
V_{I,s_l}({\bf r}) & = & E_{I,s_l}({\bf r})\ -\ 2 M_B
\ =\ \overline{\Lambda}_{I,s_l}({\bf r})\ -\ 2
\overline{\Lambda}_{{1\over 2},{1\over 2}} 
\ \ \ ,
\label{eq:potential}
\end{eqnarray}
where the isospin and spin of the ldof can take the values $I=0,1$ and
$s_l = 0,1$.\footnote{We classify these states using the quantum
  numbers of the infinite-volume continuum. For $s_l=0,1$ this is
  legitimate as there is a exact correspondence between these
  representations of O(3) and the $A_1$ and $T_1$ representations of
  H(3) \cite{Mandula:1983ut}. The continuum symmetry is O(3) as the
  presence of the infinitely massive quarks breaks O(4) to its spatial
  subgroup.}  Contributing to both of these energies are the
interactions between the ldof, the interactions between the ldof and
the static sources, and the interactions between the static sources.
As $|{\bf r}|\rightarrow\infty$, each of the contributions factorize,
leaving the contribution from two non-interacting $B$-mesons, and
therefore a vanishing potential.

The lattice calculation is a straightforward extension of the
calculation of $\overline{\Lambda}$ for the heavy hadrons.  The
energy, $ {\cal E}_{I,s_l}^{BB}({\bf r})$, of two $B$-mesons composed of
static sources and light-quarks, placed on the lattice with relative
displacement ${\bf r}$, is computed using the correlators in
Section~\ref{sec:lattice}.  The potential computed on the lattice then
becomes
\begin{eqnarray}
V_{I,s_l}^{\rm latt}({\bf r}) & = & {\cal E}_{I,s_l}^{BB}({\bf r}) \ -\ 2\ {\cal
  E}_{{1\over 2},{1\over 2}}
\ \ \ ,
\label{eq:lattpot}
\end{eqnarray}
where 
\begin{eqnarray}
  \label{eq:3}
  {\cal E}_{I,s_l}^{BB}({\bf r}) = \overline{\Lambda}_{I,s_l}({\bf r})  - \delta
  {\cal V}_{R}({\bf r}) + 2 \delta m
\ \ \ ,
\end{eqnarray}
(the subscript $R$ labels the color representation of the heavy quark
system dictated by the light quantum numbers $I$ and $s_l$).  The
residual masses of the static sources induced by interactions with the
gauge fields cancel in $V_{I,s_l}^{\rm latt}({\bf r})$.  However, a
perturbative subtraction corresponding to the differences in
interactions between the static sources in the continuum and on the
lattice, (which at leading order in the strong coupling arises from
one gluon exchange (OGE)) $\delta {\cal V}_{R}({\bf r}) = {\cal
  V}_{\rm OGE}^{\rm cont} - {\cal V}_{\rm OGE}^{\rm latt}$ remains.
Therefore, the energies measured in the lattice calculation, and the
potential between two $B$-mesons in the continuum are related by
\begin{eqnarray}
V_{I,s_l}({\bf r}) & = & 
{\cal E}_{I,s_l}^{BB}({\bf r})
\ -\ 2 {\cal  E}_{{1\over 2},{1\over 2}}
\ +\ \delta  {\cal V}_{R}({\bf r})
\ \ \ .
\label{eq:pottolattdef}
\end{eqnarray}

In the cases where the spin of the ldof is $s_l=1$, the potential receives
contributions from both a central component and a tensor component,
\begin{eqnarray}
V^{(S=1)}({\bf r}) & = & V_C(r)\ +\ \hat S_{12}\ V_T (r)\,,
\label{eq:VCVT}
\end{eqnarray}
where
\begin{eqnarray}
\hat S_{12} & = & {3\over 2}\left(\hat S_+ (\hat r_x-i\hat r_y) + \hat S_-
  (\hat r_x+i\hat r_y) + 2 \hat S_z \hat r_z \right)^2 \ -\ 2 \hat S^2
\ \ \ ,
\end{eqnarray}
$\hat S_i$ are the spin operators, and $\hat {\bf r}$ is the  
unit vector in the direction of the displacement.  It follows that the central and tensor
potentials can be determined from the potential at $x$ and $z$
displacements,
\begin{eqnarray}
V_C^{(S=1)}(r) & = & {1\over 3}\ \left(\ V^{(S=1)}(r\ \hat {\bf e}_z)
\ +\  2  V^{(S=1)}(r\ \hat {\bf e}_x)\ \right)
\nonumber\\
V_T^{(S=1)}(r) & = & {1\over 3}\ \left(\ V^{(S=1)}(r\ \hat {\bf e}_z)
\ -\   V^{(S=1)}(r\ \hat {\bf e}_x)\ \right)
\ \ \ ,
\label{eq:VCVTdecompo}
\end{eqnarray}
where $\hat {\bf e}_j$ is the unit-vector in the ``$j$'' direction.

%%%%%%%%%%%%%%%%%%%%%%%%%%%%%%%%%%%%%%%%%
\subsection{Lattice Spacing Effects in  a Finite-Volume}
\label{sec:latt-spac-finite}

The potential measured in the lattice calculation will differ from
that at infinite-volume due to the presence of image $B$-mesons
resulting from the periodic boundary conditions in the spatial
directions of the lattice.  Therefore, the single particle energies
that are extracted correspond to the energy of single particle that is
interacting with its images, located at $|{\bf r}| > L$.  In the case
of the energy of two particles interacting in a periodic cubic volume,
the potential energy, $ V^{(L)}({\bf r})$, measured includes the sum
over the contributions from the images\footnote{This is correct for
  interactions via single particle exchange but receives corrections
  that we discuss in Section \ref{sec:infin-volume-potent}.}
\begin{eqnarray}
V^{(L)}({\bf r}) & = & V({\bf r})\ +\ \sum_{ {\bf n}\ne {\bf 0} }\  V({\bf
  r}+{\bf n} L)
\,.
\label{eq:finiteVolPot}
\end{eqnarray}
When the displacement between the mesons is $|{\bf r}|>L/2$ the
interaction with the nearest image is more important than the
interaction within the volume. Consequently, we have only computed the
potential for $|{\bf r}|\leq 8$ lattice spacings on $16^3\times 32$
lattices.  After taking the continuum limit of the lattice
calculation, the finite-volume effects due to the images must be
removed to recover the infinite-volume continuum limit potential,
$V({\bf r})$. This is discussed below.

The finite lattice spacing, $b$, eliminates ultraviolet modes on the
lattice leaving $|{\bf q}| < \pi/b$, and hence the strong Coulomb potential
that exists between two static-sources due to OGE is significantly
modified for $|{\bf r}|$ less than a few $b$.  
In the heavy-quark limit, the OGE potential is spin-independent but 
does depend upon the color representation of the combined heavy quark system.

The potential between two static color ${\bf 3}$ sources combined into
a color $\overline{\bf 3}$, at a finite lattice spacing, $b$, and in a
finite-volume becomes
\begin{eqnarray}
&& {\cal V}_{\overline{\bf 3};LO}^{\rm latt}({\bf r}) \ =\ 
-{\overline{\alpha}(\mu)\over 3\pi^2 b}\ 
\int_{-{\pi}}^{\pi}\ d\tilde q_x\ 
\int_{-{\pi}}^{\pi}\ d\tilde q_y\ 
\int_{-{\pi}}^{\pi}\ d\tilde q_z\ 
G_{00}(\hat {\bf q}) \ e^{i\tilde {\bf q}\cdot \tilde {\bf r}}
\nonumber\\
&& \rightarrow 
-{\overline{\alpha}(\mu)\over 3\pi^2 b}\ 
\left({2\pi\over \tilde L}\right)^3\ 
\sum_{{\bf n}}^{|n_i|\le {\tilde L\over 2}}
\ e^{i 2 \pi {\bf n}\cdot \tilde {\bf r}/\tilde L}\ 
G_{00}\left(2\sin\left({\pi n_x\over\tilde L}\right),2\sin\left({\pi
      n_y\over\tilde L}\right),2\sin\left({\pi n_z\over\tilde
      L}\right),0\right) 
\ ,
\label{eq:finitevolA}
\end{eqnarray}
where $\tilde L = L/b$ is the spatial extent of the lattice in lattice
units, and ${\bf \tilde r} = {\bf r}/b$ is the displacement between
the static sources in lattice units.  The summation in
Eq.~(\ref{eq:finitevolA}) is over all $-\tilde L/2 < n_x, n_y, n_z <
\tilde L/2$.  This finite-volume expression has an infrared divergence
due to the ${\bf n}={\bf 0}$ mode, however, the difference between the
finite-volume OGE potentials at finite lattice spacing and in the
continuum is of the form
\begin{eqnarray}
\delta {\cal V}_{\overline{\bf 3};LO}^{(L)}({\bf r}) 
& =  & 
-{\overline{\alpha}(\mu)\over 3\pi^2 b}\ 
\left({2\pi\over \tilde L}\right)^3\ 
\left[\
 \left({\tilde L\over 2\pi}\right)^2 
\sum_{\bf n}^\infty\ \frac{e^{i 2 \pi {\bf n}\cdot \tilde {\bf
      r}/\tilde L}\  }{ |{\bf n}|^2}
-\sum_{{\bf n}}^{|n_i|\le {\tilde L\over 2}}
\ e^{i 2 \pi {\bf n}\cdot \tilde {\bf r}/\tilde L}\ 
G_{00}\left(\hat n_x,\hat n_y,\hat n_z,0\right)
\ \right]
\ ,
\nonumber\\
\label{eq:finitevolB}
\end{eqnarray}
where $\hat n_i = 2\sin\left(\pi n_i/ \tilde L\right)$, and is
well-behaved in the infrared. Spurious contributions from ill-defined
low-momentum gluon modes included in the above sums cancel to a large
extent with residual effects at most of ${\cal O}(b^2)$.  Further
discussion of this issue and the numerical evaluation of $\delta {\cal
  V}_{\overline{\bf 3};LO}^{(L)}({\bf r}) $ can be found in
Appendix~\ref{ap:A}.

The BLM procedure is again used to set the scale of the correction
factor, leading to
\begin{eqnarray}
 \delta {\cal V}_{\overline{\bf 3};NLO}^{(L)}({\bf r}) 
& = &
-{\overline{\alpha}(\mu)
\over 3\pi^2 b}\ 
\left({2\pi\over \tilde L}\right)^3\ 
\left[
 \left({\tilde L\over 2\pi}\right)^2 
\sum_{\bf n}^\infty\ \ {e^{i 2 \pi {\bf n}\cdot \tilde {\bf r}/\tilde
    L}\over |{\bf n}|^2} 
\left( 1 
\ -\ 
{\overline{\alpha}(\mu)\beta\over 4\pi}
\log\left({4\pi^2\left(|{\bf n}|^2\right)\over\mu^2
    L^2}\right)
\ \right)
\right.
\nonumber\\
&& \left.
\ -\ \sum_{{\bf n}}^{|n_i|\le {\tilde L\over 2}}
\ e^{i 2 \pi {\bf n}\cdot \tilde {\bf r}/\tilde L}\ 
G_{00}\left(\hat n_x,\hat n_y,\hat n_z,0\right)
\left( 1 
\ -\ 
{\overline{\alpha}(\mu)\beta\over 4\pi}
\log\left({|\hat {\bf n}|^2\over\mu^2
    b^2}\right)\right)
\ \right]
\nonumber\\
& = & 
-{\overline{\alpha}(\mu) \over r}
\left[\ 
{\cal A}\left(\tilde {\bf r}, \tilde L\right)
\ -\ 
{\overline{\alpha}(\mu)\beta\over 4\pi}
{\cal B}\left(\tilde {\bf r}, \tilde L, \mu\right)
\ \right]
\nonumber\\
& = & 
-{\overline{\alpha}(q^*(\tilde {\bf r}, \tilde L)) \over r}\ 
{\cal A}\left(\tilde {\bf r}, \tilde L\right)
\ +\ ...
\ .
\label{eq:finitevolBLM}
\end{eqnarray}
The coefficient functions ${\cal A}(\tilde {\bf r},\tilde L)$ and
${\cal B}(\tilde {\bf r},\tilde L, \mu)$ evaluated on DBW2 lattices,
using the techniques in Appendix~\ref{ap:A}, are given in
Table.~\ref{tab:BLMcoeffs} at the required separations.  The resulting
BLM-scale and potential shifts are given in Table.~\ref{tab:Vcorrs}
(results for the color ${\bf 6}$ OGE potential are related by $\delta
{\cal V}_{{\bf 6};(N)LO}^{(L)}({\bf r})=-1/2\,\delta {\cal
  V}_{\overline{\bf 3};(N)LO}^{(L)}({\bf r})$).
The correction factors, $\delta {\cal
  V}_{\overline{\bf 3};LO}^{(L)}({\bf r})$ and $\delta {\cal V}_{{\bf
    6};LO}^{(L)}({\bf r})$, should be added to the lattice
measurements of ${\cal E}_{I,s_l}-2{\cal E}_{\frac{1}{2},\frac{1}{2}}$
to give the potential.  At relative displacements that are large
compared with the lattice spacing, this correction factor scales as
$b^2/|{\bf r}|^3$, as expected.  However, there are still ${\cal O}(b)$ lattice
artifacts from the discretisation of the light quark sector. These can
only be eliminated using data at different lattice spacings or using
a light-quark action that is ${\cal O}(b)$-improved.
\begin{table}[!t]
\begin{ruledtabular}
\begin{tabular}{ccccc}
$\tilde {\bf r}$  &  ${\cal A}(\tilde {\bf r},16)$ &  ${\cal B}(\tilde {\bf
  r},16,b^{-1})$ 
 &  ${\cal A}(\tilde {\bf r},\infty)$ &  ${\cal B}(\tilde {\bf r},\infty,b^{-1})$\\
\hline 
$(1,0,0)$ & $+0.2656 $ & $-0.04261$  & 0.2654 & -0.0938 \\
$(1,1,0)$ & $+0.2011 $ & $-0.1791$ &  0.2102 &  -0.2236\\
$(2,0,0)$ & $+0.1203 $ & $-0.2352$ & 0.1629 &  -0.2451\\
$(2,1,0)$ & $+0.1144 $ & $-0.2130$ & 0.1370 &  -0.2750\\
$(3,0,0)$ & $+0.1206 $ & $-0.0750$ & 0.1083 &  -0.2357\\
$(4,0,0)$ & $-0.0176 $ & $-0.1975$ & 0.0665 &  -0.2048\\
$(5,0,0)$ & $+0.0592 $ & $+0.0874$ & 0.0370 &  -0.1653\\
$(6,0,0)$ & $-0.1055 $ & $-0.1359$ & 0.0176 &  -0.1261\\
$(7,0,0)$ & $+0.0409 $ & $+0.2310$ & 0.0054 &  -0.0921\\
$(8,0,0)$ & $-0.1593 $ & $-0.1247$ & 0.0020 &  -0.0647\\
\end{tabular}
\end{ruledtabular}
 \caption{The functions ${\cal A}(\tilde {\bf r},\tilde L)$ and 
 ${\cal B}(\tilde {\bf r},\tilde L,\mu)$ that contribute to the
 difference between  the finite-lattice spacing and continuum OGE
 potentials at finite-volume. }
\label{tab:BLMcoeffs}
\end{table}
\begin{table}[!t]
\begin{ruledtabular}
\begin{tabular}{ccccc}
$\tilde {\bf r}$  &  $b\,q^*(\tilde {\bf r},16)$  
&  
$b\  \delta {\cal V}_{\overline{\bf 3};LO}^{(L)}({\bf r})$ 
& 
$b\  \delta {\cal V}_{\overline{\bf 3};NLO}^{(L)}({\bf r})$ 
& 
$\delta {\cal V}_{\overline{\bf 3};NLO}^{(L)}({\bf r})$ [ MeV ] 
\\
\hline 
$(1,0,0)$ & $0.9229$ & $-0.0794(40)$ & $-0.0828(41)$ & $-163.8(8.6)$   \\
$(1,1,0)$ & $0.6406$ & $-0.0425(21)$ & $-0.0524(26)$ & $-103.7(5.4)$   \\
$(2,0,0)$ & $0.3762$ & $-0.01798(90)$ & $-0.0272(14)$ & $-53.8(2.8)$   \\
$(2,1,0)$ & $0.3941$ & $-0.01529(77)$ & $-0.0227(11)$ & $-45.0(2.4)$   \\
$(3,0,0)$ & $0.7327$ & $-0.01202(60)$ & $-0.01398(70)$ & $-27.7(1.4)$   \\
$(4,0,0)$ & $269.6$  & $+0.001319(66)$ & $-0.00254(13)$  & $-5.04(26)$   \\
$(5,0,0)$ & $2.092$  & $-0.00354(18)$ & $-0.00217(10)$ & $-4.30(23)$   \\
$(6,0,0)$ & $1.904$  & $+0.00525(26)$ & $+0.00348(17)$  & $+6.90(36)$   \\
$(7,0,0)$ & $16.87$  & $-0.001747(87)$ & $+0.000836(42)$ & $+1.654(87)$   \\
$(8,0,0)$ & $1.479$  & $+0.00595(30)$ & $+0.00473(24)$  & $+9.37(49)$   \\
\end{tabular}
\end{ruledtabular}
 \caption{Corrections to the potential between two static color ${\bf 3}$ 
   sources combined into the $\overline{\bf 3}$ representation,
  displaced by $\tilde{\bf r}$ computed on  $16^3\times 32$
   DBW2 lattices with a lattice spacing of $b=0.0997\pm 0.0015~{\rm
     fm}$. The quoted uncertainty is due to the uncertainty in $\alpha_s(2~{\rm
     GeV})$, and that of the lattice spacing.  The NLO result does not use the
   BLM scale, $q^*$, but the sum of the LO and NLO contributions.
}
\label{tab:Vcorrs}
\end{table}
%

%%%%%%%%%%%%%%%%%%%%%%%%%%%%%%
\subsection{The Lattice and Continuum Finite-Volume Potentials}
\label{sec:pots}

Using the techniques described in Sec.~\ref{sec:lattice} we have
computed the correlation functions corresponding to the energy
differences of Eq.~(\ref{eq:lattpot}).  Correlated and uncorrelated
single exponential fits to the correlation functions are performed to
determine the energy of the two B-mesons, and the Jackknife method is
used to determine the uncertainty.  Further, Jackknife is used to
determine the correlated difference in energy between the two
$B$-mesons, and twice the single $B$-meson mass.

\begin{figure}[!thp]
  \centering
  \includegraphics[width=0.32\columnwidth]{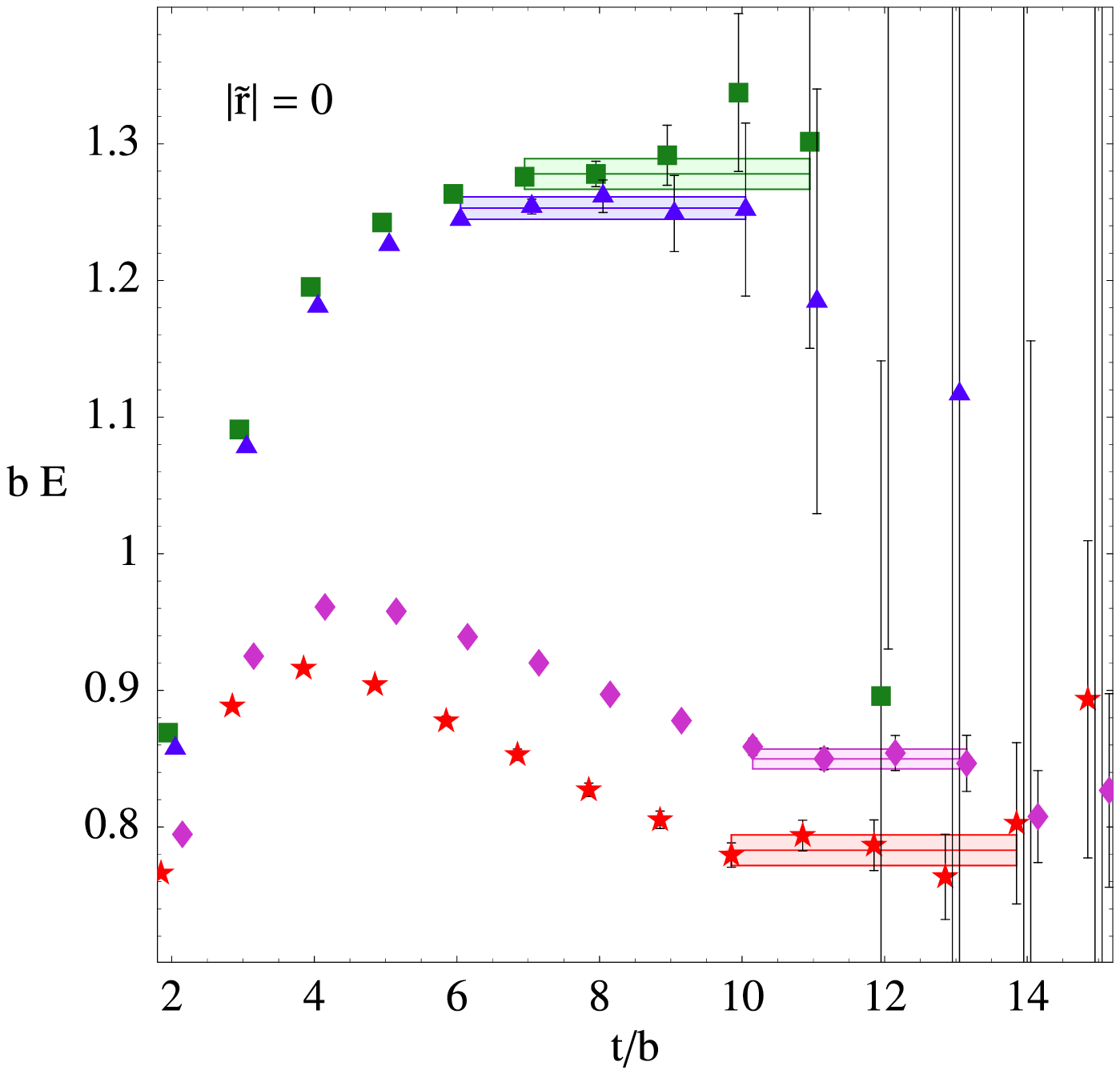}
\includegraphics[width=0.32\columnwidth]{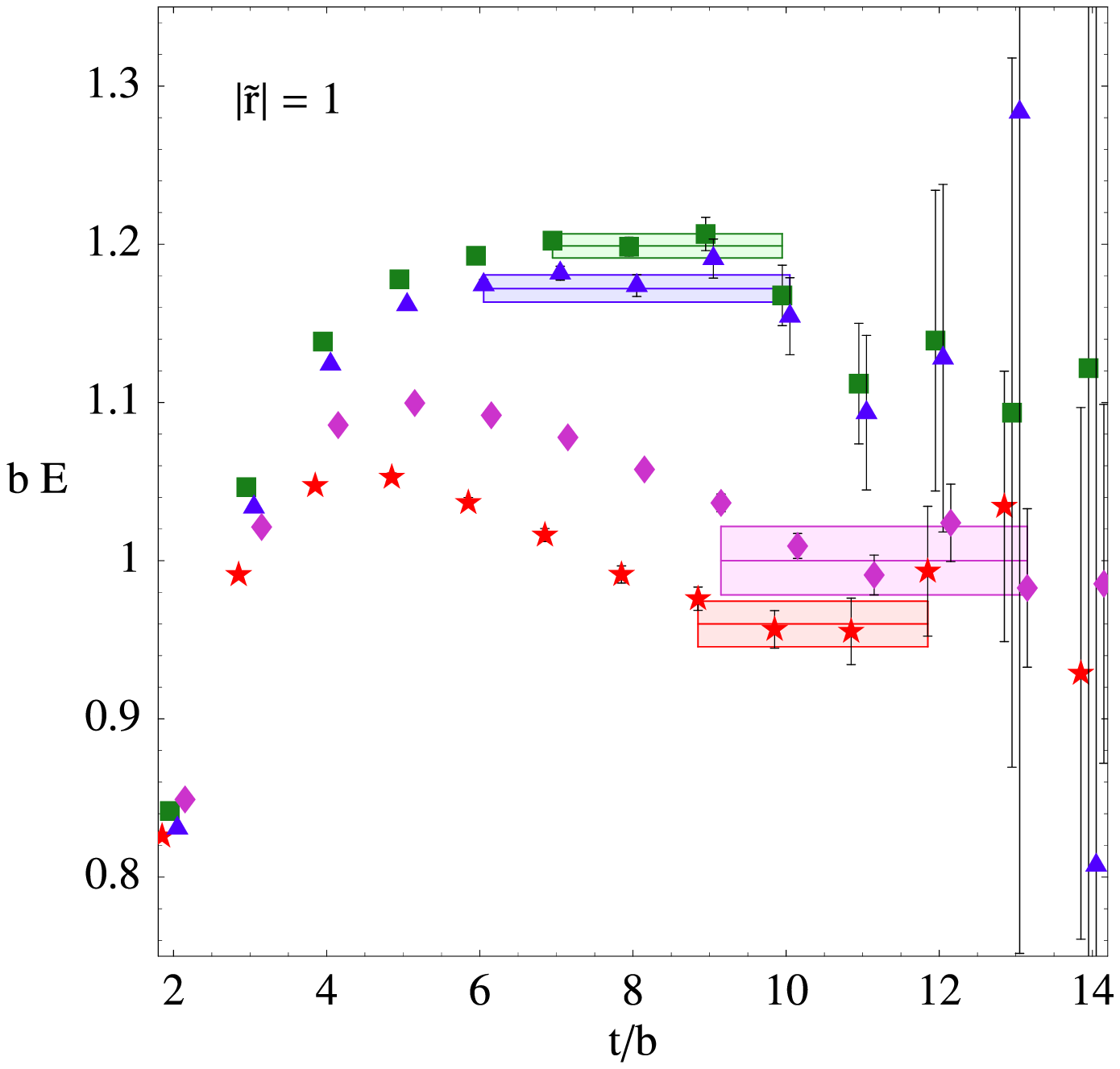}  
  \includegraphics[width=0.32\columnwidth]{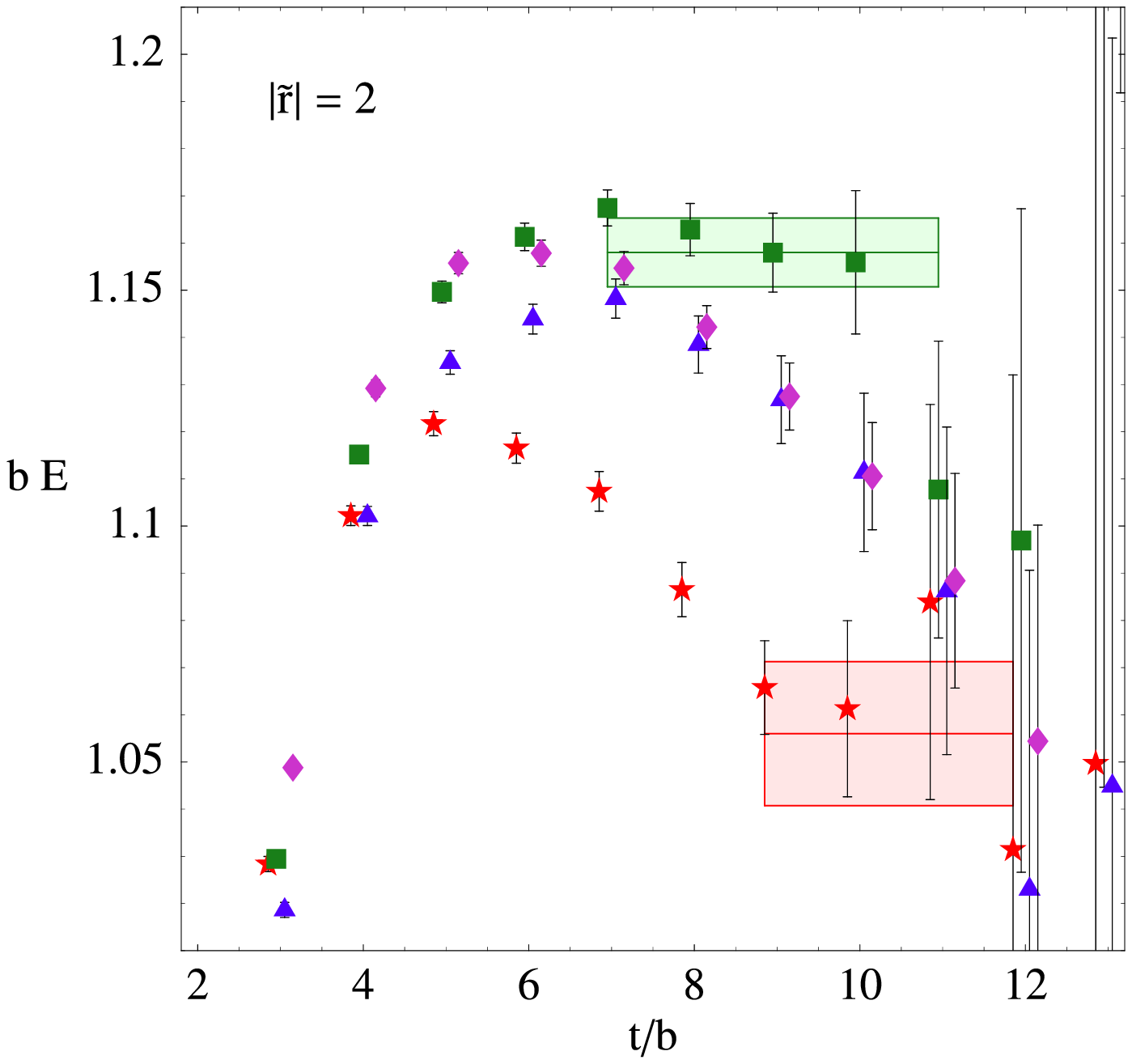} \\
\includegraphics[width=0.32\columnwidth]{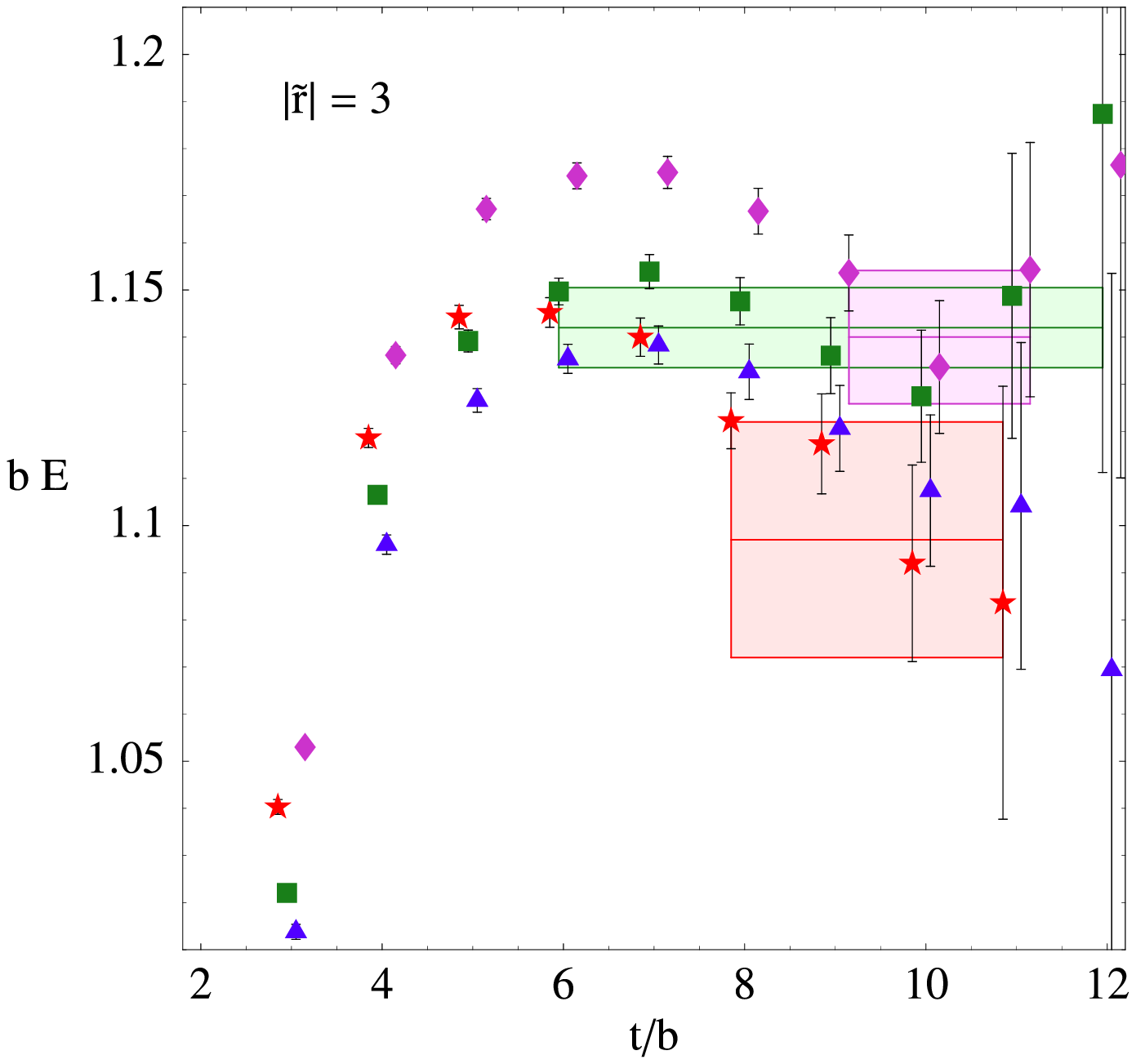} 
  \includegraphics[width=0.32\columnwidth]{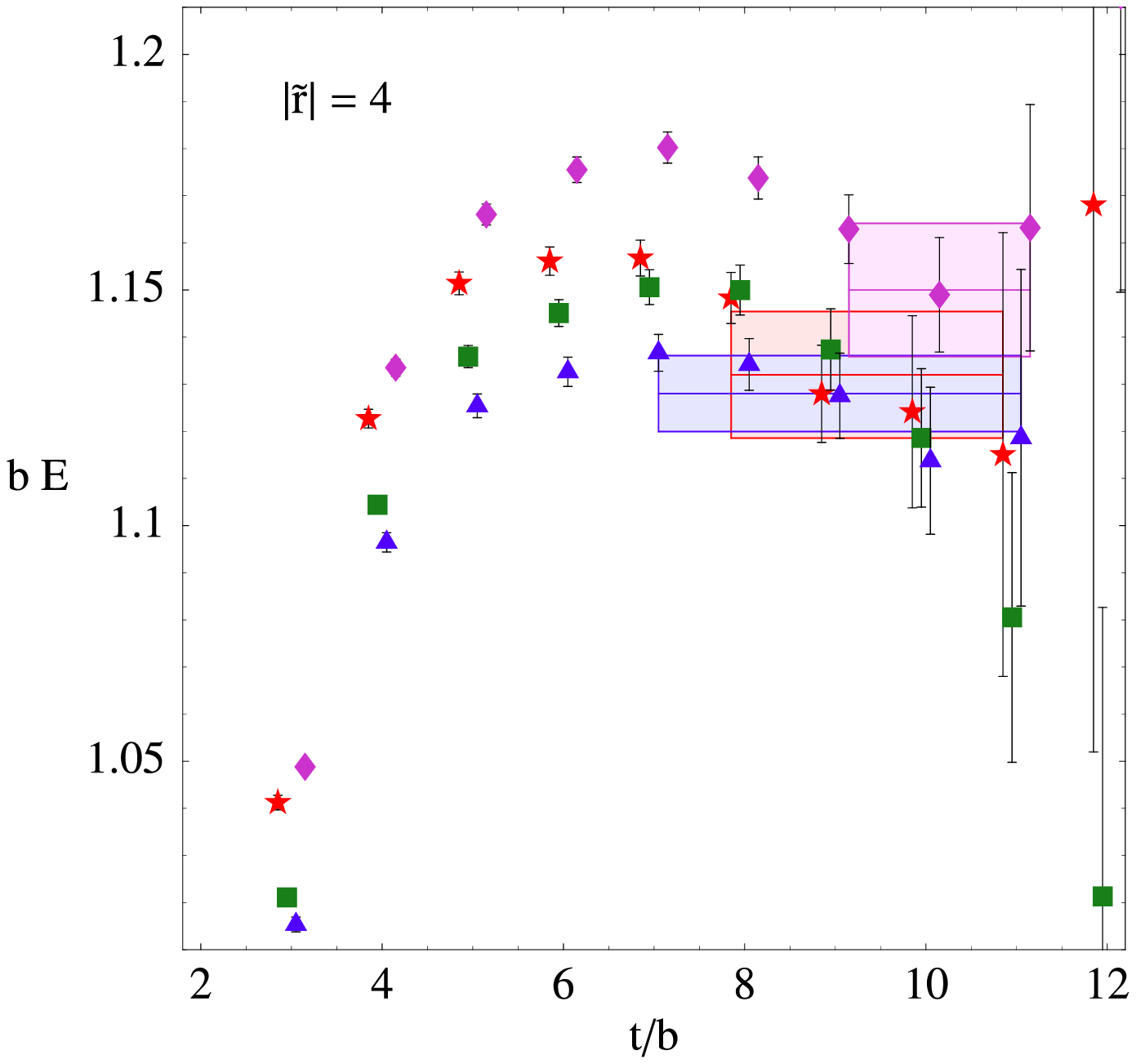}
\includegraphics[width=0.32\columnwidth]{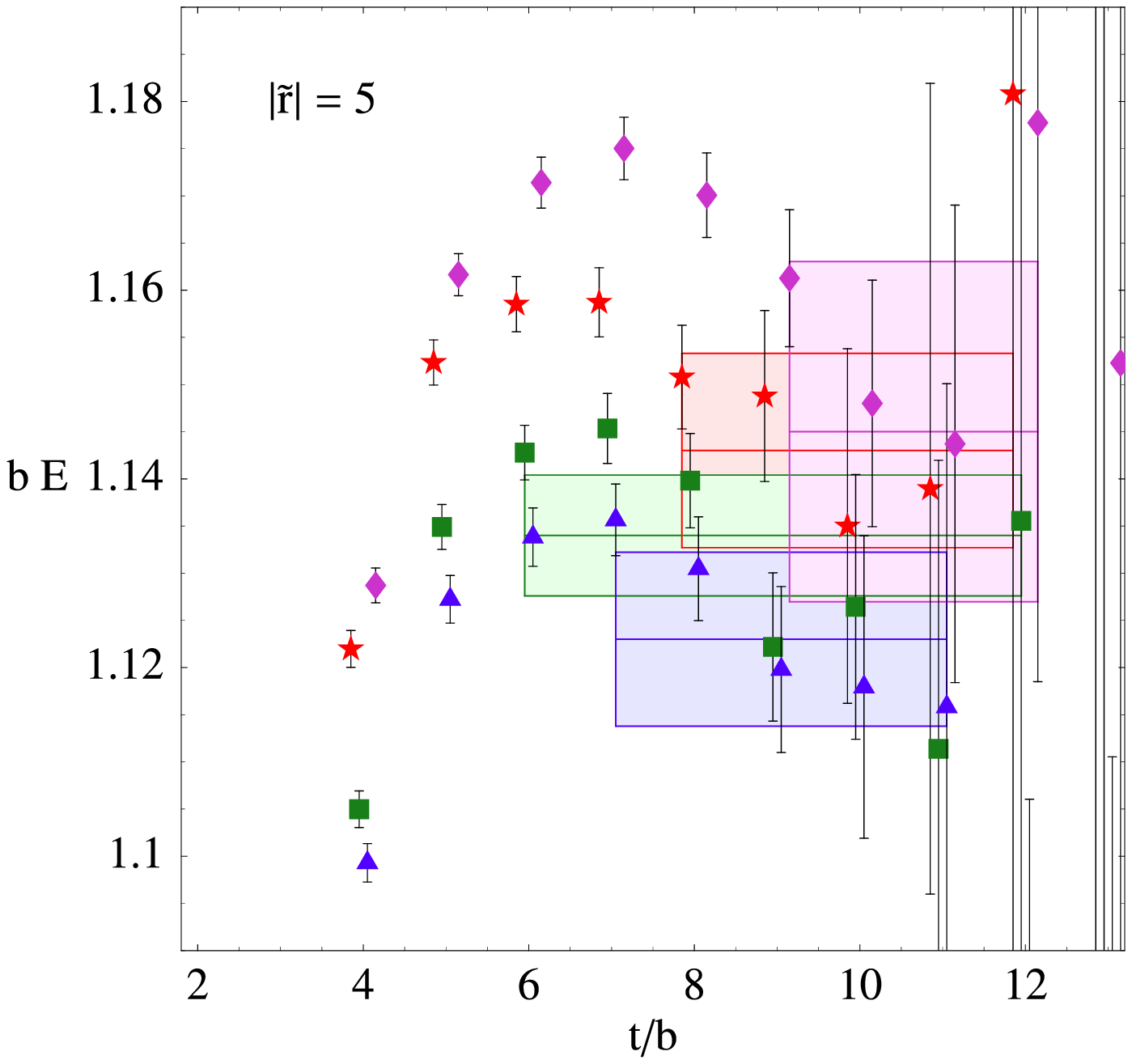}
\\
  \includegraphics[width=0.32\columnwidth]{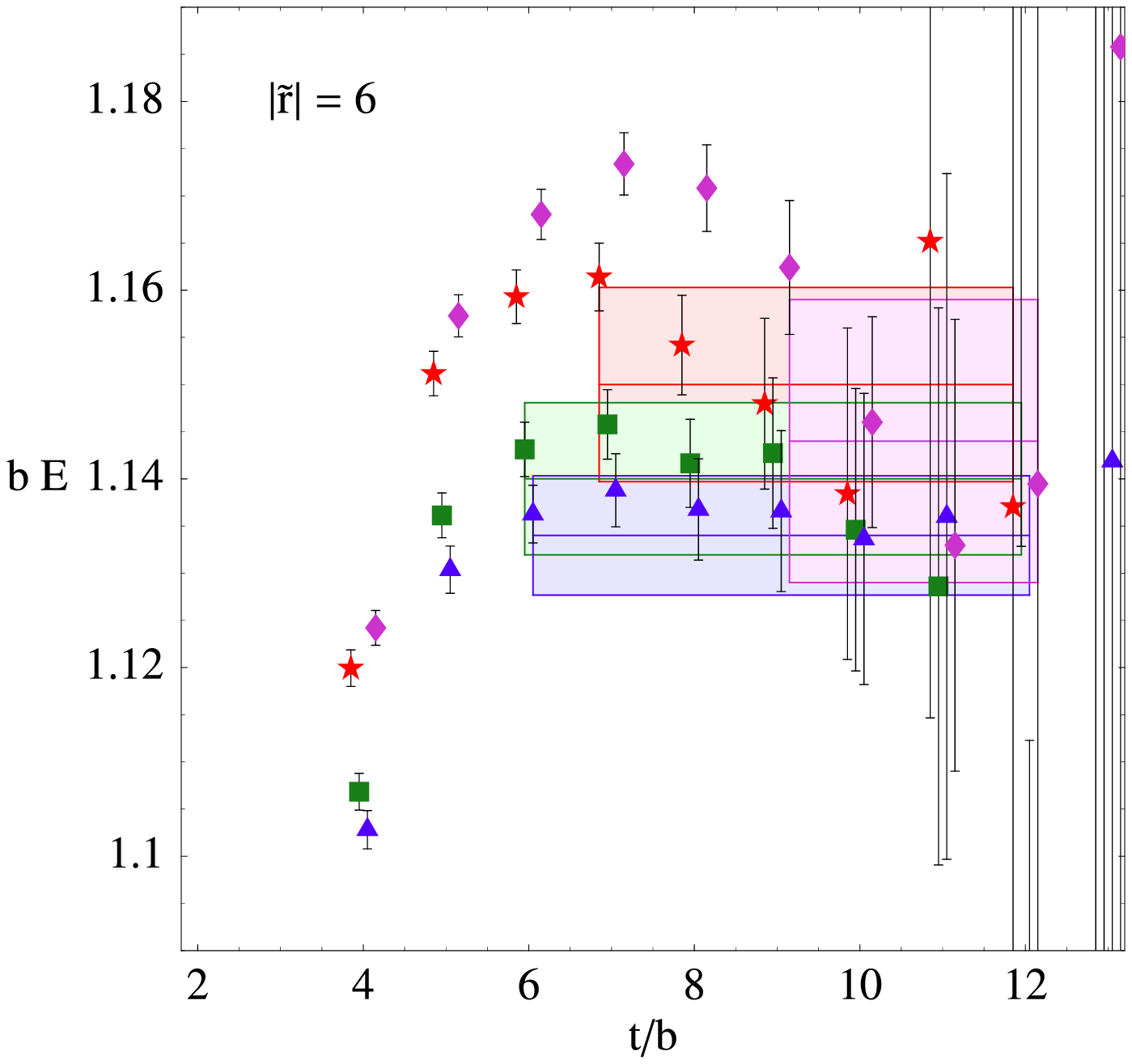}
\includegraphics[width=0.32\columnwidth]{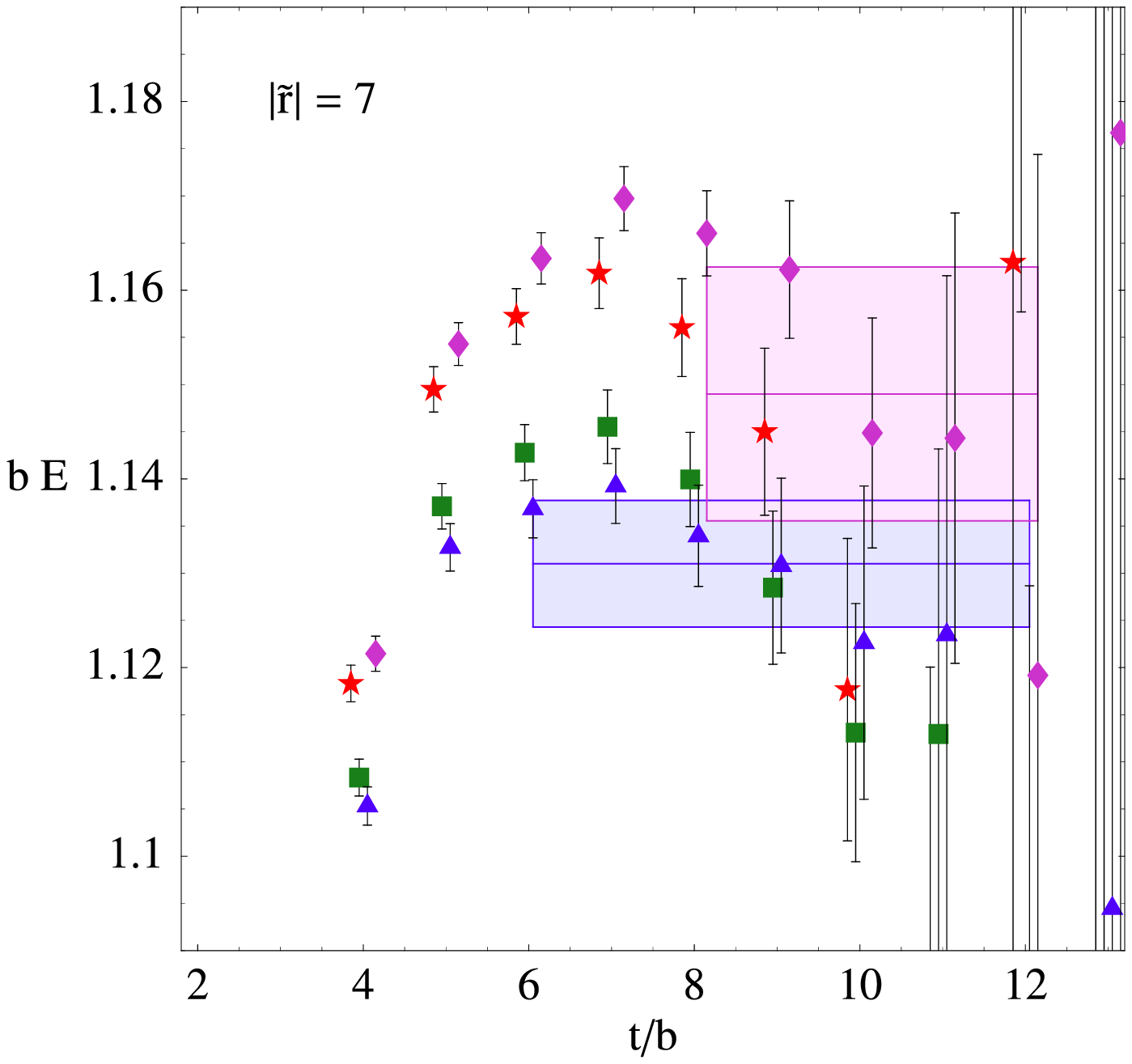}
  \includegraphics[width=0.32\columnwidth]{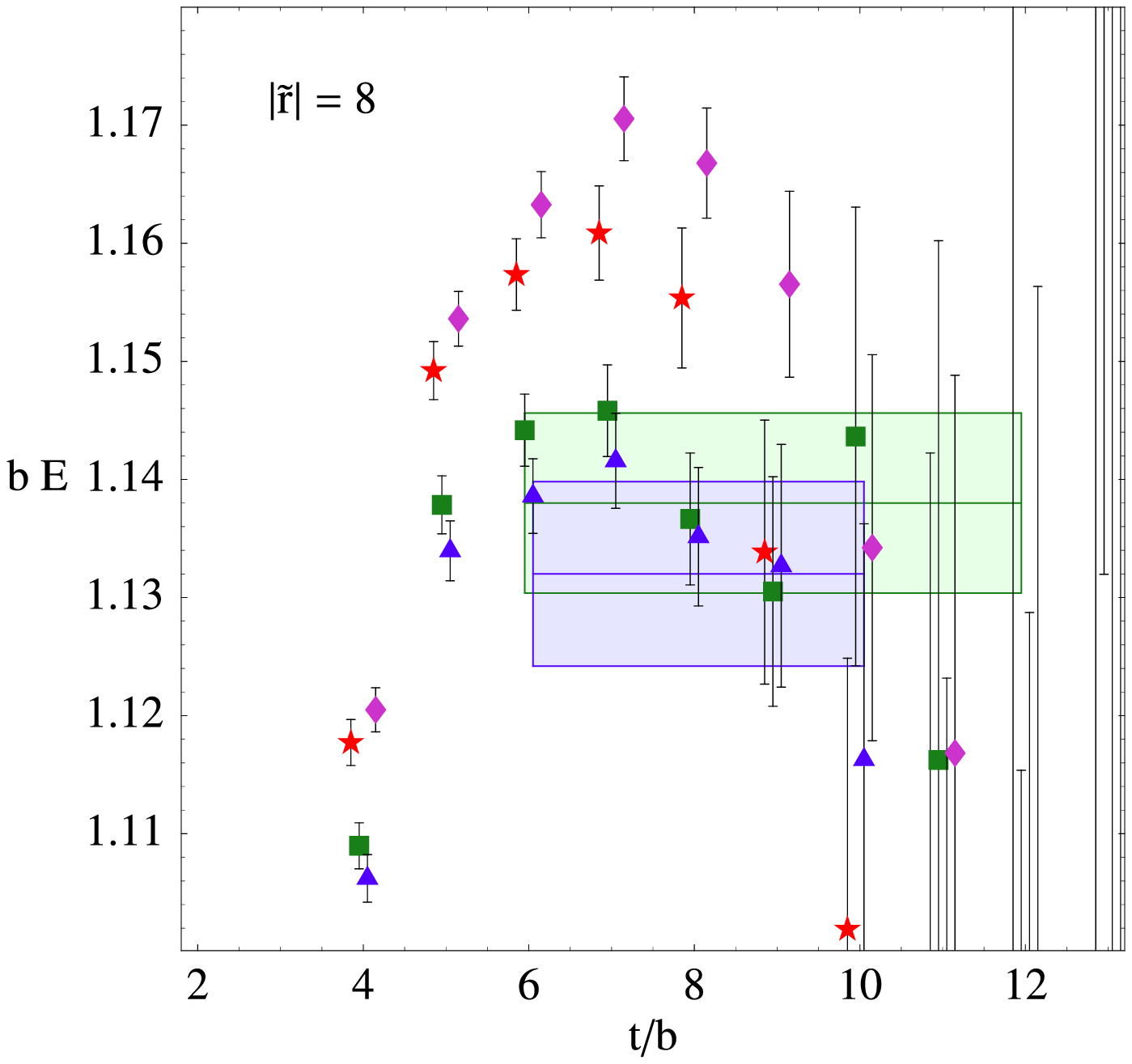}
  \caption{The effective mass plots for the central potentials at
    $|\tilde {\bf r}|=0,1,2,3,4,5,6,7,8$ for each spin-isospin
    channel.  Red stars correspond to $(I,s_l)=(0,0)$, green squares
    to $(I,s_l)=(1,0)$, blue triangles to $(I,s_l)=(0,1)$ and magenta
    diamonds to $(I,s_l)=(1,1)$.  Extracted masses and uncertainties
    are shown as the shaded regions in channels where a signal can be
    extracted, and are given in Tables~\protect\ref{tab:V00}-\protect\ref{tab:V01}.}
  \label{fig:effmassC}
\end{figure}

\begin{figure}[!thp]
  \centering
  \includegraphics[width=0.32\columnwidth]{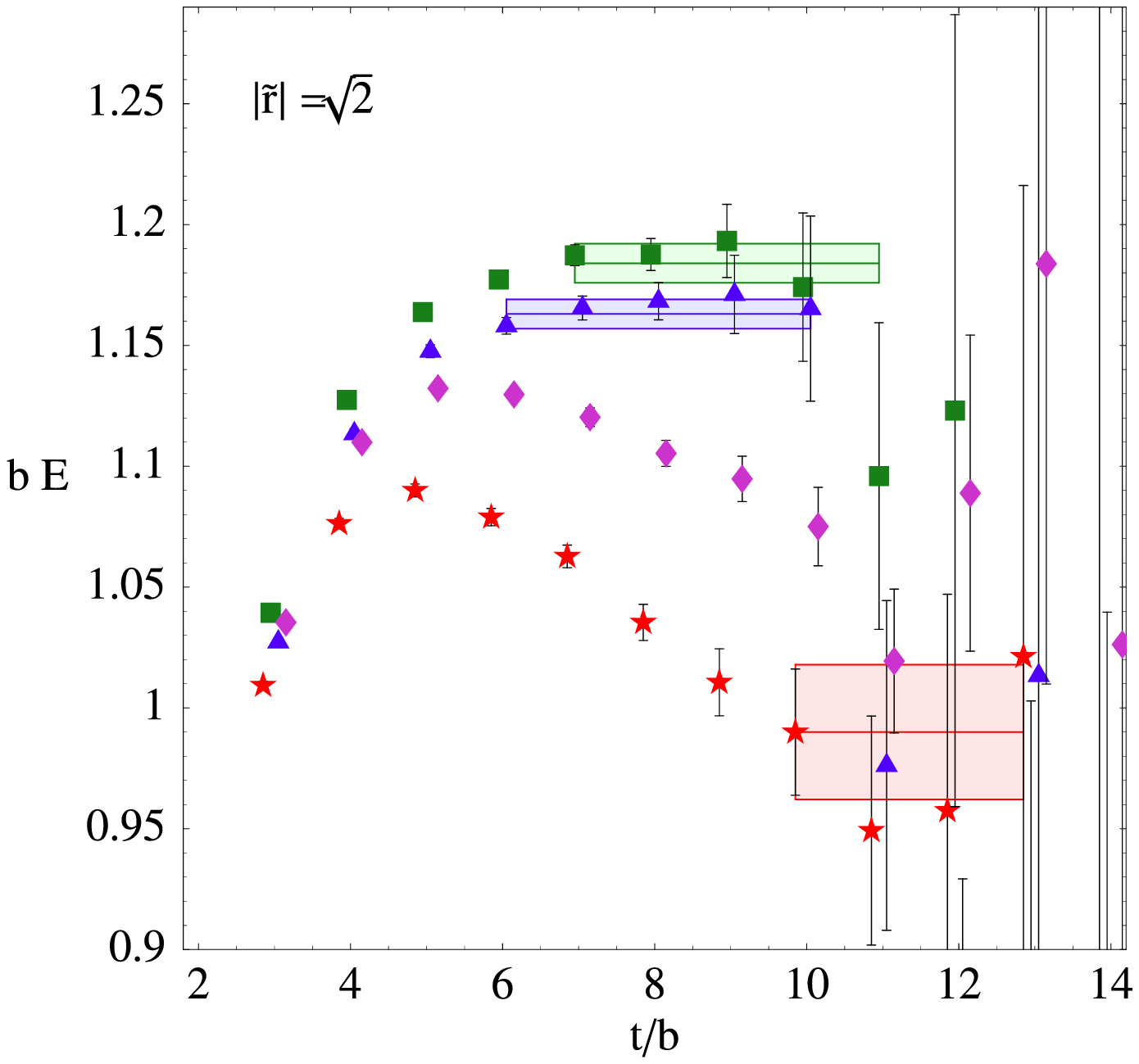}
\includegraphics[width=0.32\columnwidth]{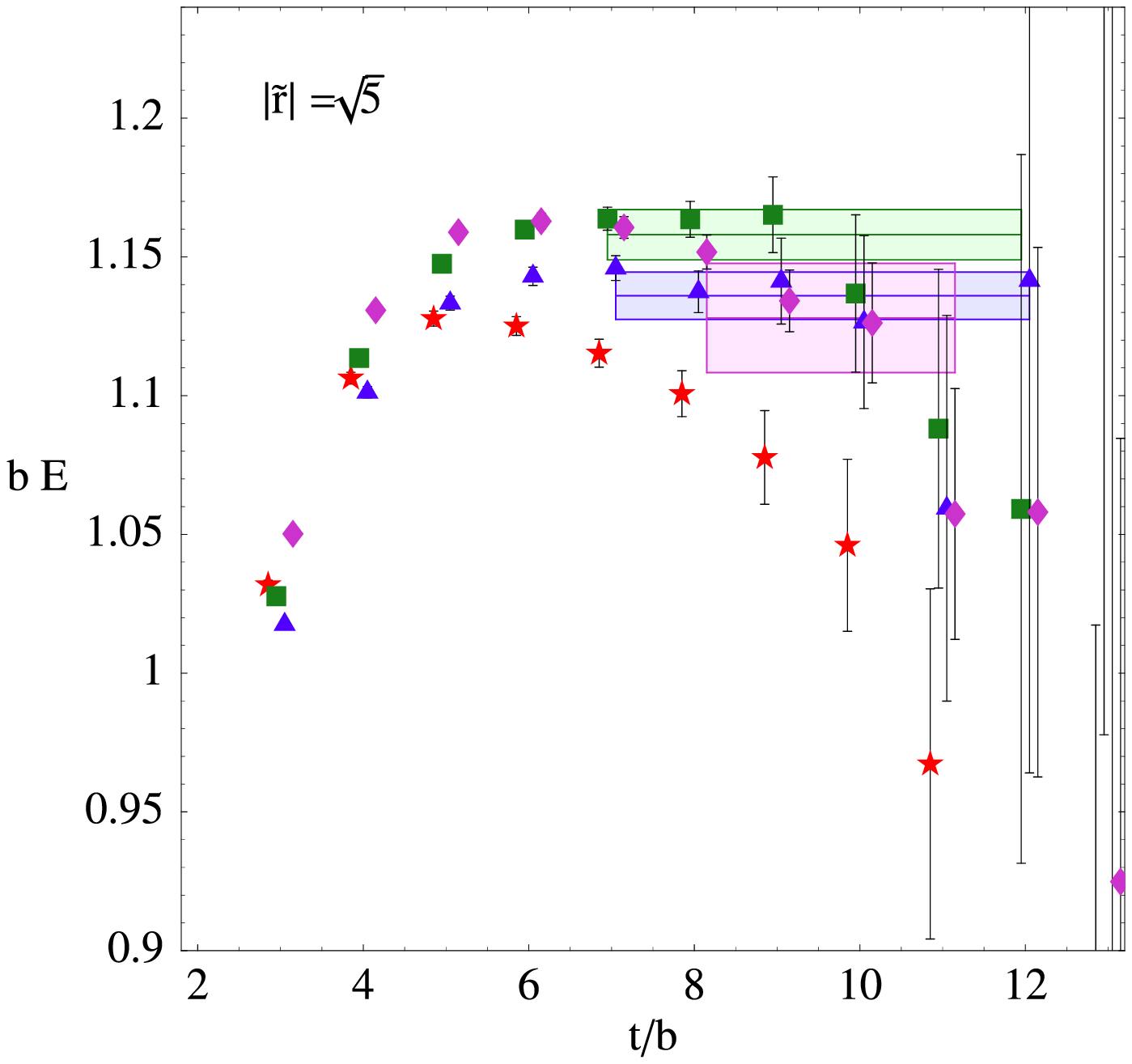}
  \caption{The effective mass plots for the potentials at displacements $\tilde{\bf
      r} = (1,1,0)$ and $(2,1,0)$ which are linear combinations of the
    central and tensor potentials. Red stars correspond to
    $(I,s_l)=(0,0)$, green squares to $(I,s_l)=(1,0)$, blue triangles
    to $(I,s_l)=(0,1)$ and magenta diamonds to $(I,s_l)=(1,1)$.
    Extracted masses and uncertainties are shown as the shaded regions
    in channels where a signal can be extracted, 
and are given in Tables~\protect\ref{tab:V00}-\protect\ref{tab:V01}.}
  \label{fig:effmassSQRTS}
\end{figure}

\begin{figure}[htbp]
  \centering
  \includegraphics[width=0.95\columnwidth]{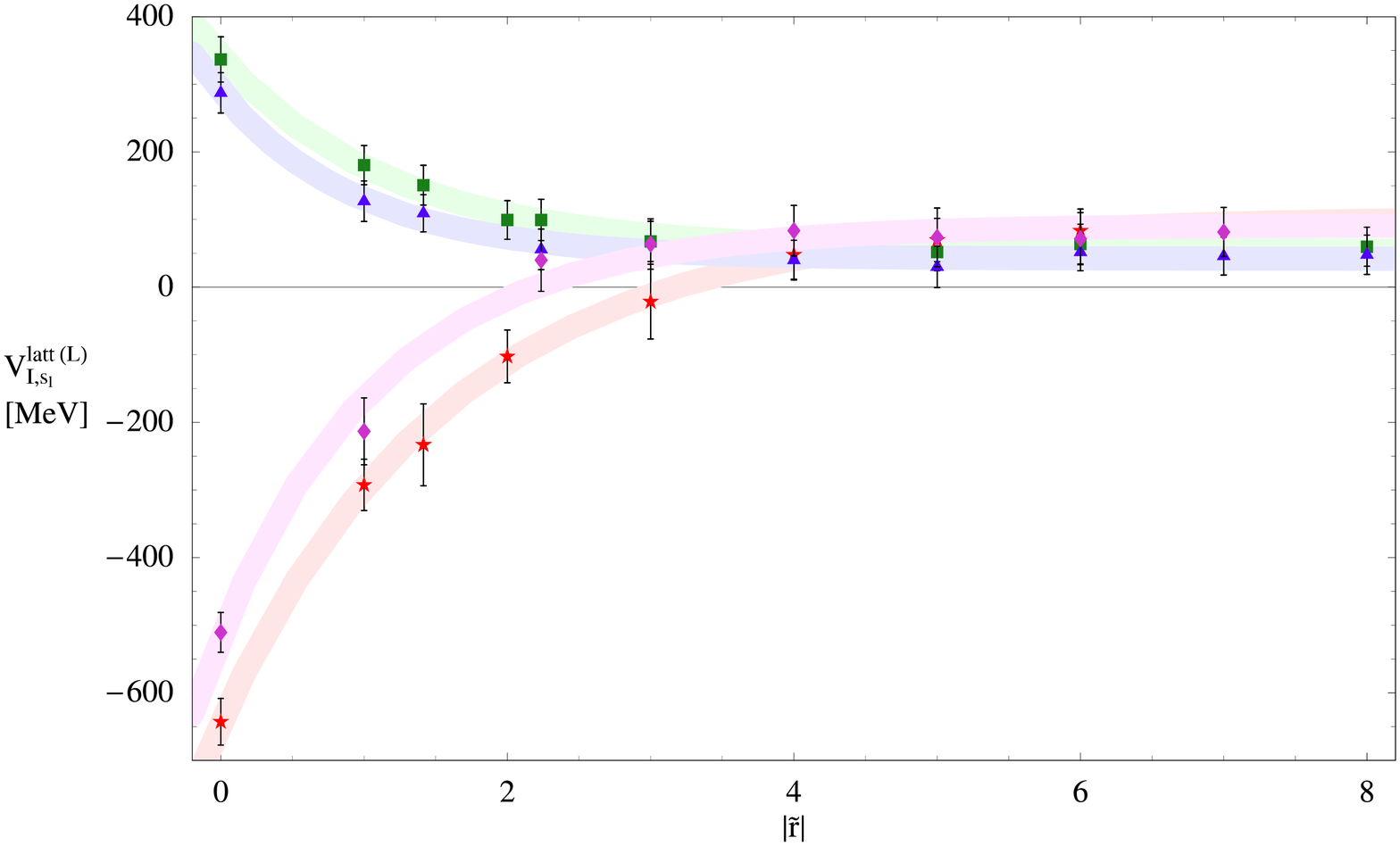}
  \caption{The finite-volume and finite-lattice spacing 
    central potentials, $V^{{\rm latt}(L)}_{I,s_l}$, 
    extracted from the lattice calculation.
    The shaded regions are simple
    fits to guide the eye. Red stars, green squares, blue triangles
    and magenta diamonds correspond to
    $(I,s_l)\ = \ (0,0),\,(1,0),\,(0,1),\,(1,1)$ respectively.} 
  \label{fig:VlattAll}
\end{figure}

\begin{figure}[htbp]
  \centering
  \includegraphics[width=0.95\columnwidth]{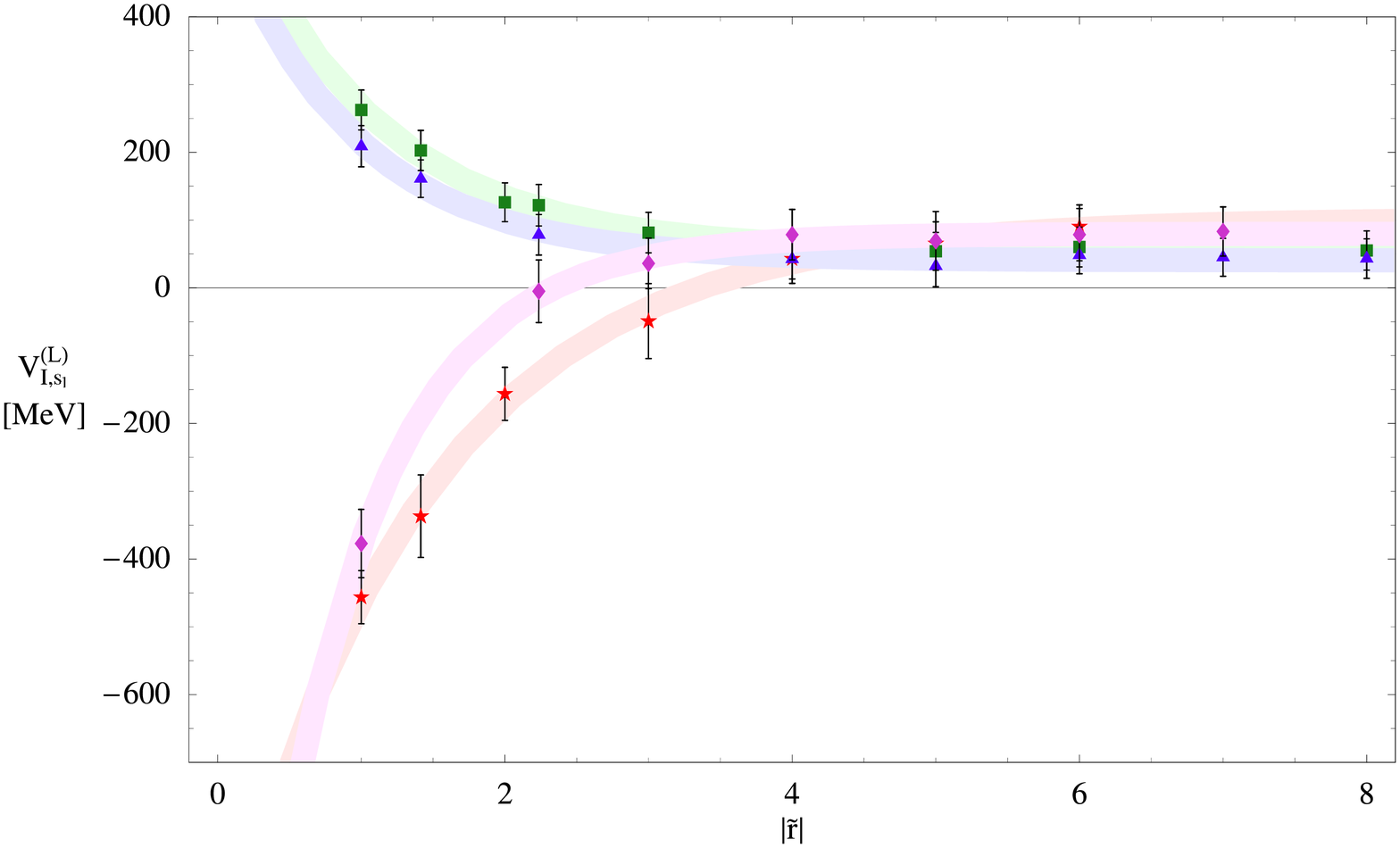}
  \caption{The finite-volume  central potentials extracted from the
    lattice calculation 
    including the leading lattice correction to OGE, $V^{(L)}_{I,s_l}$.
    The shaded regions are
    simple fits to guide the eye. Red stars, green squares, blue
    triangles and magenta diamonds correspond to
    $(I,s_l)\ = \ (0,0),\,(1,0),\,(0,1),\,(1,1)$ respectively.} 
  \label{fig:VlattAllShifted}
\end{figure}

The effective mass plots for the correlators defining the central
potentials at $|\tilde {\bf r}|=0,1,2,3,4,5,6,7,8 $ are shown in
Fig.~\ref{fig:effmassC} and those for the potentials at displacements
of $\tilde {\bf r}=(1,1,0)$ and $(2,1,0)$ are shown in
Fig.~\ref{fig:effmassSQRTS}.  It is not possible to further decompose
these latter potentials into the central and tensor components,
without additional information.  However, given that the tensor
potentials at all the other displacements are found to be very small,
it is not unreasonable to assume that they are also small for these
displacements, and therefore we can assume that they provide a good
determination of the central potentials alone.  For a number of
combinations of $I$, $s_l$ and $\tilde{\bf r}$, it was not possible to
extract a signal and these points are omitted in the effective mass
plots and tables below.
\begin{table}[!ht]
  \centering
  \begin{ruledtabular}
    \begin{tabular}{c|cccc}
      $\tilde{r}$ & $b\ {\cal E}^{BB}_{00}(\tilde{r})$ & $b\ V^{{\rm latt}(L)}_{I=S=0}(\tilde{r})$ & $b\
      V^{(L)}_{I=S=0}(\tilde{r})$ & $V^{(L)}_{I=S=0}(\tilde{r})$ [MeV] \\
      \hline
      0          & $0.783(11)(02)$ & $-0.325(13)(10)$ & $-\infty$ & $-\infty$  \\
      1          & $0.960(12)(08)$ & $-0.148(14)(13)$ & $-0.231(15)(13)$ & $-456(30)(25)$ \\
      $\sqrt{2}$ & $0.990(27)(07)$ & $-0.118(28)(12)$ & $-0.170(28)(12)$ & $-337(56)(24)$ \\
      2          & $1.056(13)(08)$ & $-0.052(15)(13)$ & $-0.079(15)(13)$ & $-156(30)(25)$ \\
      $\sqrt{5}$ & --- & ---          & ---  & ---                            \\
      3          & $1.097(20)(15)$ & $-0.011(21)(18)$  &  $-0.025(21)(18)$  & $-49(42)(36)$ \\
      4          & $1.132(12)(06)$ & $0.024(14)(12)$ & $0.022(14)(11)$   & $43(28)(23)$  \\
      5          & $1.143(09)(05)$ & $0.035(12)(11)$   & $0.033(12)(11)$  & $65(23)(22)$  \\
      6          & $ 1.150(09)(05)$ & $0.042(12)(11)$   & $0.046(12)(11)$  & $90(23)(22)$  \\
      7          & --- & ---          & ---         & ---   \\
      8          & --- & ---          & ---  & ---   \\
    \end{tabular}
  \end{ruledtabular}
  \caption{The extracted lattice and continuum potentials in the $I=S=0$
    channel.  The first uncertainty is statistical, while the second 
    is systematic.} 
  \label{tab:V00}
\end{table}
\begin{table}[!ht]
  \centering
  \begin{ruledtabular}
    \begin{tabular}{c|cccc}
      $\tilde{r}$ & $b\ {\cal E}^{BB}_{10}(\tilde{r})$ & $b\ V^{{\rm latt}(L)}_{I=1,S=0}(\tilde{r})$ & $b\
      V^{(L)}_{I=1,S=0}(\tilde{r})$ & $V^{(L)}_{I=1,S=0}(\tilde{r})$ [MeV] \\
      \hline
      0          & $1.278(10)(05)$ & $+0.170(12)(11)$ & $+\infty$ & $+\infty$  \\
      1          & $1.199(07)(03)$ & $+0.091(10)(10)$ & $+0.133(10)(10)$ & $+262(21)(21)$ \\
      $\sqrt{2}$ & $1.184(08)(01)$ & $+0.076(11)(10)$ & $+0.102(11)(10)$ & $+203(22)(20)$ \\
      2          & $1.158(07)(02)$ & $+0.050(10)(10)$ & $+0.064(10)(10)$ & $+126(20)(20)$ \\
      $\sqrt{5}$ & $1.158(09)(01)$ & $+0.050(12)(10)$ & $+0.062(12)(10)$ & $+122(23)(20)$  \\
      3          & $1.142(06)(06)$ & $+0.034(10)(12)$  &  $+0.041(10)(12)$  & $+82(19)(23)$ \\
      4          & --- & --- & ---   & ---  \\
      5          & $1.134(05)(04)$ & $+0.026(09)(11)$   & $+0.027(09)(11)$  & $+54(18)(21)$  \\
      6          & $1.140(07)(04)$ & $+0.032(10)(11)$   & $+0.030(10)(11)$  & $+60(20)(21)$  \\
      7          & --- & ---          & ---         & ---   \\
      8          & $1.138(07)(03)$ & $+0.030(10)(10)$& +$0.028(10)(10)$  & $+55(20)(21)$   \\
    \end{tabular}
  \end{ruledtabular}
  \caption{The extracted lattice and continuum potentials in the $I=1, S=0$
    channel.  The first uncertainty is statistical, while the second
    is systematic.} 
  \label{tab:V10}
\end{table}
\begin{table}[!ht]
  \centering
  \begin{ruledtabular}
    \begin{tabular}{c|cccc}
      $\tilde{r}$ & $b\ {\cal E}^{BB}_{11}(\tilde{r})$ & $b\ V^{{\rm latt}(L)}_{I=S=1}(\tilde{r})$ & $b\
      V^{(L)}_{I=S=1}(\tilde{r})$ & $V^{(L)}_{I=S=1}(\tilde{r})$ [MeV] \\
      \hline
      0          & $0.850(06)(04)$ & $-0.258(09)(11)$ & $-\infty$ & $-\infty$  \\
      1          & $1.000(12)(18)$ & $-0.108(14)(21)$ & $-0.191(15)(21)$ & $-377(30)(41)$ \\
      $\sqrt{2}^*$ &  --- & --- & --- & --- \\
      2          & --- & --- & --- & --- \\
      $\sqrt{5}^*$ & $1.128(18)(08)$ & $+0.020(20)(13)$ & $-0.003(20)(13)$ & $-05(38)(25)$   \\
      3          & $1.140(10)(10)$ & $+0.032(12)(14)$  &  $+0.018(12)(14)$  & $+36(25)(28)$ \\
      4          & $1.150(10)(10)$ & $+0.042(12)(14)$ & $+0.040(12)(14)$   & $+78(25)(28)$  \\
      5          & $1.145(10)(15)$ & $+0.037(12)(18)$   & $+0.035(12)(18)$  & $+69(25)(36)$  \\
      6          & $1.1544(12)(09)$ & $+0.036(14)(13)$   & $+0.040(14)(13)$  & $+79(28)(27)$  \\
      7          & $1.149(10)(09)$ & $0.041(12)(13)$ & $0.042(12)(13)$  & $+83(25)(27)$   \\
      8          & --- & ---          & ---  & ---   \\
    \end{tabular}
  \end{ruledtabular}
  \caption{The extracted lattice and continuum central potentials in the $I=S=1$
    channel.  The first uncertainty is statistical, while the second is
    systematic.  The  asterisks attached to the $r=\sqrt{2}$ and $\sqrt{5}$
    potentials indicate that these are the sum of the central and
    tensor contributions.}  
  \label{tab:V11}
\end{table}
\begin{table}[!ht]
  \centering
  \begin{ruledtabular}
    \begin{tabular}{c|cccc}
      $\tilde{r}$ & $b\ {\cal E}^{BB}_{01}(\tilde{r})$ & $b\ V^{{\rm latt}(L)}_{I=0,S=1}(\tilde{r})$ & $b\
      V^{(L)}_{I=0,S=1}(\tilde{r})$ & $V^{(L)}_{I=0,S=1}(\tilde{r})$ [MeV] \\
      \hline
      0          & $1.253(08)(02)$ & $+0.145(11)(10)$ & $+\infty$ & $+\infty$  \\
      1          & $1.172(07)(05)$ & $+0.064(10)(11)$ & $+0.106(10)(11)$ & $+209(21)(22)$ \\
      $\sqrt{2}^*$ & $1.163(06)(01)$ & $+0.055(10)(10)$ & $0.081(10)(10)$ & $+161(19)(20)$ \\
      2          & --- & --- & --- & --- \\
      $\sqrt{5}^*$ & $1.136(08)(03)$ & $+0.028(11)(10)$ & $+0.040(11)(10)$ & $+78(22)(21)$   \\
      3          & --- & ---  &  ---  & --- \\
      4          & $1.128(07)(04)$ & $+0.020(10)(11)$ & $+0.021(10)(11)$   & $+42(20)(21)$  \\
      5          & $1.123(07)(06)$ & $+0.015(10)(11)$   & $+0.016(10)(12)$  & $+32(20)(23)$  \\
      6          & $1.134(06)(02)$ & $+0.026(10)(10)$   & $+0.024(10)(10)$  & $+48(19)(20)$  \\
      7          & $1.149(10)(09)$ & $0.041(12)(13)$ & $0.042(12)(13)$  & $+83(25)(27)$   \\
      8          & $1.131(06)(03)$ & $0.023(10)(10)$ & $0.021(10)(10)$  & $+41(19)(21)$   \\
    \end{tabular}
  \end{ruledtabular}
  \caption{The extracted lattice and continuum central potentials in the $I=0 ,S=1$
    channel.  The first uncertainty is statistical, while the second is
    systematic.  The  asterisks attached to the $r=\sqrt{2}$ and $\sqrt{5}$
    potentials indicate that these are the sum of the central and
    tensor contributions.
} 
  \label{tab:V01}
\end{table}
After applying the perturbative one-loop matching discussed above, we
determine the various finite-volume potentials.  The central
potentials extracted from the lattice calculation in each of the
spin-isospin channels are given in Tables~\ref{tab:V00}-\ref{tab:V01}.
The lattice central potentials, $V^{{\rm latt}(L)}_{I,s_l}$, are shown
in Fig.~\ref{fig:VlattAll}, and the central potentials with the
leading order finite-lattice spacing correction included (as discussed
above), $V^{(L)}_{I,s_l}$, are shown in
Fig.~\ref{fig:VlattAllShifted}.  In each channel there is a clean
signal for the central potentials, with two or more of the
displacements having potentials that are clearly non-zero. For the
$s_l=1$ channels the tensor potentials are found to be consistent with
zero and are smaller than $V_T\sim 40~{\rm MeV}$ in both cases for the
entire range of displacements (the tensor potentials were also found
to small and poorly determined in Ref.~\cite{Michael:1999nq}).  At
large distances, this is not consistent with our expectations from the
NN system at the physical value of $m_\pi$, but may result from the
relevant $B^{(\ast)}B^{(\ast)} M$ couplings ($M$ represents the various mesons)
being small or giving rise to cancellations, or from the unphysically
large pion mass.  Further studies of this issue are warranted.  The
measurements at $\tilde{\bf r}=0$ contain no information about the
continuum potentials, which diverge as
$\sim\overline{\alpha}(r^{-1})/r$. As discussed in Sections
\ref{sec:hadron-spectrum} and \ref{sec:exotics}, the lattice energies
measured for coincident $B$-mesons, in fact, determine the energies of
the $\Lambda_b$ and $\Sigma_b$ and their exotic partners.
%

%%%%%%%%%%%%%%%%%%%%%%%%%%%%%%
\subsection{Potentials with $t$-Channel Quantum Numbers}

Up until this point we have classified the potentials between the
B-mesons in terms of the $s$-channel quantum numbers, the total
isospin and spin of the ldof.  These potentials are extracted from the
energies calculated on the lattice by subtracting twice the $B$-meson
mass.  The statistical and systematic uncertainty in determining the
B-meson mass propagates through to all four central potentials,
leading to larger uncertainties in the potential than from the
calculation of the energy of the two $B$-mesons alone.  Motivated by
the success of traditional nuclear physics phenomenological potentials
constructed from the exchange of mesons in the $t$-channel, we have
formed linear combinations of the $s$-channel potentials to give
potentials with well-defined spin and isospin quantum numbers that can
be identified with the exchange of one or more hadrons.  The central
potential can be decomposed as
\begin{eqnarray}
  \label{eq:tchannelpots}
  V_{I,s_l}(|{\bf r}|) = V_{1} (|{\bf r}|) \ +\  
\sigma_1\cdot\sigma_2\  V_{ \sigma }(|{\bf r}|) 
\ +\ \sigma_1\cdot\sigma_2\,\tau_1\cdot\tau_2\,V_{\sigma\tau} (|{\bf r}|) 
\ +\  \tau_1\cdot\tau_2\ V_{ \tau} (|{\bf r}|)
\ \ ,
\end{eqnarray}
where it is straightforward to show that, in terms of the s-channel
central potentials,
\begin{eqnarray}
  \label{eq:Vrels}
  V_{1} &=& \frac{1}{16}\left(V_{0,0} + 3 V_{0,1} + 3 V_{1,0} + 9
  V_{1,1}\right) \,,
\nonumber\\
V_{{\sigma}} &=& \frac{1}{16}\left( -V_{0,0} + V_{0,1} - 3 V_{1,0} + 3
V_{1,1} \right) \,,
\nonumber\\
V_{{\sigma\tau}} &=& {1\over 16} \left( V_{0,0} - V_{0,1} - V_{1,0} +
  V_{1,1} \ \right) \,,
\nonumber\\
V_{{\tau}} &=& \frac{1}{16}\left(-V_{0,0} - 3 V_{0,1} + V_{1,0} + 3 V_{1,1}\right)
\,.
\end{eqnarray}
An important point to observe is that the three potentials,
$V_{{\sigma}}$, $V_{{\sigma\tau}}$ and $V_{{\tau}}$ can be extracted
from the lattice calculation without reference to the $B$-meson mass.
This is not true for $ V_{1}$.  It was pointed out in
Ref.~\cite{Beane:2002nu} that the $\eta^\prime$ double-pole that is
present in quenched calculations will dominate the interactions
between nucleons and $B$-mesons at long-distances.  However, we see
that this can only contribute to $V_{{\sigma}}$, and not to the other
three potentials.  Further, the exchange of a single $\pi$ will
contribute only to the $V_{{\sigma\tau}}$ potential.  Therefore, the
potential $V_{{\tau}}$ does not receive contributions from hairpins
nor from $\pi$-exchange, and does not depend upon the $B$-meson mass
extraction from the lattice calculation. It is expected to be clean,
and determined by short-range and medium range interactions.

\begin{table}[!ht]
  \centering
  \begin{ruledtabular}
    \begin{tabular}{c|cccc}
      $\tilde{r}$ & $b\ V^{{\rm latt}(L)}_{\sigma}(\tilde{r})$ & $b\
      V^{(L)}_{\sigma}(\tilde{r})$ & $V^{(L)}_{\sigma}(\tilde{r})$ [MeV] \\
      \hline
 0 & -0.050(17)(06)& $-\infty$ & $-\infty$ \\
 1 & -0.0174(33)(08)& -0.0337(34)(08) & -66.8(6.8)(1.7) \\
 $\sqrt{2}^* $ & -0.0081(25)(02) & -0.0178(25)(02) & -35.1(5.1)(0.3) \\
%2 & -0.0050(82)(42) & -0.0144(82)(42) & -29(16)(08) \\
2 & --- & --- & --- \\
$ \sqrt{5}^* $ & -0.0016(25)(09) & -0.0049(25)(09) & -9.7(5.0)(1.8) \\
 3 & +0.0035(13)(00) & +0.0008(12)(00) & +1.6(2.6)(0.1) \\
 4 & +0.0048(13)(08) & +0.0035(13)(08) & +6.9(2.6)(1.5) \\
 5 & +0.0054(12)(05) & +0.0044(12)(05) & +8.8(2.4)(1.0) \\
 6 & --- & --- & --- \\
 7 & +0.0054(18)(02) & +0.0054(18)(02) & +10.7(3.6)(0.3) \\
 8 & +0.0047(16)(03) & +0.0053(16)(03) & +10.5(3.1)(0.6)
       \end{tabular}
  \end{ruledtabular}
  \caption{The extracted lattice and continuum central potential
    $V^{(L)}_{\sigma}$, defined in Eq.~(\protect\ref{eq:Vrels}).   
    The first uncertainty is statistical, while the second is
    systematic.  The asterisks attached to the $r=\sqrt{2}$ and $\sqrt{5}$
    potentials indicate that these are the sum of the central and
    tensor contributions. } 
  \label{tab:Vsig}
\end{table}
\begin{table}[!ht]
  \centering
  \begin{ruledtabular}
    \begin{tabular}{c|cccc}
      $\tilde{r}$ & $b\ V^{{\rm latt}(L)}_{\sigma \tau}(\tilde{r})$ & $b\
      V^{(L)}_{\sigma\tau}(\tilde{r})$ & $V^{(L)}_{\sigma \tau}(\tilde{r})$ [MeV] \\
      \hline
 0 & -0.0539(56)(07)& $-\infty$ & $-\infty$ \\
 1 & -0.0228(24)(08)& -0.0383(25)(08) & -75.9(5.2)(1.5) \\
 $\sqrt{2}^* $ & -0.0158(21)(02) & -0.0256(22)(02) & -50.8(4.4)(0.3) \\
2 & -0.0064(22)(07) & -0.0115(22)(07) & -22.7(4.3)(1.3) \\
$ \sqrt{5}^* $ & -0.0062(42)(05) & -0.0105(42)(05) & -20.7(8.4)(1.0) \\
 3 & -0.0003(23)(12) & -0.0029(22)(12) & -5.7(4.5)(2.3) \\
 4 & +0.00205(82)(50) & +0.00158(82)(50) & +3.1(1.6)(1.0) \\
 5 & +0.00343(09)(32) & +0.0030(09)(03) & +6.0(1.9)(0.6) \\
 6 & --- & --- & --- \\
 7 & +0.00294(55)(07) & +0.00309(55)(07) & +6.1(1.1)(0.1) \\
 8 & --- & --- & ---
       \end{tabular}
  \end{ruledtabular}
  \caption{The extracted lattice and continuum central potential $V^{(L)}_{\sigma
      \tau}$, defined in Eq.~(\protect\ref{eq:Vrels}).  
    The first uncertainty is statistical, while the second is
    systematic.  The asterisks attached to the $r=\sqrt{2}$ and $\sqrt{5}$
    potentials indicate that these are the sum of the central and
    tensor contributions. } 
  \label{tab:Vsigtau}
\end{table}
\begin{table}[!ht]
  \centering
  \begin{ruledtabular}
    \begin{tabular}{c|cccc}
      $\tilde{r}$ & $b\ V^{{\rm latt}(L)}_{ \tau}(\tilde{r})$ & $b\
      V^{(L)}_{\tau}(\tilde{r})$ & $V^{(L)}_{ \tau}(\tilde{r})$ [MeV] \\
      \hline
 0 & -0.0390(49)(03)& $-\infty$ & $-\infty$ \\
 1 & -0.0136(39)(02)& -0.0292(40)(02) & -57.7(7.8)(0.4) \\
 $\sqrt{2}^* $ & -0.0033(31)(05) & -0.0131(31)(05) & -25.9(6.3)(1.0) \\
2 & +0.00539(66)(02) & +0.00029(70)(02) & +0.58(1.4)(0.0) \\
$ \sqrt{5}^* $ & +0.00587(53)(15) & +0.00161(57)(15) & +3.2(1.2)(0.3) \\
 3 & +0.00787(71)(12) & +0.00525(73)(12) & +10.4(1.4)(0.2) \\
 4 & +0.00767(61)(22) & +0.00720(61)(22) & +14.2(1.2)(0.4) \\
 5 & +0.00664(97)(04) & +0.00623(97)(04) & +12.3(1.9)(0.1) \\
 6 & --- & --- & --- \\
 7 & +0.00509(50)(07) & +0.00525(50)(07) & +10.4(1.0)(0.1) \\
 8 & +0.00455(62)(24) & +0.00544(62)(24) & +10.8(1.2)(0.5)
       \end{tabular}
  \end{ruledtabular}
  \caption{The extracted lattice and continuum central potential $V^{(L)}_{
      \tau}$, defined in Eq.~(\protect\ref{eq:Vrels}).  
    The first uncertainty is statistical, while the second is
    systematic.  The asterisks attached to the $r=\sqrt{2}$ and $\sqrt{5}$
    potentials indicate that these are the sum of the central and
    tensor contributions. } 
  \label{tab:Vtau}
\end{table}
\begin{table}[!ht]
  \centering
  \begin{ruledtabular}
    \begin{tabular}{c|cccc}
      $\tilde{r}$ & $b\ V^{{\rm latt}(L)}_{ 1}(\tilde{r})$ & $b\
      V^{(L)}_{ 1}(\tilde{r})$ & $V^{(L)}_{1}(\tilde{r})$ [MeV] \\
      \hline
 0 & -0.117(16)(03)& $-\infty$ & $-\infty$ \\
 1 & -0.0616(96)(25)& -0.0520(96)(25) & -103(19)(04) \\
 $\sqrt{2}^* $ & -0.0193(74)(04) & -0.0210(74)(04) & -42(15)(0.7) \\
2 & -0.0135(58)(18) & -0.0101(58)(18) & -20(12)(4) \\
$ \sqrt{5}^* $ & -0.0020(78)(10) & -0.0019(78)(10) & -4(16)(2) \\
 3 & +0.0030(43)(17) & +0.0037(43)(17) & +7.4(8.4)(3.3) \\
 4 & +0.0093(45)(10) & +0.0094(45)(10) & +16.6(8.9)(2.0) \\
 5 & +0.0069(19)(01) & +0.0064(19)(01) & +12.6(3.8)(0.1) \\
 6 & --- & --- & --- \\
 7 & +0.0074(43)(14) & +0.0057(43)(14) & +11.2(8.4)(2.8) \\
 8 & +0.0051(29)(07) & +0.0055(29)(07) & +10.8(5.7)(1.4)
       \end{tabular}
  \end{ruledtabular}
  \caption{The extracted lattice and continuum central potential
    $V^{(L)}_{1}$, defined in Eq.~(\protect\ref{eq:Vrels}).   
    The first uncertainty is statistical, while the second is
    systematic.  The asterisks attached to the $r=\sqrt{2}$ and $\sqrt{5}$
    potentials indicate that these are the sum of the central and
    tensor contributions. } 
  \label{tab:Vone}
\end{table}
\begin{figure}[htbp]
  \centering\hspace*{-5mm}
  \includegraphics[width=1.05\columnwidth]{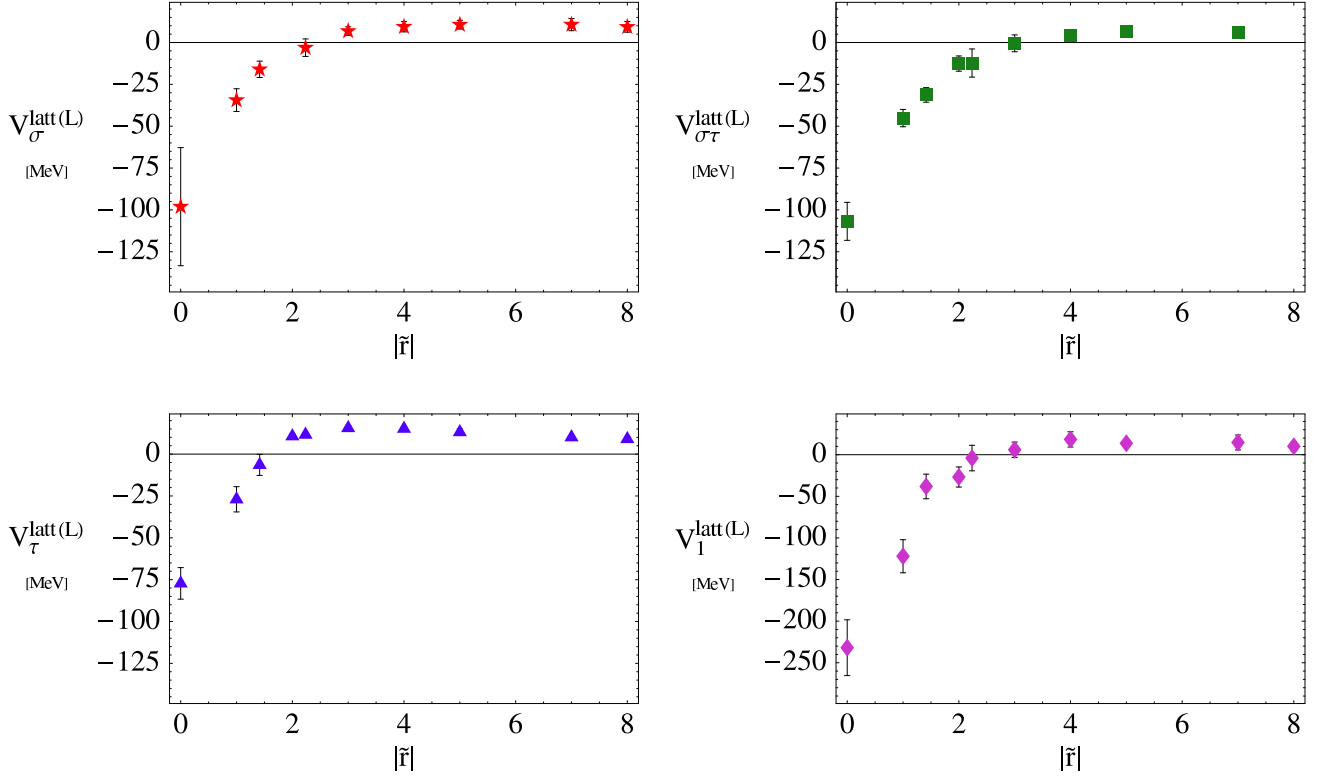}
  \caption{The central finite-volume lattice potentials $V^{{\rm
      latt}(L)}_{ \sigma, \sigma \tau, \tau, 1}$, as defined in
    Eq.~(\protect\ref{eq:Vrels}).  The statistical and systematic
    errors have been added in quadrature.  }
  \label{fig:Vtchannellatt}
\end{figure}
\begin{figure}[!thp]
  \centering
  \includegraphics[width=0.32\columnwidth]{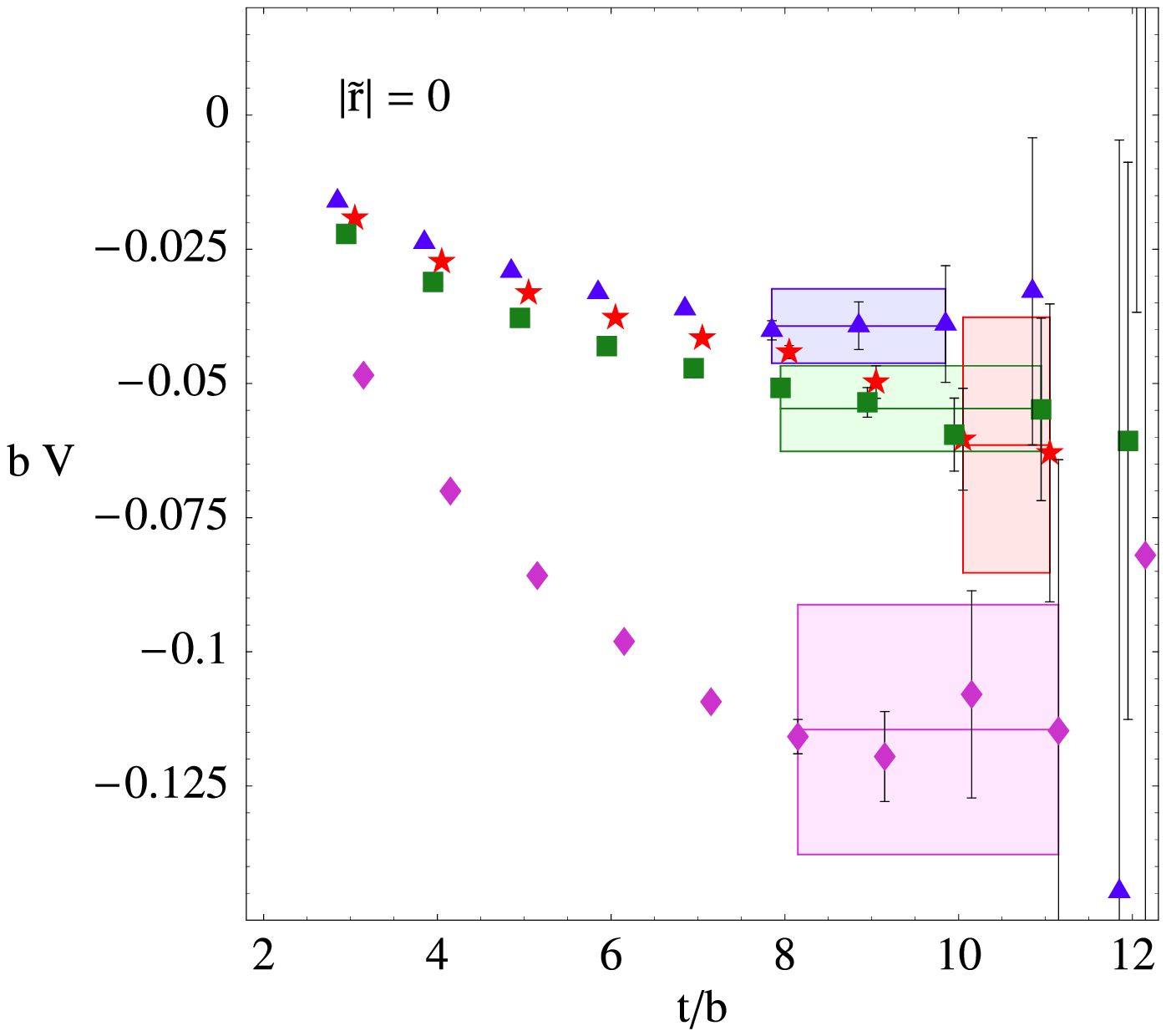}
\includegraphics[width=0.32\columnwidth]{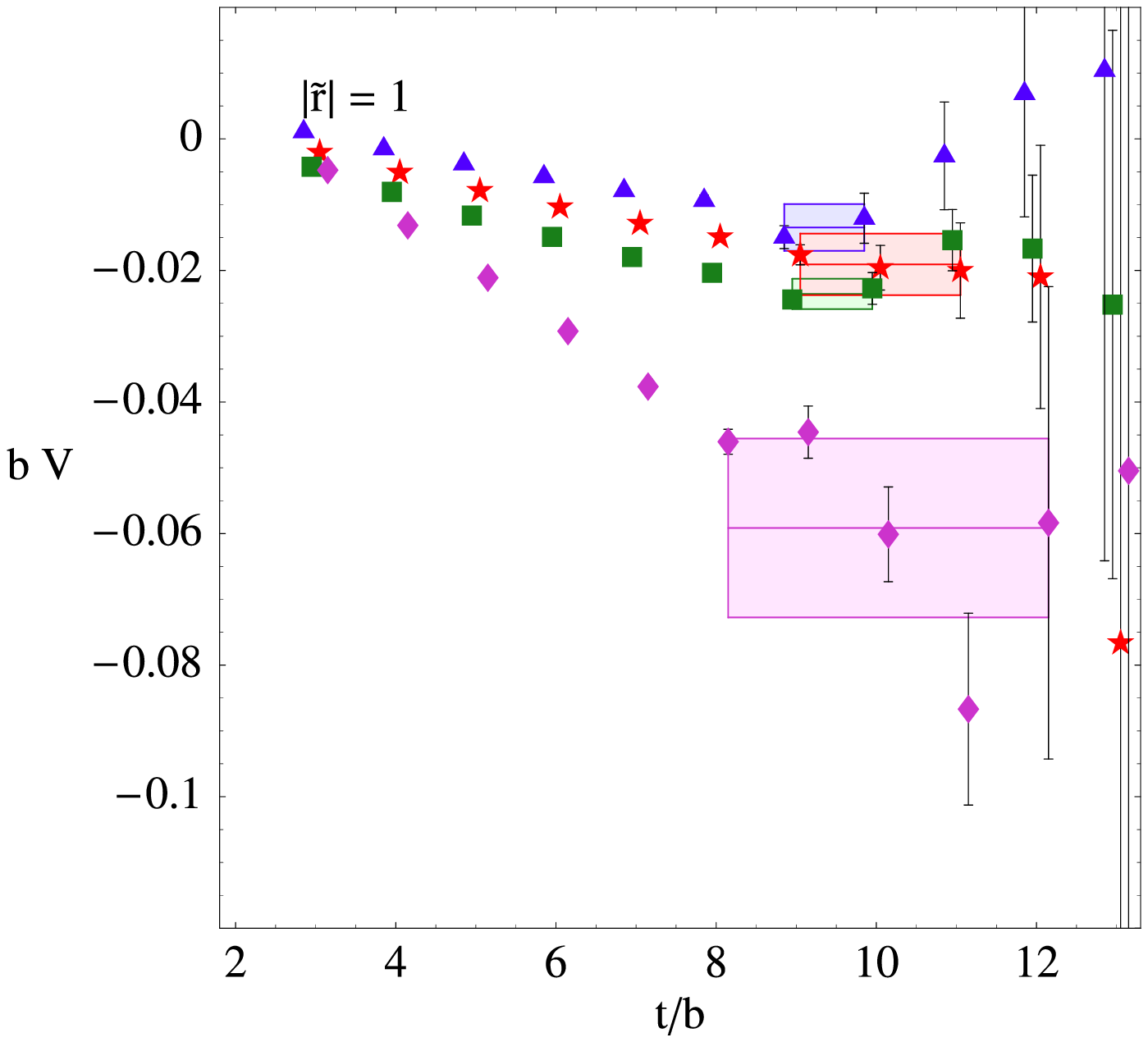}  
  \includegraphics[width=0.32\columnwidth]{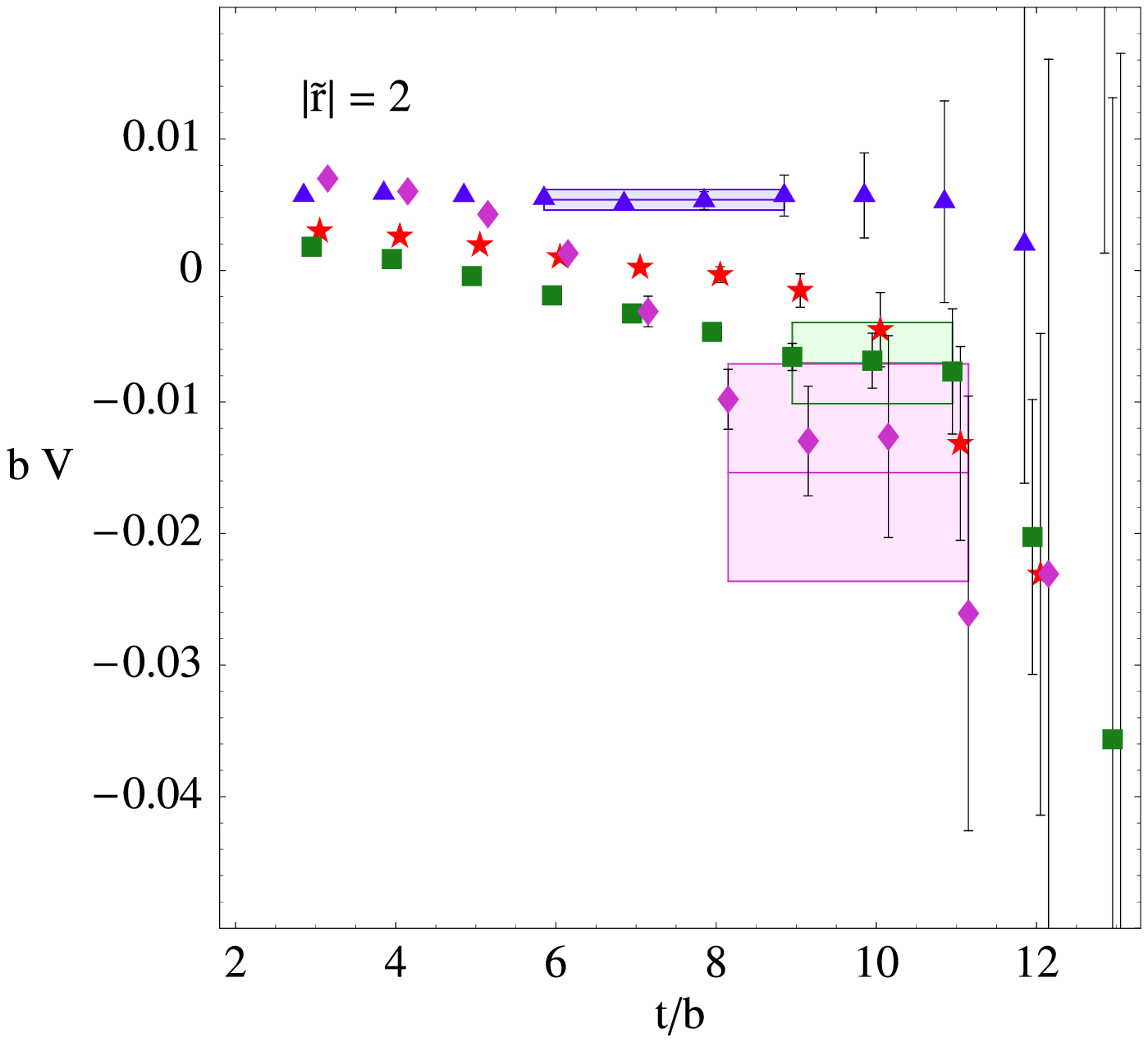} \\
\includegraphics[width=0.32\columnwidth]{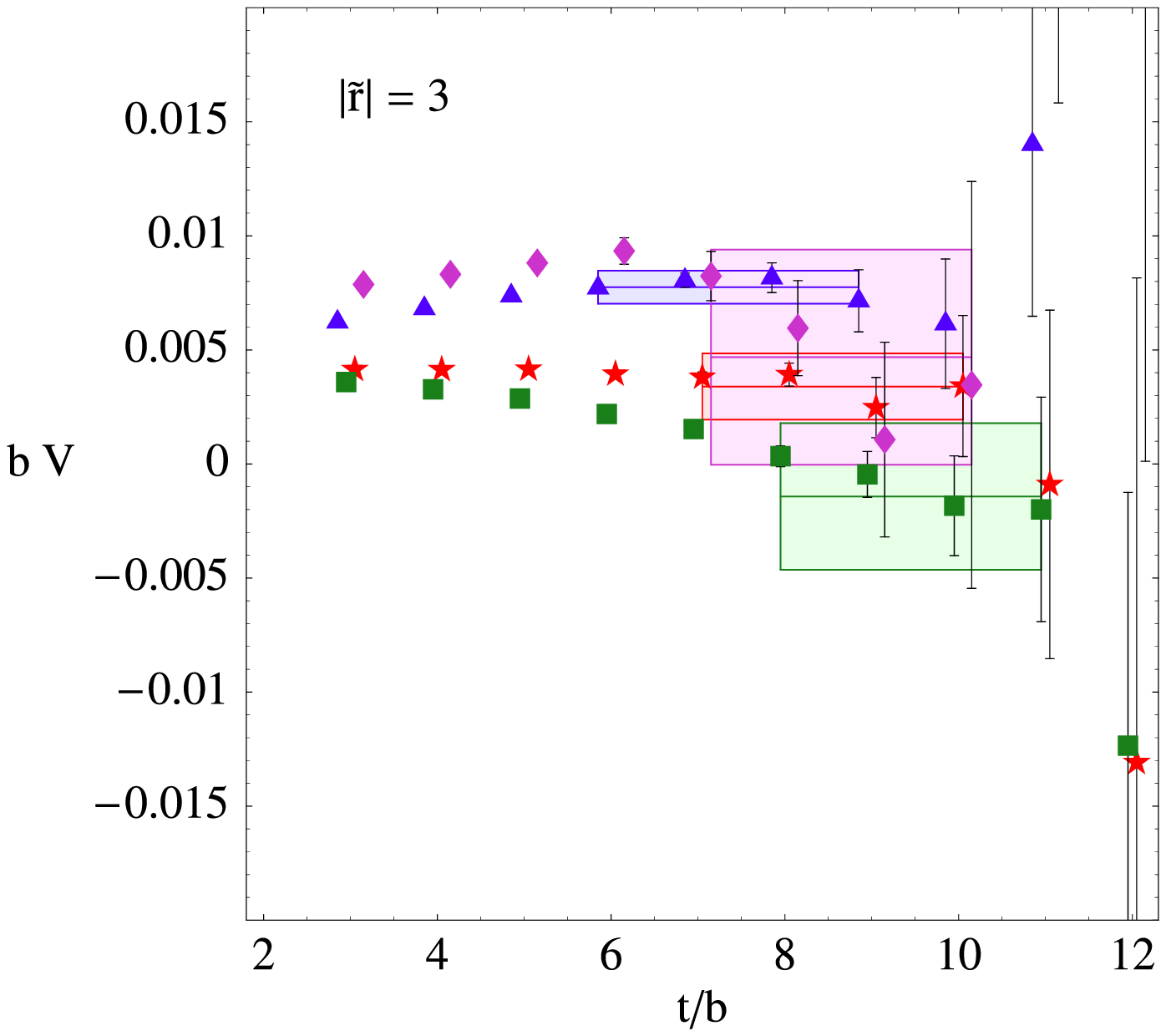} 
  \includegraphics[width=0.32\columnwidth]{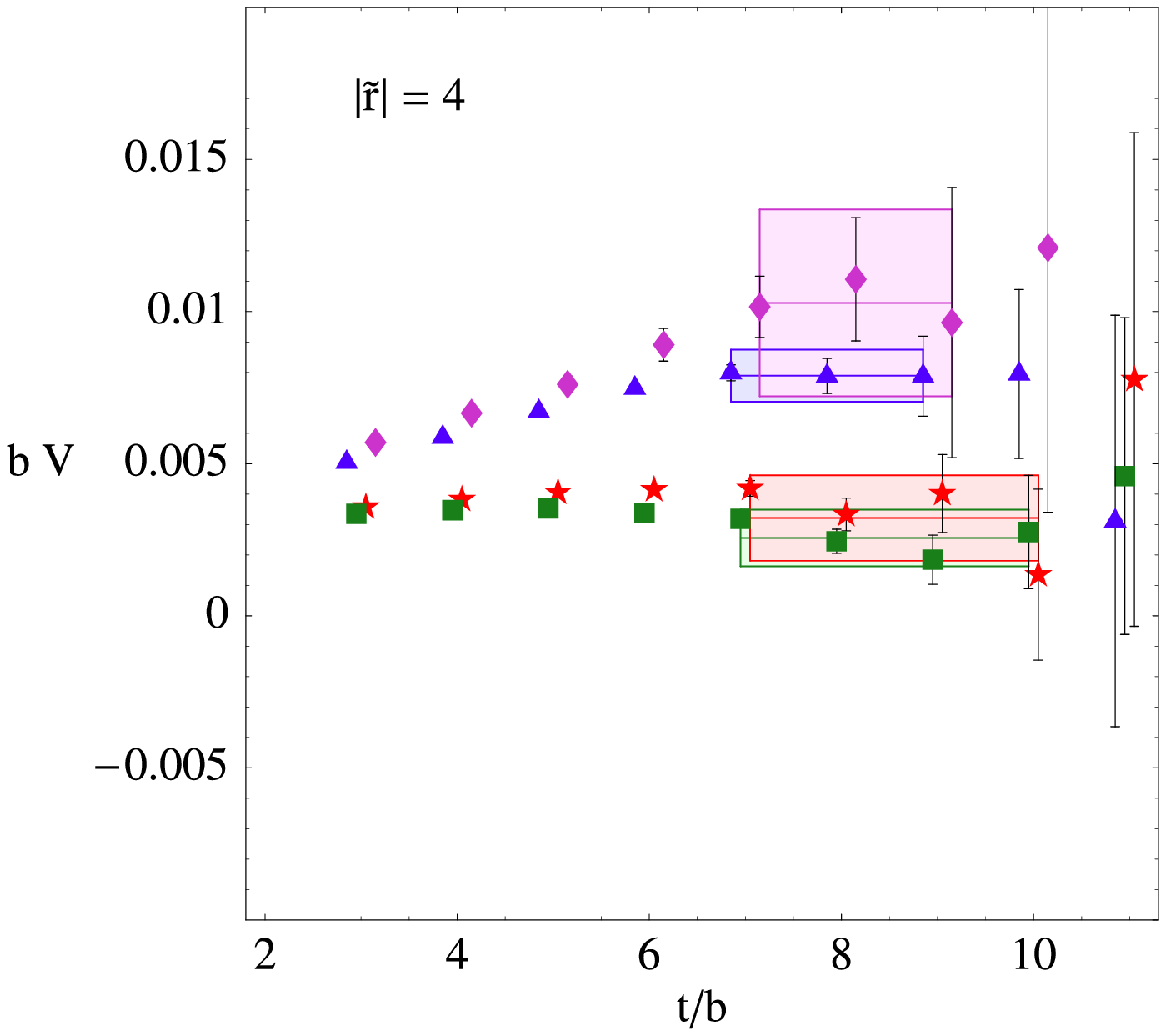}
\includegraphics[width=0.32\columnwidth]{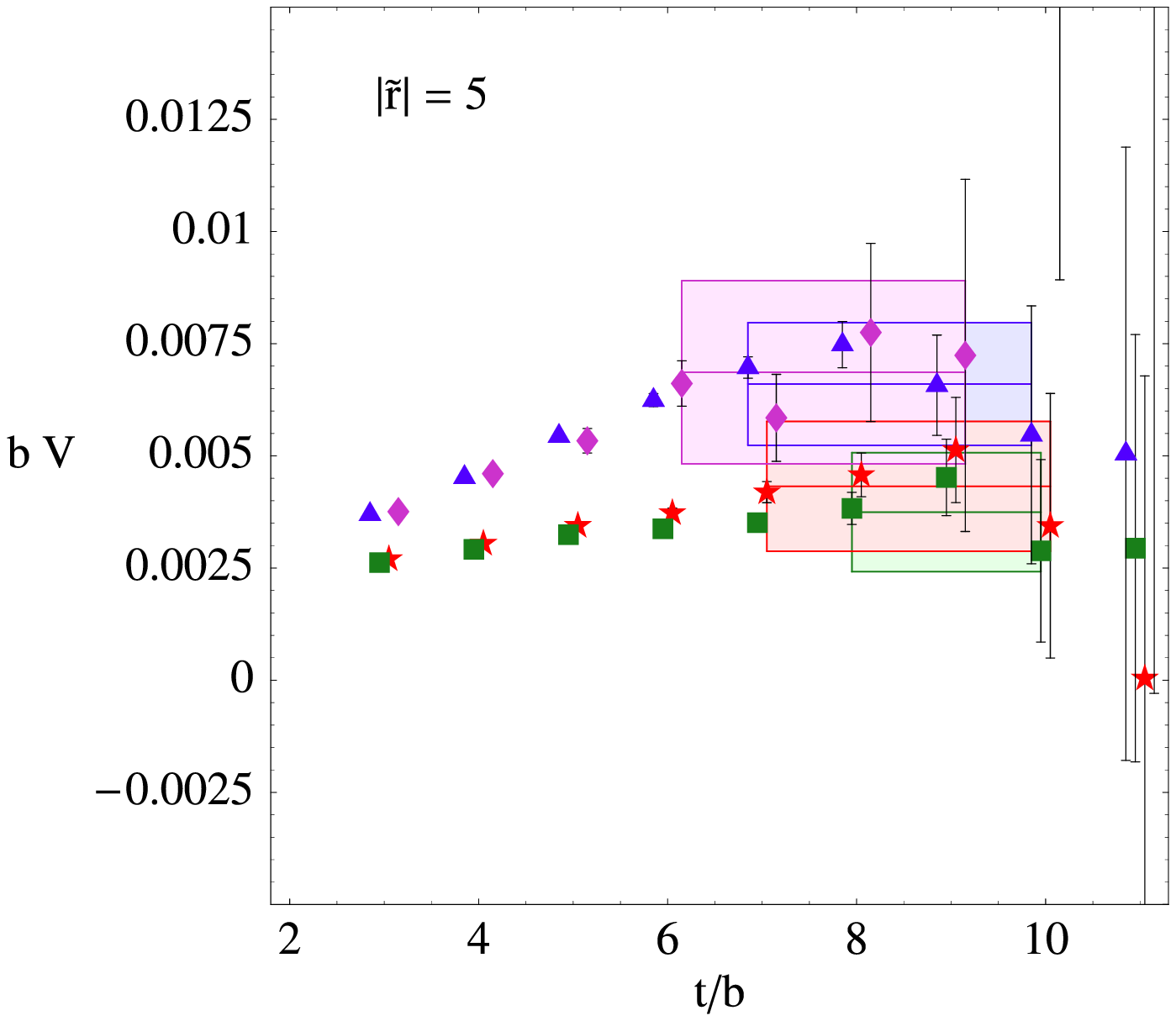}
\\
  \includegraphics[width=0.32\columnwidth]{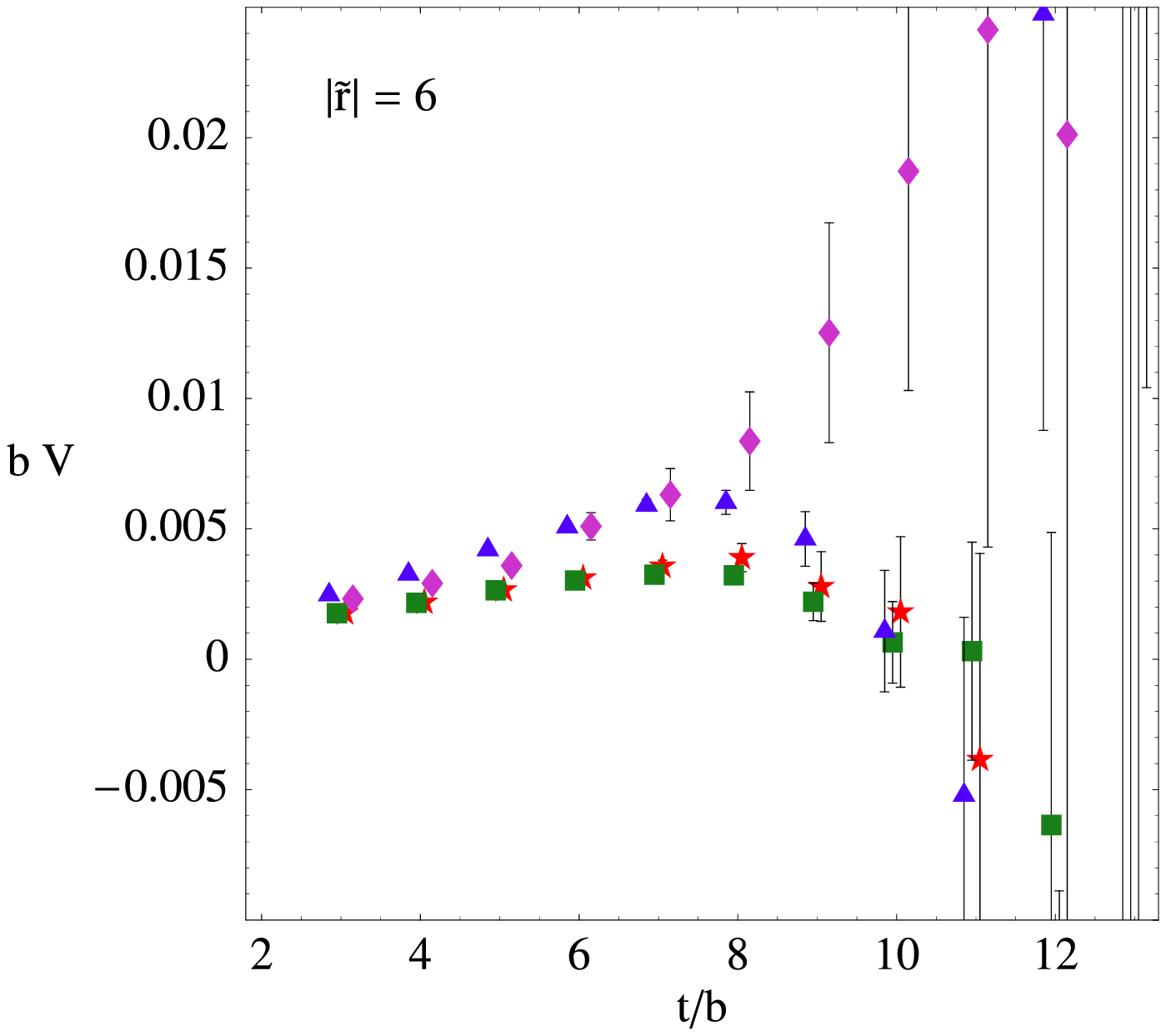}
\includegraphics[width=0.32\columnwidth]{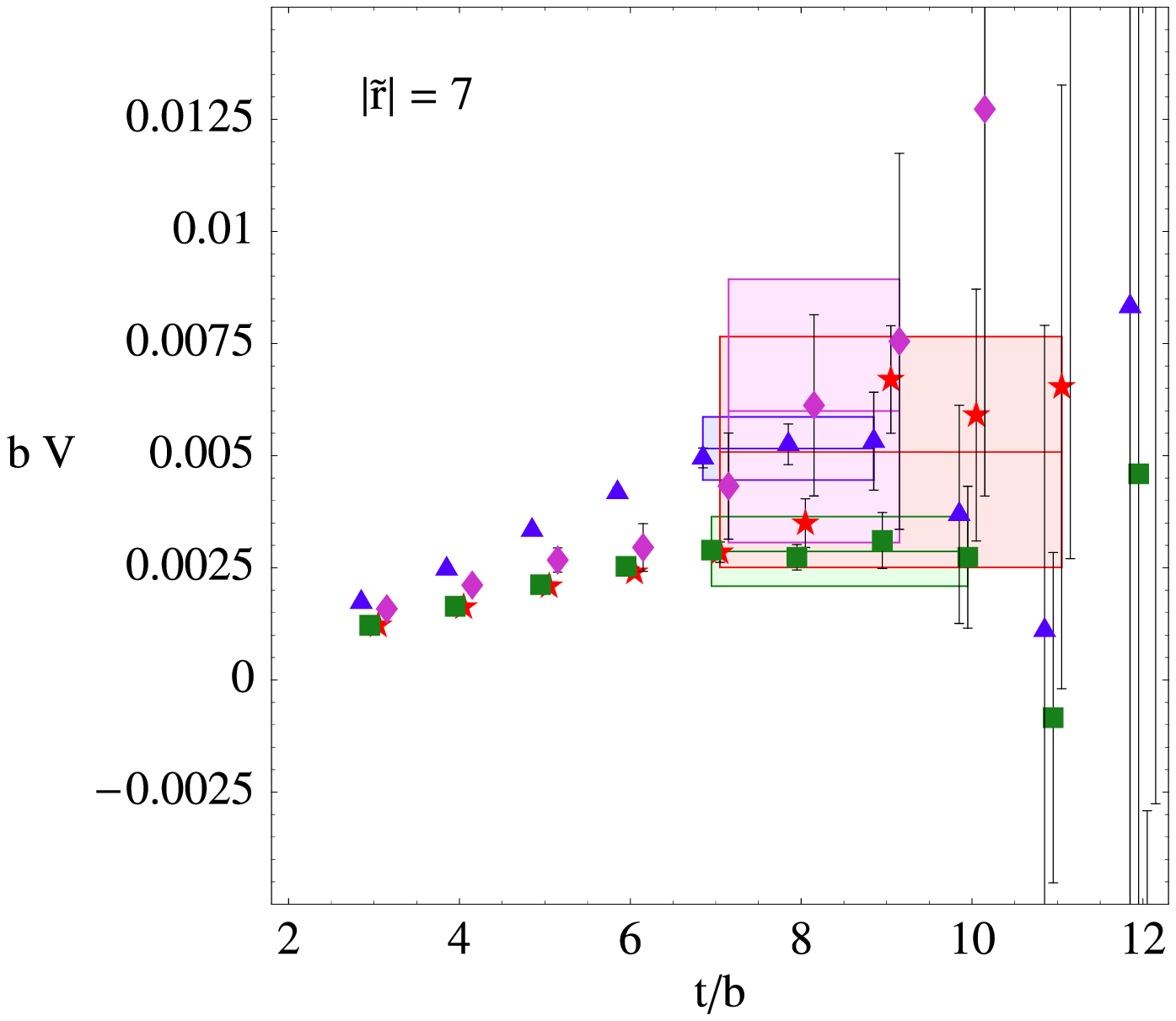}
  \includegraphics[width=0.32\columnwidth]{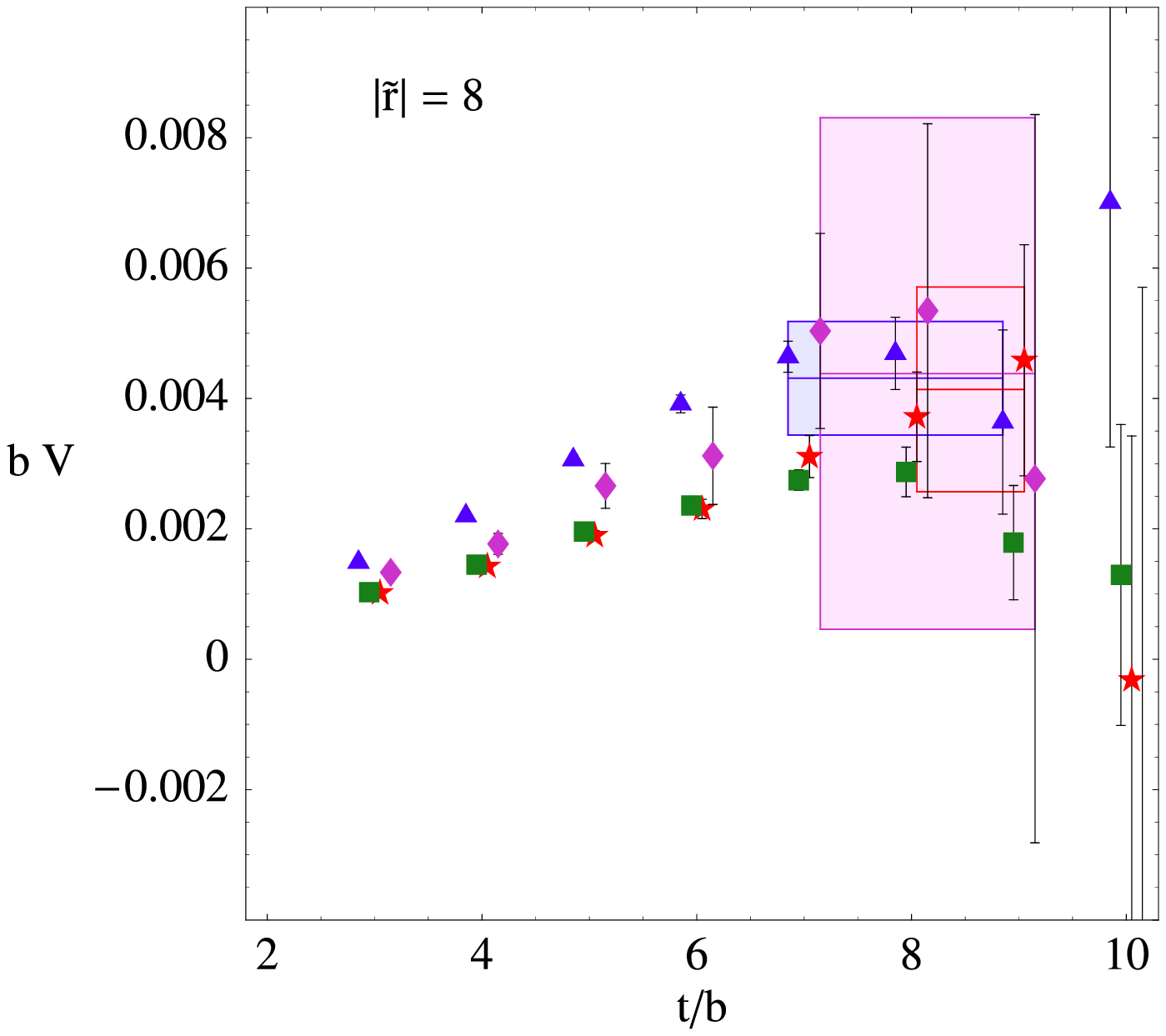}
  \caption{The effective mass plots for the central potentials at
    $|\tilde {\bf r}|=0,1,2,3,4,5,6,7,8$ for each $t$-channel potential.  Red
    stars correspond to $V_\sigma$, green squares to $V_{\sigma\tau}$,
    blue triangles to $V_\tau$ and magenta diamonds to $V_{1}$.
    Extracted masses and uncertainties are shown as the shaded regions
    in channels where a signal can be extracted.}
  \label{fig:effmassCRPES}
\end{figure}
\begin{figure}[!thp]
  \centering
  \includegraphics[width=0.32\columnwidth]{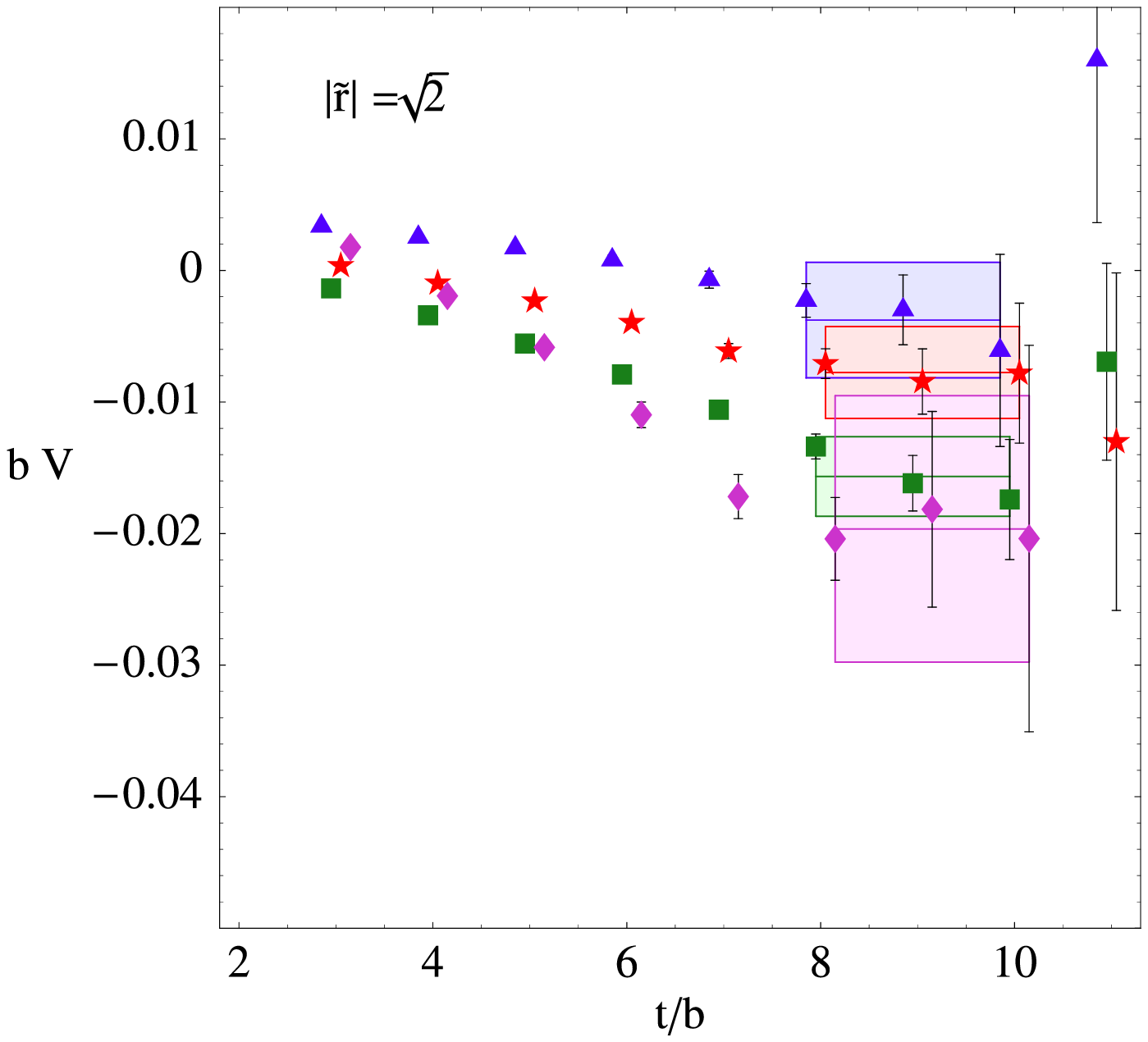}
\includegraphics[width=0.32\columnwidth]{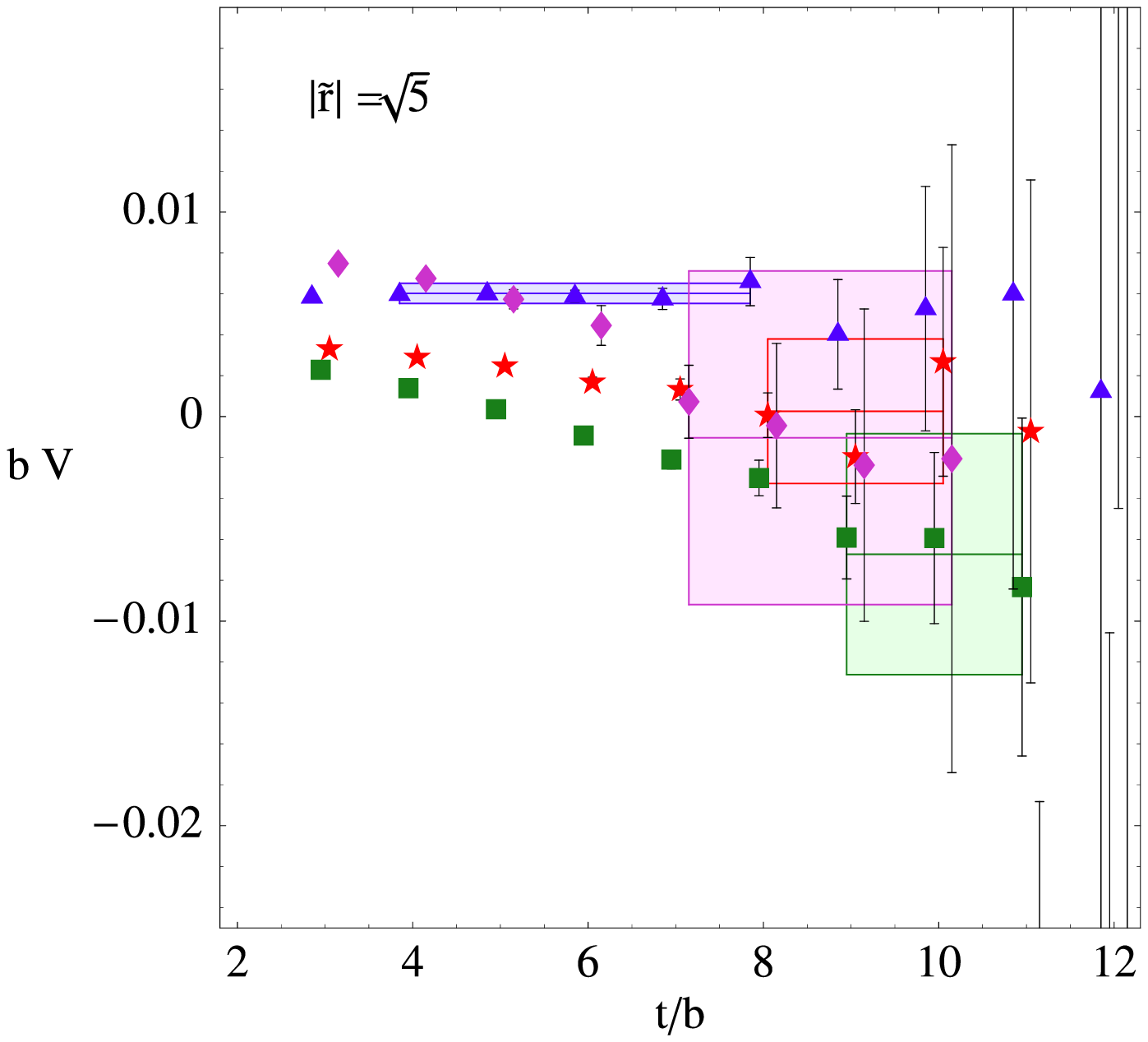}
  \caption{The effective mass plots for the potentials at displacements $\tilde{\bf
      r} = (1,1,0)$ and $(2,1,0)$ Red stars correspond to $V_\sigma$,
    green squares to $V_{\sigma\tau}$, blue triangles to $V_\tau$ and
    magenta diamonds to $V_{1}$.  Extracted masses and uncertainties
    are shown as the shaded regions in channels where a signal can be
    extracted.}
  \label{fig:effmassSQRTSRPES}
\end{figure}
\begin{figure}[htbp]
  \centering
  \includegraphics[width=1.0\columnwidth]{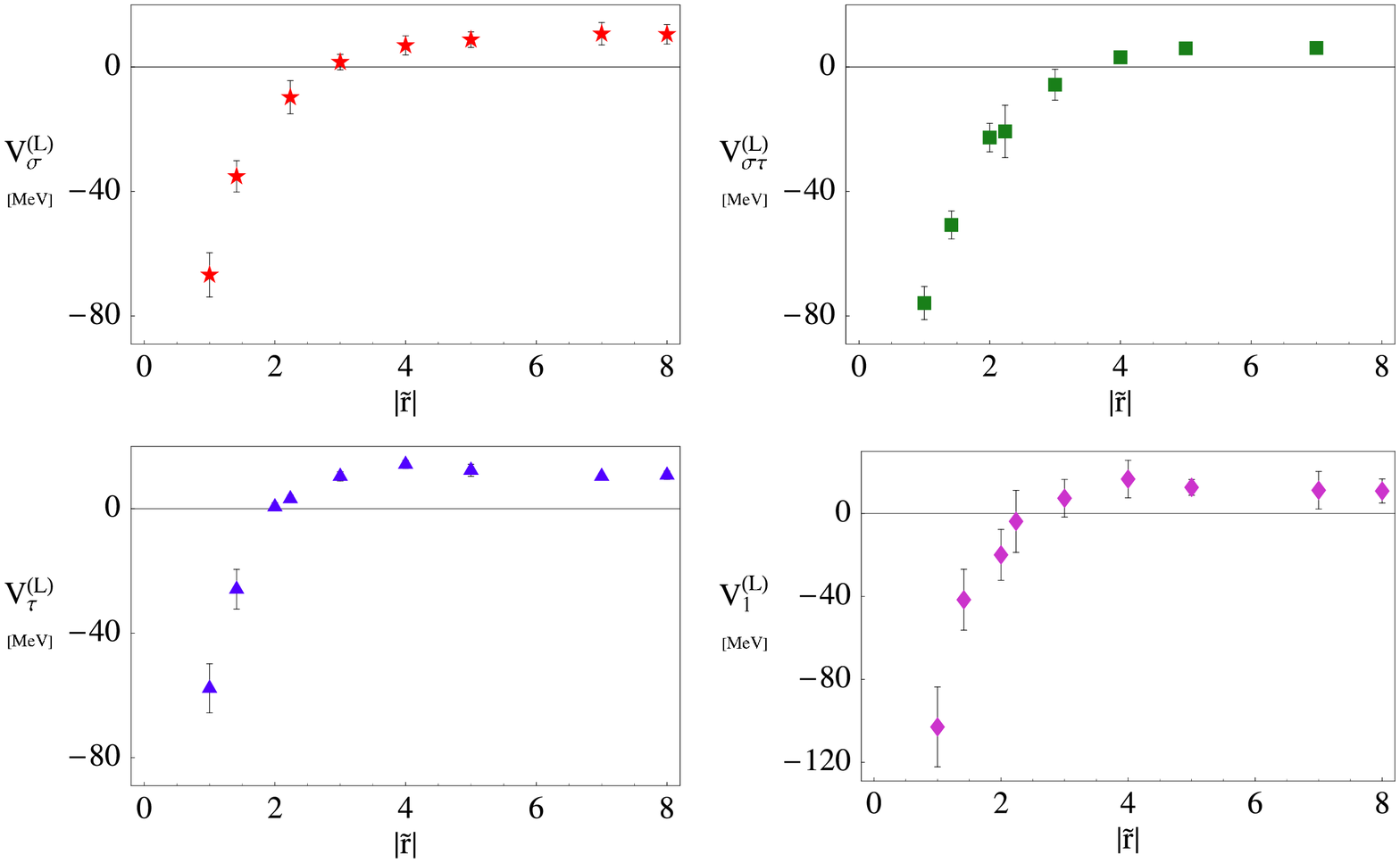}
  \caption{The central finite-volume potentials $V_{\sigma,\sigma
      \tau, \tau, 1}$, as defined in Eq.~(\protect\ref{eq:Vrels}) (the
    lattice potentials plus the leading lattice spacing corrections to
    OGE).  The statistical and systematic errors have been added in
    quadrature.  }
  \label{fig:Vtchannel}
\end{figure}
The finite-volume potentials calculated on the lattice, $V^{{\rm
    latt}(L)}_{ {\sigma},{\tau},{\sigma\tau},{1}} (\tilde{\bf r})$,
are given in Tables~\ref{tab:Vsig}-\ref{tab:Vone}, and are shown in
Fig.~\ref{fig:Vtchannellatt}.  Effective mass plots for these
potentials are shown in Figs. \ref{fig:effmassCRPES} and
\ref{fig:effmassSQRTSRPES}. The potentials corrected for the
finite-lattice spacing contributions to OGE between the heavy quarks
are also given in Tables~\ref{tab:Vsig}-\ref{tab:Vone}, and are shown
in Fig.~\ref{fig:Vtchannel}.  The uncertainties in these potentials
are seen to be significantly smaller than those of $V_{I,s_l}$, in
part due to the fact that the $B$-meson mass extraction, and its
associated uncertainty, does not contribute.  The most striking
potential is $V_{\tau}$; it is clear that this potential is of shorter
range than $V_{\sigma}$ and $V_{\sigma\bm\tau}$, due to the absence of
OPE and OHE.

%%%%%%%%%%%%%%%%%%%%%%%%%%%%%%%%%%%%%%
\subsection{Extrapolation to Infinite-Volume Potentials}
\label{sec:infin-volume-potent}

The final stage of analysis is to use the extracted finite-volume
potentials in either the $s$- or $t$-channels to determine the
infinite-volume forms. At short distances, $|{\bf r}|\alt
\Lambda_\chi^{-1}$ ($|\tilde{\bf r}|\alt2$ for our analysis), the
infinite-volume extrapolation must be done empirically, fitting
functions with the correct long-distance behavior to the results of
lattice calculations in multiple volumes. In principle, this
extrapolation can be performed systematically for larger $|{\bf r}|$
as effective field theory describes the potential in this regime.
Here we explore how the matching to EFT can be implemented, focusing
on the isovector potentials.

In QCD, the long range pieces of the infinite-volume, $t$-channel
isovector potentials are expected to have the form
\begin{eqnarray}
  \label{eq:Vtausigma}
  V_{\sigma\tau}^{(\infty)}({\bf r}) & \stackrel{|{\bf
      r}|\to\infty}{\longrightarrow} & \ \frac{g^2\ 
    m_\pi^2}{24\pi f_\pi^2} \ \frac{e^{-m_\pi
      |{\bf r}|}}{|{\bf r}|} \ +\ 
V_{\sigma\tau}^{(2\pi)}(|{\bf r}|) 
\ +\ \ldots
\end{eqnarray}
\begin{eqnarray}
  \label{eq:Vtau}
  V_{\tau}^{(\infty)}({\bf r}) &\stackrel{|{\bf r}|\to\infty}{\longrightarrow} &
\ \frac{g_\rho^2}{4\pi} \ 
\frac{e^{-m_\rho |{\bf r}|}}{|{\bf r}|} \ +\  
V_\tau^{(2\pi)}(|{\bf r}|) \ +\ \ldots
\end{eqnarray}
where $V_\tau^{(2\pi)}$ and $V_{\sigma\tau}^{(2\pi)}$ are the two pion
exchange potentials defined in Ref.~\cite{Kaiser:1997mw} (with nucleon
couplings replaced by the relevant couplings of the $B$ sector),
$f_\pi\sim132$~MeV, $g$ is the chiral coupling of pions to heavy
mesons occurring in the heavy meson chiral perturbation theory
Lagrangian
\cite{Burdman:1992gh,Wise:1992hn,Yan:1992gz,Booth:1994hx,Sharpe:1995qp},
and $g_\rho$ is a phenomenological $BB\rho$ coupling. The ellipses
denote contributions suppressed at large separations.

Ideally, lattice determinations of the meson masses and potentials at
long distances could be used to fit the couplings in the above
equations (the two pion contributions contain additional parameters).
However, a number of issues complicate this analysis. 
The quenched nature of our calculations introduces artefacts as in this
case, the $\eta^\prime$ meson remains degenerate with the pions but has
a modified propagator \cite{Bernard:1992mk,Sharpe:1992ft},
\begin{eqnarray}
  \label{eq:7}
  G_{\eta^\prime}(q^2) &=& \frac{i}{q^2-m_\pi^2+i\epsilon}
  +\frac{i(M_0^2-\alpha_\Phi q^2)}{(q^2-m_\pi^2+i\epsilon)^2}, 
\end{eqnarray}
($M_0$ and $\alpha_\Phi$ are couplings occurring in the quenched
chiral Lagrangian \cite{Bernard:1992mk,Sharpe:1992ft}) that produces
unphysical components of the potential. In particular, both of the
isovector $t$-channel potentials receive contributions from
one-pion--one-$\eta^\prime$ exchange that are longer range than the
two pion exchange contributions. These contributions are calculable,
but involve additional low energy constants.  Additional issues are
introduced by the unphysically large quark mass used in our
calculations with the identification of the dominant contribution in
$V_{\tau}$ depending on the quark mass. At the physical mass, single
$\rho$ exchange is sub-dominant to two pion exchange, however here,
$m_\rho<2m_\pi$ so it is $\rho$-exchange that persists to the longest
distance.  In our calculation, $2m_\pi\sim\Lambda_\chi$ and, in both
channels, the two Goldstone boson exchange contributions are
indistinguishable from short distance contributions not describable in
EFT.  Finally, the formula for the potential at finite-volume in
Eq.~(\ref{eq:finiteVolPot}) is valid only for single particle exchange
and is significantly modified if two or more particle exchange effects
are included at infinite-volume; the two particles can interact with
sources in different periodic copies.

\begin{figure}[!t]
  \centering
  \includegraphics[width=0.46\columnwidth]{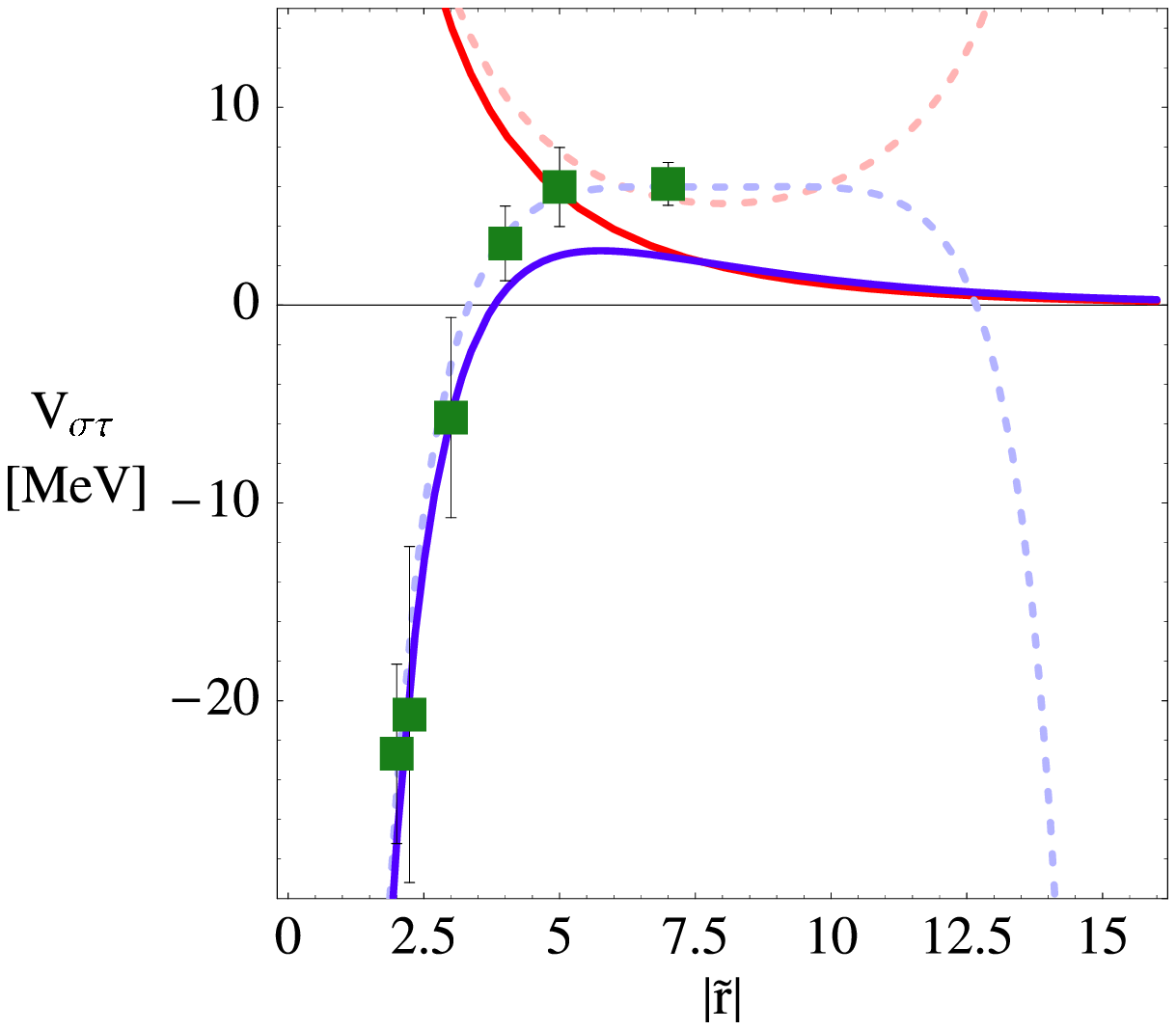}
\hspace{8mm}
  \includegraphics[width=0.46\columnwidth]{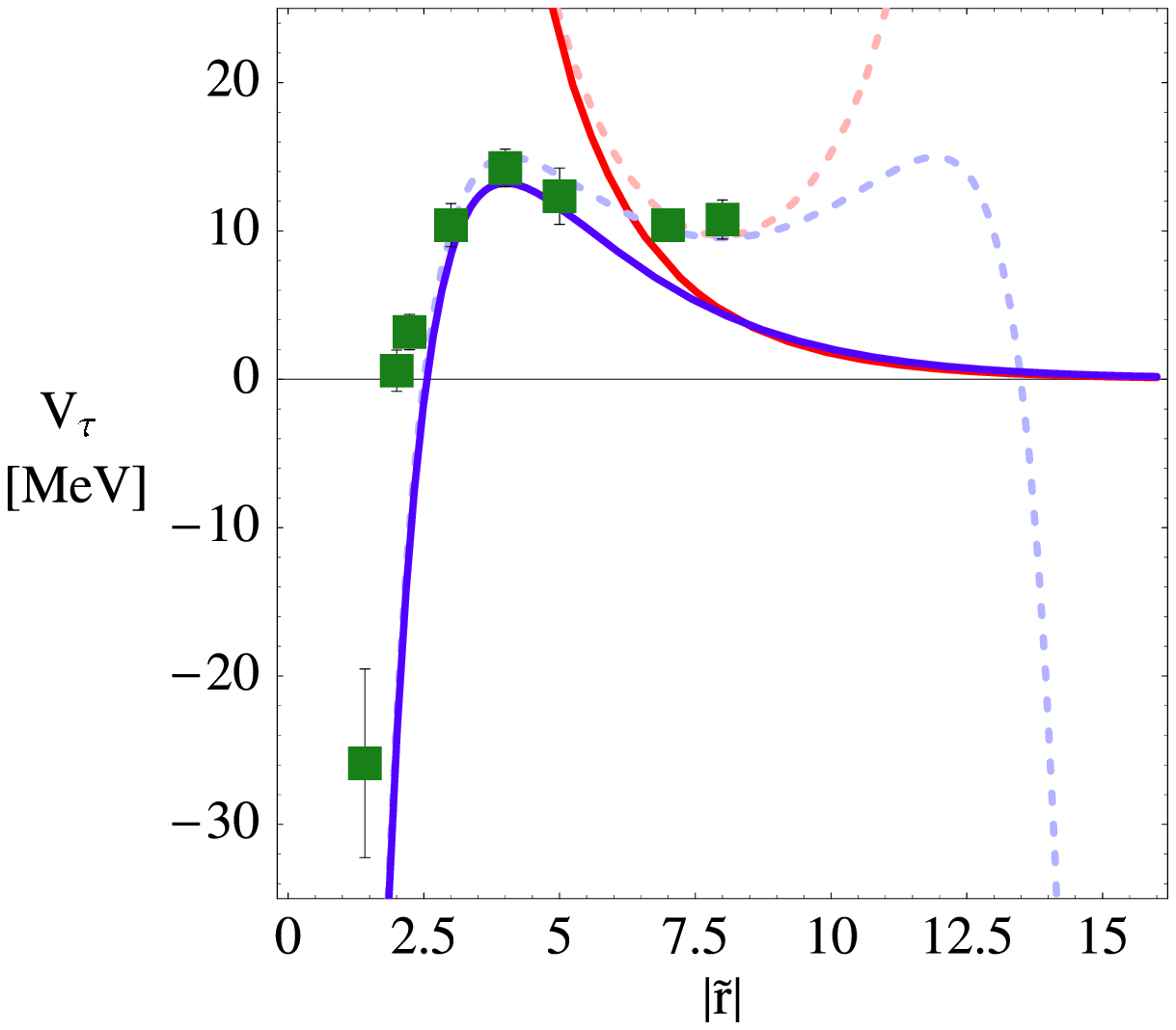}
  \caption{Fits to the finite-volume isovector $t$-channel
    potentials. The dashed lines correspond to the finite-volume fits
    to the lattice data, and the solid curves are the infinite-volume
    extrapolations.}
  \label{fig:FVfits}
\end{figure}

Since only the longest range contribution to the potential in each channel
can be identified, we fit our results at large separations, $|{\bf
  r}|>\Lambda_\chi^{-1}$, using the finite-volume versions (computed
using Eq.~(\ref{eq:finiteVolPot})) of the simplified infinite-volume
potentials,
\begin{eqnarray}
  \label{eq:Vtausigmasimp}
  V_{\sigma\tau}^{(\infty)}({\bf r}) &\stackrel{|{\bf
      r}|\to\infty}{\longrightarrow} & \ \frac{g^2\ 
    m_\pi^2}{24\pi f_\pi^2} \ \frac{e^{-m_\pi
      |{\bf r}|}}{|{\bf r}|} \ +\ 
\alpha^\prime_\chi \ \frac{e^{-\Lambda_\chi |{\bf r}|}}{|{\bf r}|}
\ \ \ ,
\end{eqnarray}
\begin{eqnarray}
  \label{eq:Vtausimp}
  V_{\tau}^{(\infty)}({\bf r}) &\stackrel{|{\bf r}|\to\infty}{\longrightarrow} &
\ \frac{g_\rho^2}{4\pi} \ \frac{e^{-m_\rho |{\bf r}|}}{|{\bf r}|} \ +\  
\alpha_\chi \ \frac{e^{-\Lambda_\chi |{\bf r}|}}{|{\bf r}|}
\ \ \ .
\end{eqnarray}
Using the measured values and uncertainties of $m_\pi$ and $m_\rho$
and the physical value of $f_\pi$ we first determine the couplings $g$
and $g_\rho$ by setting $\alpha_\chi=\alpha_\chi^\prime=0$ and fitting
the finite-volume potentials at the two largest
separations.\footnote{Simple fits using the infinite-volume long range
  behaviour were considered in Ref.~\cite{Michael:1999nq}.}  These
fits are shown by the dashed red curves in Fig.~\ref{fig:FVfits} and
the resulting couplings are found to be
\begin{eqnarray}
  \label{eq:8}
  g_\rho= 2.17\pm 0.08\,, &\quad\quad &
  g = 0.57\pm 0.06\,.
\end{eqnarray}
These couplings are stable under decreasing the minimum separation toward the
point
where the finite-volume potential crosses zero however the $\chi^2$ of
the fit worsens. Having determined these parameters, we reconstruct
the infinite-volume potentials that are shown in the figure as the
solid red lines.  

If the couplings $\alpha_\chi^{(\prime)}$ in
Eqs.(\ref{eq:Vtausigmasimp}) and (\ref{eq:Vtausimp}) are included as fit parameters, we
obtain instead
\begin{eqnarray}
  \label{eq:9}
  g_\rho= 3.02\pm 0.09\,, &\quad\quad &
  g = 0.69\pm 0.03\,,  
\end{eqnarray}
with the finite-volume fits and their infinite-volume reconstructions
shown as the dashed- and solid- blue curves in Fig.~\ref{fig:FVfits}.
In this case we have set the minimum separation, $r_{\rm min}$, used in
our fits to be $2b$ for $V_{\sigma\tau}$ and $3b$ for $V_\tau$ although
the fits vary only slightly under changes of $r_{\rm min}$ from $b$ to
$5b$. Averaging the two sets of extractions, we find
\begin{eqnarray}
  \label{eq:10}
  g_\rho= 2.6\pm 0.1\pm0.4\,, &\quad\quad &
  g = 0.63\pm 0.05\pm0.06\,, 
\end{eqnarray}
where the second error is an estimate of systematic errors determined
by differences between the two fits and variation of the fit range.
These numbers represent our best estimates of the couplings but we
caution that we are currently unable to investigate the full
systematics of this determination. Further refinement would require
lattice calculations at a range of different volumes, lattice spacings
and quark masses.

In both isovector channels, the agreement of the two infinite-volume
extractions at large separations suggests that the long range piece of
the extraction is robust. The pion coupling, $g$, is related to the
forward limit of $\langle B|\ j^a_{\mu 5}\ |B^\ast\rangle$, the matrix element of
the isovector axial-vector current,
through PCAC and the value we extract is consistent with
direct determinations of the quenched axial coupling: 0.42(4)(8)
\cite{deDivitiis:1998kj}, 0.69(18) \cite{Abada:2002xe}, 0.48(3)(11)
\cite{Abada:2003un}, 0.517(16) \cite{Negishi:2006sc}. We note that
extraction of this coupling from the potential does not require
renormalisation of axial current and suffers from different systematic
effects. Agreement between the two procedures is encouraging.

The isoscalar channels suffer from more severe unphysical artefacts in
quenched QCD and we are not able to extract meaningful information
from the long distance potentials. For $V_\sigma$, EFT predicts a long
range single Goldstone boson exchange potential \cite{Beane:2002nu}
\begin{eqnarray}
  \label{eq:11}
  V_\sigma^{(\infty)}(|{\bf r}|) 
&\stackrel{|{\bf
      r}|\to\infty}{\longrightarrow} & \frac{g_0^2\ 
    m_\pi^2}{24\pi f_\pi^2} \left[(1-\alpha_\Phi)\frac{e^{-m_\pi
      |{\bf r}|}}{|{\bf r}|} - \frac{M_0^2-\alpha_\Phi m_\pi^2}{2m_\pi}e^{-m_\pi
      |{\bf r}|} \right]
\end{eqnarray}
($g_0$ is the $\eta^\prime$ axial-coupling occurring in the quenched
heavy-meson chiral Lagrangian \cite{Booth:1994hx,Sharpe:1995qp}) with
a long distance exponential tail dominating. Unlike the other
channels, the suppression of the sub-leading contribution is not
exponential and our data is insufficient to resolve these pieces.
$V_1$ is not determined by single particle exchange (though many
phenomenological approaches in the nucleon sector include exchange of
the $\sigma(550)$ resonance) and in this channel our data are
particularly poor. In both isoscalar channels two-$\eta^\prime$
exchange is also present and enhanced compared to two-pion, and
one-$\eta^\prime$--one-pion- exchange, further polluting the signals.
Larger volumes and multiple quark masses will be needed to perform
extractions of couplings in these channels.

%%%%%%%%%%%%%%%%%%%%%%%%%%%%%%%%%%%%%%%%%%%%%%%%%%%%%%%%%
\section{Discussion}
\label{sec:resdisc}

\noindent 

We have studied the potentials between two $B$-mesons in the heavy
quark limit.  The calculations were performed on $16^3\times 32$
quenched lattices with a spatial length of $\sim 1.6~{\rm fm}$, and
with a quark mass such that $m_\pi\sim 400~{\rm MeV}$.  The leading
lattice space corrections to the one-gluon-exchange potential between
the two heavy quark propagators in the finite-volume were included in
order to extract the physical potential between $B$-mesons in the
continuum but at finite-volume.  We find clear evidence of repulsion
between the $B$-mesons in the $I\ne s_l$ channels and attraction in
the $I=s_l$ channels.  Three of the four potentials defined with
$t$-channel spin-isospin quantum numbers have significantly smaller
uncertainties than the potentials defined with $s$-channel quantum
numbers.  From the large separation behaviour of these potentials at
finite-volume, $B$-meson couplings to the $\pi$ and $\rho$ were
extracted.

This calculation can be improved in a number of areas but shows that a
rigorous first principles calculation of the $B$-meson potential is
achievable in the near future.  The next stage of our study will
progress from unphysical quenched QCD to fully dynamical QCD. This is
mandatory for connection to the real world but will also significantly
simplify the analysis of the long range potential using EFT. To separate
the different components of the potential in the short-, intermediate-
and long- range regimes requires multiple volumes and quark masses.
Finally calculations at a number of different lattice spacing are
required to control the remaining discretisation effects. Completion
of this ambitious program will provide deep insight into the $BB$
system and, ultimately, nuclei.

%%%%%%%%%%%%%%%%%%%%%%%%%%%%%%%%%%%%%%%%%%%%%%%%%%%%%%%%%%%%%%%
\acknowledgments

\noindent 

We would like to thank Silas Beane for his involvement in the initial
stages of this project.  We would like to thank the {\it Institute for
  Nuclear Theory} for kindly allowing us to use some of their
workstations to perform the contractions.  We also thank the computing
support group of the Departments of Physics and Astronomy at the
University of Washington, for installing and maintaining the {\it
  Deuteronomy} cluster with which the majority of this work was
performed.  We thank R.~Edwards for help with the QDP++/Chroma
programming environment~\cite{Edwards:2004sx} with which the
calculations discussed here were performed.  The work of WD and MJS is
supported in part by the U.S.~Dept.~of Energy under Grant
No.~DE-FG03-97ER4014.  The work of KO is supported in part by DOE
contract DE-AC05-06OR23177 under which Jefferson Science Associates,
LLC currently operates JLab.

\appendix

%%%%%%%%%%
%%%%%%%%%%%      Appendix     %%%%%%%%%%%%%%%%%%
%%%%%%%%%%%%%%%%%%%%%%%%%%%%%%%%%%%%%%%%%%%%%%%%%%
\section{Potentials from Lattice Wavefunctions}
\label{sec:ill-def-pot}

Recently, it has been claimed that the nucleon-nucleon potential can
be extracted from the lattice wavefunctions of two
nucleons~\cite{Ishii:2006ec}, extending the technique that CP-PACS has
successfully used to determine $I=2$ $\pi\pi$ scattering
parameters~\cite{Aoki:2005uf}.  In this appendix, we question the
validity of that calculation.

An interpolating field for the nucleon is,
\begin{eqnarray}
\hat {\cal O}_1({\bf x},t)_\alpha^i & = & \epsilon_{abc} \ q^{i,c}_\alpha \ 
\left( q^{a,T}C\gamma_5\tau_2 q^b \right) ({\bf x},t)
\ \ \ ,
\end{eqnarray}
where $i$ is a Dirac-index, $\alpha$ is an isospin index, and $a,b,c$
are color indices.  This operator has a non-zero overlap with a
nucleon momentum-eigenstate
\begin{eqnarray}
\langle 0|\hat {\cal O}_1({\bf 0},0)_\alpha^i |N_\beta^j({\bf p})\rangle
& = & Z_N({\bf p})\ \delta_{\alpha\beta}\ \delta^{ij} 
\ \ \ .
\end{eqnarray}
The two-nucleon correlation function measured on the lattice in
Ref.~\cite{Ishii:2006ec} is
\begin{eqnarray}
G_{NN}({\bf x},{\bf y},t) & = & \langle 0 | 
\hat {\cal O}_1({\bf x},t)_{\alpha}^{i} 
\hat {\cal O}_1({\bf y},t)_{\beta}^{j} 
\overline{J}(0) |0\rangle
\nonumber\\
& = & 
\sum_n\  \langle 0 | 
\hat {\cal O}_1({\bf x},0)_{\alpha}^{i} 
\hat {\cal O}_1({\bf y},0)_{\beta}^{j} 
|\psi_n\rangle
\langle\psi_n| \overline{J}(0) |0\rangle \ {e^{-E_n t}\over 2 E_n}
\ \ \ ,
\end{eqnarray}
where $\overline{J}$ is a wall-source on the initial time-slice
$t_0=0$, and $|\psi_n\rangle$ are the eigenstates of the Hamiltonian
in the finite-volume. In particular, $|\psi_n\rangle$ are states of
definite baryon number ($B=2$), isospin and transformation under the
hyper-cubic group. Setting $\langle\psi_n| \overline{J}(t_0) |0\rangle
= A_n(t_0)$, at long times the correlation function becomes
\begin{eqnarray}
G_{NN}({\bf x},{\bf y},t) & \rightarrow &
 A_0(0)\ \langle 0 | 
\hat {\cal O}_1({\bf x},0)_{\alpha}^{i} 
\hat {\cal O}_1({\bf y},0)_{\beta}^{j} 
|\psi_0\rangle\ {e^{-E_0 t}\over 2 E_0}
\ \ \ ,
\end{eqnarray}
and Ref.~\cite{Ishii:2006ec}  asserts that $\langle 0 |
\hat {\cal O}_1({\bf x},t_0)_{\alpha}^{i} \hat {\cal O}_1({\bf
  y},t_0)_{\beta}^{j} |\psi_0\rangle = \langle 0 | N({\bf
  x},t_0)_{\alpha}^{i} N({\bf y},t_0)_{\beta}^{j} |\psi_0\rangle\equiv
\Phi^{ij}_{\alpha\beta}$ is the Bethe-Salpeter wavefunction (their
Eq.~(4)).  From this definition, they generate a nucleon-nucleon
potential via
\begin{eqnarray}
V(r) & = & E \ +\ {1\over 2 \mu} {\nabla^2 G_{NN}\over G_{NN}}
\ =\ E \ +\ {1\over 2 \mu} {\nabla^2 \Phi_{NN}\over \Phi_{NN}}
\ \ \ ,
\end{eqnarray}
where $\Phi_{NN}$ is the projection of the Bethe-Salpeter wavefunction
onto definite isospin and transformation under the hyper-cubic group.
$\mu$ is the reduced mass of the two-nucleon system.  However, this
identification of the Bethe-Salpeter wavefunction is incorrect and the
most general form for the matrix element is
\begin{eqnarray}
\langle 0 | \hat {\cal O}_1({\bf x},t_0)_{\alpha}^{i} 
\hat {\cal O}_1({\bf y},t_0)_{\beta}^{j} 
|\psi_0\rangle 
& = &  
Z_{NN}^{(S,I)}(|{\bf r}|)\ 
\langle 0 | N({\bf x},t_0)_{\alpha}^{i} N({\bf y},t_0)_{\beta}^{j}
|\psi_0\rangle
+\ldots
\, ,
\label{eq:full}
\end{eqnarray}
where $Z_{NN}^{(S,I)}(|{\bf r}|)$ is an unknown function that depends
upon the spin, isospin and structure of the composite sink, $\hat
{\cal O}_1({\bf x},t_0)_{\alpha}^{i} \hat {\cal O}_1({\bf
  y},t_0)_{\beta}^{j}$, and where ${\bf r}={\bf x}-{\bf y}$. The
ellipsis denotes additional contributions from the tower of states of
the same global quantum numbers. With this complete form of the matrix
element, it is not possible to determine the potential from $G_{NN}$
without additional information. In the limit $|{\bf r}|\to\infty$,
$Z_{NN}^{(S,I)}(|{\bf r}|)\to 1$ and the additional terms in
Eq.~(\ref{eq:full}) containing $p>2$ particles are suppressed.
Consequently the scattering parameters can be rigorously extracted as
has been done for the case of the pion~\cite{Aoki:2005uf}.

%%%%%%%%%%%%%%%%%%%%%%%%%%%%%%%%%%%%%
\section{Finite Lattice Spacing Correction to the Potential}
\label{ap:A}

The finite lattice spacing correction to the 
potential (in the $\overline{\bf 3}$ color channel) is given by,
\begin{eqnarray}
\delta V_{QQ;\overline{\bf 3}}^{(L)}({\bf r}) 
& =  & 
-{\overline{\alpha}(\mu)\over 3\pi^2 b}\ 
\left({2\pi\over \tilde L}\right)^3\ 
\left[\
 \left({\tilde L\over 2\pi}\right)^2 
\sum_{\bf n}^\infty\ \ \frac{e^{i 2 \pi {\bf n}\cdot \tilde {\bf r}/\tilde L}}
{|{\bf n}|^2}
\ -\ 
\sum_{{\bf n}}^{|n_i|\le {\tilde L\over 2}}
\ e^{i 2 \pi {\bf n}\cdot \tilde {\bf r}/\tilde L}\ 
G_{00}\left(\hat n_x,\hat n_y,\hat n_z,0\right)
\ \right]
\,,
\nonumber\\
\label{eq:finitevolC}
\end{eqnarray}
arising from the difference between continuum and lattice one gluon
exchange evaluated at finite-volume. 

The lattice contribution to this expression is simple to evaluate for
the DBW2 action (the full form of the improved gluon propagator is
given in Ref.~\cite{Weisz:1983bn}), however calculating the continuum
contribution is somewhat subtle. Difficulties arise in both the
infrared and ultraviolet regimes. Both the lattice and continuum
finite-volume sums are IR divergent, however provided both are
regulated in the same way a sensible result ensues; the simplest
procedure is to omit the zero-mode\footnote{Any regularization is
  equally valid as the mode expansion of the perturbative gluon
  propagator is intrinsically ill-defined in the IR region.
  Differences in IR regularization lead to ${\cal O}(b^2)$ differences
  in the perturbative corrections to the potentials, parametrically
  smaller than the effects of the Wilson fermion discretisation used
  herein.}.

While the continuum contributions to ${\cal A}$ and ${\cal B}$,
defined in Eq.~(\ref{eq:finitevolBLM}), are strictly UV convergent,
that convergence is highly oscillatory.  Computing these contributions
is simplified by the use of the Poisson summation formula which allows
the sum to be rewritten as
\begin{eqnarray}
  \label{eq:1}
  \sum_{{\bf n}\ne0} \frac{e^{2\pi i {\bf n}\cdot\tilde{\bf
        r}/\tilde{L}}}{|{\bf n}|^2}
&=&
x^2\sum_{{\bf n}\ne0} \frac{e^{2\pi i {\bf n}\cdot\tilde{\bf
        r}/\tilde{L}}}{|{\bf n}|^2(|{\bf n}|^2+x^2)}
+ \sum_{{\bf n}\ne0} \frac{e^{2\pi i {\bf n}\cdot\tilde{\bf
        r}/\tilde{L}}}{(|{\bf n}|^2+x^2)}
\nonumber\\
&=&
x^2\sum_{{\bf n}\ne0} \frac{e^{2\pi i {\bf n}\cdot\tilde{\bf
        r}/\tilde{L}}}{|{\bf n}|^2(|{\bf n}|^2+x^2)}
-\frac{1}{x^2} + \frac{L\pi}{|\tilde{{\bf r}}|} e^{-2\pi x
  |\tilde{{\bf r}}|/L} + \pi \sum_{{\bf m}\ne0} \frac{e^{-2\pi
    |{\bf m}+\tilde{{\bf r}}/\tilde{L}| x}}
{|{\bf m}+\tilde{{\bf r}}/\tilde{L}|}
\ \ \ ,
\end{eqnarray}
which is independent of the value of $x$.  The sums on the rhs of this
expression are more convergent than that on the lhs and can be
numerically evaluated reliably.  Similar techniques allowed us to deal
with the analogous differences defining the function ${\cal B}$.

In the limit that $|\tilde{\bf r}| \to 0$, the continuum contribution
is singular, leading to a correction factor of
\begin{eqnarray}
\delta V_{QQ;\overline{\bf 3}}^{(L)}({\bf r}) 
&\rightarrow &
-{2\ \overline{\alpha}(\mu)\over 3 \ r}\ +\ ...
\ \ \ ,
\end{eqnarray}
which is nothing other than the strong Coulomb interaction between the
heavy quarks in the continuum.  In the continuum limit and
infinite-volume limit, $b \ll r \ll L$, the leading correction factor
is found to be
\begin{eqnarray}
  \label{eq:2}
\delta V_{QQ;\overline{\bf 3}}^{(L)}({\bf r}) 
&\rightarrow &
-\frac{\overline{\alpha}(r^{-1})\ b^2}{ 6 \ r^3}
\left(1+12c_1-12c_1^3\right)\ + \ \ldots
\,,
\end{eqnarray}
where we have used the BLM procedure to set the scale.  This improved
perturbative shift can be eliminated for suitable choices of $c_1$.
Clearly, the L\"uscher-Weisz-improved value of $c_1={1\over 12}$ maximally
improves the lattice calculation.  That is to say that, neglecting the
small $c_1^3$ contribution, the correction factor that must be applied
to the lattice calculation in order to recover the continuum
potentials is minimized by L\"uscher-Weisz-improvement.

%%%%%%%%%%%%%%%%%%%%%%%%%%%%%%%%

\end{document}